\documentclass[preprint2]{aastex}

\shorttitle{Calibration of the LAXPC}
\shortauthors{}

\begin{document}

\title{Calibration of the Large Area X-ray Proportional Counter (LAXPC) instrument on-board AstroSat}


\author{}


\author{H. M. Antia \altaffilmark{1}, J. S. Yadav \altaffilmark{1},
P. C. Agrawal \altaffilmark{2}, Jai Verdhan Chauhan \altaffilmark{1},
R. K. Manchanda \altaffilmark{3}, Varsha Chitnis \altaffilmark{1},
Biswajit Paul \altaffilmark{4}, Dhiraj Dedhia \altaffilmark{1},
Parag Shah \altaffilmark{1}, V. M. Gujar \altaffilmark{1},
Tilak Katoch \altaffilmark{1}, V. N. Kurhade \altaffilmark{1},
Pankaj Madhwani \altaffilmark{1}, T. K. Manojkumar \altaffilmark{1},
V. A. Nikam \altaffilmark{1}, A. S. Pandya \altaffilmark{1},
J. V. Parmar \altaffilmark{1}, D. M. Pawar \altaffilmark{1},
Mayukh Pahari \altaffilmark{5},
Ranjeev Misra \altaffilmark{5}, K. H. Navalgund \altaffilmark{6},
R. Pandiyan \altaffilmark{6}, K. S. Sharma \altaffilmark{6},
K. Subbarao \altaffilmark{6}}
\affil{$^1$ Tata Institute of Fundamental Research, Homi Bhabha Road, Mumbai 400005, India}
\affil{$^2$ UM-DAE Centre of Excellence for Basic Sciences, University of Mumbai, Kalina, Mumbai 400098, India} 
\affil{$^3$ University of Mumbai, Kalina, Mumbai 400098, India}
\affil{$^4$ Department of Astronomy \& Astrophysics, Raman Research Institute,  Bengaluru 560080 India}
\affil{$^5$ Inter-University Centre for Astronomy and Astrophysics, Pune 411007, India}
\affil{$^6$ ISRO Satellite Centre, HAL Airport Road, Bengaluru 560017, India}




\begin{abstract}
We present the calibration and background model for the Large Area X-ray
Proportional Counter (LAXPC) detectors on-board AstroSat. LAXPC instrument
has three nominally identical detectors to achieve large collecting area.
These detectors are independent of each other and in the event analysis
mode, they record the arrival time and energy of each photon that is
detected. The detectors have a time-resolution of 10 $\mu$s and a dead-time
of about 42 $\mu$s. This makes LAXPC ideal for timing studies.
The energy resolution
and peak channel to energy mapping were obtained from calibration on ground
using radioactive sources coupled with GEANT4 simulations of the detectors.
The response matrix was further refined from observations of the Crab after launch.
At around 20 keV the energy resolution of detector
is 10--15\%, while the combined effective area of the 3 detectors is
about 6000 cm$^2$. 
\end{abstract}


\keywords{Instrumentation: detectors; Space vehicles: instruments}

\section{Introduction}

The Large Area X-ray Proportional Counter (LAXPC) instrument aboard the
Indian Astronomy mission AstroSat consists of 3 co-aligned large area proportional counter
units for X-ray timing and spectral studies over an energy range of
3--80 keV \citep{agr06,yad16a}. AstroSat was launched on September 28, 2015 with
5 major astronomy payloads \citep{agr06,sin14}. Apart from LAXPC, these are the Ultra Violet
Imaging Telescope (UVIT), Soft X-ray Telescope (SXT), Cadmium-Zinc-Telluride
Imager (CZTI) and the Scanning Sky Monitor (SSM). The first four
instruments are co-aligned so as to point to the same source.
LAXPC instrument is described in detail by \citet{agr17}.
The LAXPC detectors have a collimator
with about $1^\circ\times1^\circ$ field of view. Each LAXPC detector is independent and
can operate in event analysis mode where time of arrival of each photon is recorded
to a time resolution of 10 $\mu$s giving an unprecedented sensitivity
to a wide variety of timing phenomena.
The detectors are filled with Xenon-methane mixture at 2 atmospheric pressure and detection
volume has a depth of 15 cm, which gives large sensitivity at energies
up to 80 keV.

This paper presents details of calibration for LAXPC detectors. The preliminary
calibration for the energy scale and energy resolution was carried out in
thermovac chamber on ground using radioactive sources. These observations
were compared with GEANT4 \citep{geant03} simulation of the detectors to obtain preliminary
response matrix, as well as background and field of view. These calibrations
were refined by observations of Crab and other known astronomical calibrators
after launch.

The rest of the paper is organized as follows: Section 2 gives an overview
of the instrument. Section 3 describes ground calibration for energy resolution
and energy scale. Section 4 describes the GEANT4 simulations and resulting
response matrix. Section 5 describes on-orbit calibration using Crab and Cas A
sources. Section 6 describes attempts to characterize the background.
Section 7 describes the long term performance of LAXPC detectors in orbit.
Section 8 gives the summary of calibration.

\section{The LAXPC detectors}

\begin{figure}
\epsscale{.95}
\plotone{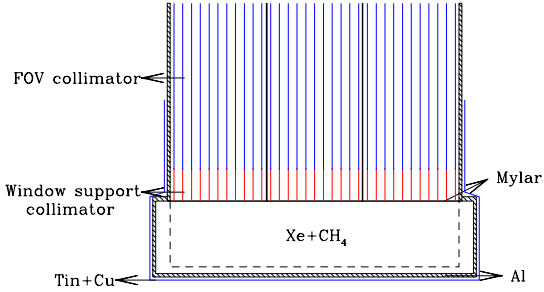}
\caption{Schematic diagram showing the LAXPC detector. The dashed lines
in the chamber mark the active volume of the detector covering the main
anodes.}
\label{fig:detec}
\end{figure}

\begin{figure}
\epsscale{.95}
\plotone{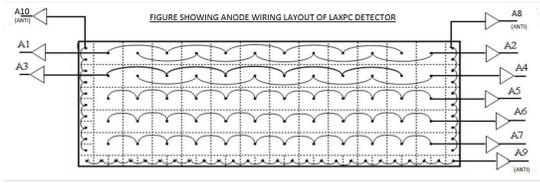}
\caption{Configuration of anodes in LAXPC detectors. The figure shows projection
on a plane perpendicular to the length of anode cells ($39\times16.5$ cm).
The cells in main-anodes are labelled as C1 to C12 from right to left in the
figure.}
\label{fig:anode}
\end{figure}

The LAXPC instrument consists of 3 nominally identical proportional
counters each with a geometric collection area of $100\times36$ cm$^2$.
The detectors are labelled as LAXPC10, LAXPC20, LAXPC30 (LX10, LX20, LX30 in brief).
Figure~\ref{fig:detec} shows a schematic diagram of the LAXPC detector
including the collimators and shield.
Each detector has 5 anode layers, each of 12 anode cells of size
$100\times3\times3$ cm. The top two layers are divided into 2 parts with
alternate cells connected. These gives 7 main anodes, A1, A2 in the top
layer, A3, A4 in the second layer from the top and A5, A6, A7 in the
remaining three layers. The main anodes are surrounded on 3 sides with veto
cells of thickness 1.5 cm. The veto-anode A9 covers the bottom area
($100\times 39$ cm), while veto-anodes A8, A10 cover the two sides
($100\times 15$ cm). The configuration of anodes is shown in Fig.~\ref{fig:anode}.
The two sides ($36\times15$ cm) perpendicular to the length of the cells do not have any veto
anode.
The cells are labelled as C1, C2, $\ldots$, C11, C12 from right to left in
the figure. Thus the cell C1 is adjacent to the veto-anode A8, while the cell
C12 is adjacent to the veto-anode A10.

The entire volume of the detectors, LAXPC10 and LAXPC20 is filled with a mixture
of Xenon (90\%) and Methane (10\%) at a pressure of about 2 atmospheres. The
LAXPC30 has a mixture of Xenon (84.4\%), Methane (9.4\%) and Argon (6.2\%)
at a pressure of about 2 atmospheres. The reason for using a different
gas mixture in LAXPC30 was that after filling the first two detectors it was
noticed that the energy resolution at 60 keV was about 13\% with the high
voltage employed to get the required energy range. Hence it was decided to
add a small amount of Argon in LAXPC30 as that is expected to improve the energy
resolution \citep{rao87}.
The low energy threshold is kept at
about 3 keV. The top of the detector is sealed by a 50 $\mu$m thick aluminised Mylar
window which allows X-rays with energy $>3$ keV to pass through and dictates
the low energy threshold of the detectors. Only events with larger energy
can trigger the electronics. The Mylar window only covers the active
detector volume. The veto-anodes on the two sides are covered by aluminium
box and shield and are not expected to get incident photons coming from
the top.

The LAXPC detectors are similar to the Rossi X-ray Timing Explorer,
Proportional Counter Array (RXTE-PCA) which was launched in 1995
\citep{jah06}.
The main differences are the increased pressure and depth of the detector,
which gives high effective area at higher energies ($>30$ keV). At the same time a larger
detector volume increases the background in LAXPC, thus making it difficult
to study faint sources. Higher pressure in LAXPC also degrades the energy
resolution to some extent. Another major difference is the availability of
event mode data which gives the arrival time and energy of each photon
that is detected.
This allows a detailed timing studies for all sources that are observed.
Apart from these, there are differences in the
nature and distribution of veto layers which may also contribute to differences
in the background.

An ADC converts the signal to 10 bits digital form, giving
1024 energy channels. Considering an energy resolution of about 10\%, the
number of channels can be reduced to 512 or 256 by combining 2 or 4 consecutive
channels. This appears to be desirable as the output shows some fluctuations
between even and odd channels. These fluctuations are of a few percent
in magnitude and the ratio of counts in consecutive channels
are independent of time. These can be seen in residuals after fitting the
spectrum.
For LAXPC10 and LAXPC30, 512 channels are
adequate, while for LAXPC20 we need to go down to 256 channels.
This truncated range is implemented in the
LaxpcSoft software, while other level-2 pipeline
software\footnote{http://www.tifr.res.in/\~{}astrosat\_laxpc/software.html}
produces 1024 channel output for all detectors.

The Mylar window is supported against the gas pressure by a 
collimator of height 7.5 cm made of square geometry aluminium cells, termed as the
window support collimator. The field of view collimator of height 37 cm is
placed above the window support collimator. Both these
collimators have aligned gaps
with a pitch of 7.0 mm. The field of view collimator has a tin sheet
sandwiched between Copper and Aluminium using epoxy. This gives a field of
view of about $0.9^\circ\times0.9^\circ$.
Since aluminium in
window support collimator is almost transparent to higher energy X-rays,
The field of view increases at high energies as only the field of view
collimator of length 37 cm is effective.
The detector is
covered by a tin shield of thickness 1 mm, coated with Copper (50 $\mu$m) on
five sides.
On the sides of the detector the cover extends from the bottom to 23 cm above the top of the detector
thus covering the lower part of the collimator housing.
This shield is effective in cutting off the background from low energy
photons and charged particles.

The outputs from the 7 main anodes and 3 veto anodes are fed to 10 Charge
Sensitive Pre-Amplifiers (CSPAs).
The high voltage is supplied to all main anodes (A1--A7) at the same point
and its value is, therefore, the same. The veto anodes have a lower voltage
but it cannot be controlled separately.
The overall high voltage can be adjusted from ground in
appropriate steps. But the high voltage of an individual anode cannot be
adjusted. Hence, the gain equalization is achieved by varying the gain of
the corresponding CSPA of the anode. This
gain equalization is done on ground by matching the 30 keV Xe K-escape peak in
background or Am$^{241}$ source for each individual anode. The CSPA gain
cannot be adjusted after launch. The high voltage is switched off when
the satellite is passing through the South Atlantic Anomaly (SAA) region.

If more than one anode is triggered within the time resolution, the signal
is simultaneously recorded in all these anodes.
In order to reject background events the processing electronics is designed
to reject any event that satisfies the following criteria:
\begin{enumerate}
\item Any event that triggers any of the veto anodes (A8--A10).
\item Any event that deposits more than the upper limit of about 80 keV
in any anode.
\item Any event that is recorded in more than two main anodes (A1--A7).
\item Any event that is recorded in two main anodes (A1--A7) and the energy
in both anodes is not in the Xenon K X-rays range covering about 25--35 keV.
In this case if energy in any one or both anodes is in the K X-ray range
then the event is accepted, energies in the two anodes are added and the
event is recorded as a single event of combined energy.
\end{enumerate}
This logic is found to be effective in reducing the background by
about 99\% (Section 4.4).

Thus the events that are accepted are either single-events where only one
main anode is triggered or double-events where two main anodes are triggered
but energy in at least, one of them is in the Xe K X-ray range.
In both these cases, the energy deposited in these anodes should be below
the upper energy threshold. Since the threshold is applied to each anode
separately, the double-events can exceed the energy threshold by up to
about 35 keV. Further, since the detector response is not linear, the
double-events may not be recorded at the same channel as that of a single-event with
the same total energy. The lower and upper threshold for Xe K X-ray range can be
set remotely for each detector, but it applies to all anodes in a detector.

The energy resolution of detector can degrade with time if impurities
accumulate in the gas. On ground this could happen by diffusion through
Mylar window, but in the vacuum in space the rate of accumulation of
impurities would be small. To take care of impurities, a gas purification
system is included in the detector, which may be used from time to time
to maintain the energy resolution close to the optimum value.

A System Time Base Generator (STBG) provides a stable and accurate time reference
for each of the three LAXPC detectors. The time reference has a resolution
of 10 $\mu$s, but has a slow drift. The STBG time is correlated to UTC time
from an on-board SPS time to correct for the drift which is of the order of
1 part in $10^5$ or about 1 sec in a day. Extensive tests were carried out
on ground and the corrected time is found to be
satisfactory to the required accuracy even if SPS time is not available for
an hour. In practice, SPS time is almost continuously available, thus
giving the required timing stability. The STBG also provides time reference
for other science payloads on AstroSat.

The LAXPC detectors have two main modes of operation, the Event Analysis (EA) mode
and the Fast Counter (FC) mode. In the event analysis mode, which is the default, the timing of each event is
recorded along with the information about the anode ID where it is recorded
as well as the energy (Channel No.). Each event gives rise to 5 bytes of data.
In this mode the dead-time of the detector
is estimated to be about 42 $\mu$s \citep{yad16b}. The LAXPC processing electronics is
a non-paralysable system, so an event occurring during the dead time since the
previous event is simply lost. Thus with an increasing event rate it will
reach a saturating rate equal to the inverse of the dead-time. The dead-time was
measured on ground using an X-ray gun to give high count rate. The measured
count rate from the processing electronics, was compared with that obtained
from a commercial MCA (MCA8000A) which gives the dead-time corrected count
rate. The dead-time can also be measured by taking a Fourier transform of
time series for a bright source, which would give a broad peak close to the inverse of the dead-time
in the power spectrum. This can be compared with that expected
for a dead-time corrected Poisson level power \citep{zh95}.
This yields an estimate of dead-time of around 42 $\mu$s \citep{yad16b}.
It is also possible to estimate the dead-time by comparing the observed
count rate and the count rate estimated from the slope of $\delta t$
plot which shows the distribution of time-interval between two
consecutive events, which should show an exponential behaviour for
events with Poisson distribution. This technique also gives a value of
about 42 $\mu$s.
There is also a smaller dead-time of about 35 $\mu$s associated with rejected events, e.g., those which exceed the
Upper Level Discrimination (ULD) threshold or the ones that trigger the veto anode or those that trigger more than two anodes simultaneously. 

The event mode operation also generates a Broad-Band Counting (BBC) mode data
simultaneously. These data contain counts in a predefined time-bin in
four different energy bins in three layers, as well as some other counts for diagnostic
purpose. The four energy bins are 3--6 keV, 6--18 keV, 18--40 keV and
40--80 keV. The layer 1 combines anodes A1 and A2, the layer 2 combines anodes
A3 and A4, while the layer 3 combines A5, A6 and A7.
The time-bin can be set from 16 ms to 2048 ms in steps increasing
by a factor of 2, with a default value of 128 ms. Since these counters are 10 or 11
bits deep, they may overflow for bright sources if a time bin size
towards the long end of the allowed range in selected.
For example, for the Crab the counters may overflow if
time-bin greater than 256 ms is used. It may be noted that in all cases
the event mode data is always available to get the correct light curve
in any energy and time bin.
Apart from genuine events
the BBC mode also has counters for rejected events, e.g., those exceeding
the ULD in the main anodes (A1--A7), or those that trigger veto anodes.

In the FC mode the counts in only the top
layer in a fixed time-bin of 160 $\mu$s are recorded in 4 energy bins.
In this mode the rejection of events using veto anodes and mutual coincidence
is suppressed and all events registered in anodes A1 or A2 are counted.
The approximate energy
bins are 3--6 keV, 6--8 keV, 8--12 keV, and 12--20 keV. These counters are 8 bit deep.
This mode may be useful for bright
or flaring sources where the count rate is very high. In this case the
dead-time of the detector is about 10 $\mu$s and hence much higher count
rates can be recorded.

Apart form these, there is also the Anti-Bypass (AB) mode where the rejection
of events through veto anodes is suppressed. This is used for on-board
calibration using a radioactive Am$^{241}$ source which is placed in the
veto anode A8. This is used to check the energy resolution and shift in gain.
Even in default mode a fraction ($1/128$) of events in veto anodes are recorded and can be
used for checking the energy resolution and gain, if sufficiently long
observation is available.
If the resolution degrades then we can perform purification of detector gas, while the shift
in gain can be compensated by a change in high voltage. It is important to
keep the gain close to the ground setting to ensure that the double-events
are handled correctly. Energy spectrum of background or faint source also
shows a peak around 30 keV due to Xe K fluorescence X-rays, which can also
be used to monitor shift in gain.

\section{Calibration on ground}

To obtain the energy resolution and gain of detectors, three radioactive
sources at different energies in the range of LAXPC detectors were used.
The calibration was performed in a thermovac chamber and measurements
were repeated at three temperatures of $10^\circ$ C, $20^\circ$ C and $30^\circ$ C
to study the temperature dependence of detector response. The following
3 sources were used:
\begin{enumerate}
\item Fe$^{55}$ with energy of 5.9 keV: These X-rays are absorbed in the
top layer and hence only the two top anodes A1 and A2 register these events.
Hence, the anodes A3--A7 are only calibrated using energy beyond 20 keV.

\item Cd$^{109}$ with energies of 22.1 keV (54.5\%), 21.9 keV (28.8\%),
24.9 keV (13.7\%) and 88 keV (3.0\%): The first two peaks cannot be
resolved by the detector, while the third one gives a small feature
at the high end of the main peak, which can be fitted with some effort.
The last peak is beyond the range of the detector, though because of
finite resolution a small low energy tail of the peak, as well as the
contribution from double events, or when the Xe K fluorescence X-ray escapes the
detector should, in principle, be detected. However, the resulting peak
is too weak and was not detected. At 22 keV about 50\% of photons are absorbed
in the top layer, but there are significant counts in all layers.
The ratio of counts in different layers is determined by the gas density
and we could use this to estimate the density in each detector, which
is required to generate the detector response.

\item Am$^{241}$ with energy of 59.6 keV: This source also gives multiple
peaks because of loss of energy due to Xe K fluorescence X-rays escaping the detector.
The detector logic is built to add contributions in two anodes if at least one 
of them is in the range of Xe K fluorescence X-rays (25--35 keV). Hence, we get
additional peaks at 29.8 keV (due to escape of Xe K$_\alpha$) and 25.2--26.0 keV (due to
escape of Xe K$_\beta$). The Xe K$_\alpha$ X-rays account for about 81\% of Xe K X-rays,
thus giving a dominant peak at 29.8 keV. The second peak due to K$_\beta$ X-rays
is barely resolvable with the detector resolution.
Apart from these, the 59.6 keV peak is also split into two
parts, one coming from single events where all energy is deposited in
a single anode, and the second coming from double events where the energy
is split between two anodes and the contribution is added by the
detector logic. Since the detector gain is not strictly linear, but has
a small quadratic term the channel number of a double events does not coincide
with that for a single event, but the two peaks cannot be resolved properly,
though they can be fitted with some effort. The contribution of single and
double events can be separated by analyzing the single and double events
separately. Since the quadratic term is negative
the single event peak occurs at lower channel as compared to that from
double events. The energy resolution of the two peaks are also different.
The energy resolution of the double peak is determined by energy resolution
at 29.8 keV as dominant component is from two 29.8 keV components. Thus
the absolute energy resolution of 59.6 keV double event peak is $\sqrt{2}$
times that for 29.8 keV and the relative energy resolution is $1/\sqrt{2}$
times that for 29.8 keV. On the other hand, the resolution of peak due
to single events is much worse as compared to that for double events.
By fitting these two peaks around 59.6 keV, it is possible to estimate
the quadratic term in the energy to channel number mapping.
\end{enumerate}

To estimate the detector background, the counts are recorded without any
source, before and after the source measurements are done. The background
count rate is then subtracted from that for the source to get the
contribution from the source.

In order to perform the calibration inside the thermovac chamber, an $x$--$y$
motion platform for movement of radioactive sources above the field of view
collimator was designed. This can hold all three sources and expose one or
more of them at a time. The $x$-$y$ motion platform can function in
vacuum and its movements and source on/off status can be remotely
controlled. The movement of sources is controlled by a program which
moves the source with a predetermined pattern in $x$ or $y$ directions.
The pattern was chosen to cover the entire area of the detector over
about 2 hours. To study behavior of each cell, in some cases the
source was moved along the length of each of the 12 cells.

To check the long term stability of detectors, periodic measurements were
made on ground over a period of several months to check for drift in peak
channel position. A slow decrease in peak channel with time was found due
to accumulation of impurities in gas. The original position was restored
when the gas was purified. Similar monitoring is continuing after launch.
All detectors were filled with gas by 2011 and no leak was noticed
before launch. This will put a limit of less than 0.5\% per year for
the leak rate before launch for all three detectors.

Apart from energy resolution, the timing characteristics of the detectors
were also tested on ground. Fourier transform of the time series did not
show any peak at frequencies less than 2 kHz. When strong source from
X-ray gun was used the Fourier transform showed peak around 15 kHz which
is expected from the dead-time effect. This is confirmed from the
observations in orbit \citep{yad16b}.
The distribution of the time interval between two consecutive events in the
LAXPC detectors follow the expected exponential behavior for Poisson distribution.
It also shows a cutoff at 50 $\mu$s, in agreement with the measured dead time.

The operation of Fast Counter (FC) mode was also tested on ground using radioactive
sources as well as X-ray gun, and again
no peaks were found in the Fourier transform up to the Nyquist frequency
of 3125 Hz. The nominal boundary of different energy bins were also tested
by comparing the FC mode data with spectrum in EA mode for the same source.
The energy bin boundaries were found to be correct to within about 1 keV.

\subsection{Energy Resolution of LAXPC}

\begin{figure}
\epsscale{.95}
\plotone{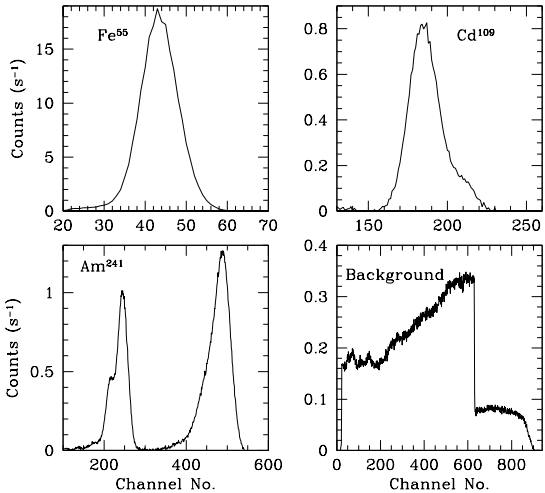}
\caption{The spectra for the three radioactive sources after correcting for the
background and the background spectrum for LAXPC30 as observed on ground.}
\label{fig:spec4}
\end{figure}

\begin{figure}
\epsscale{.95}
\plottwo{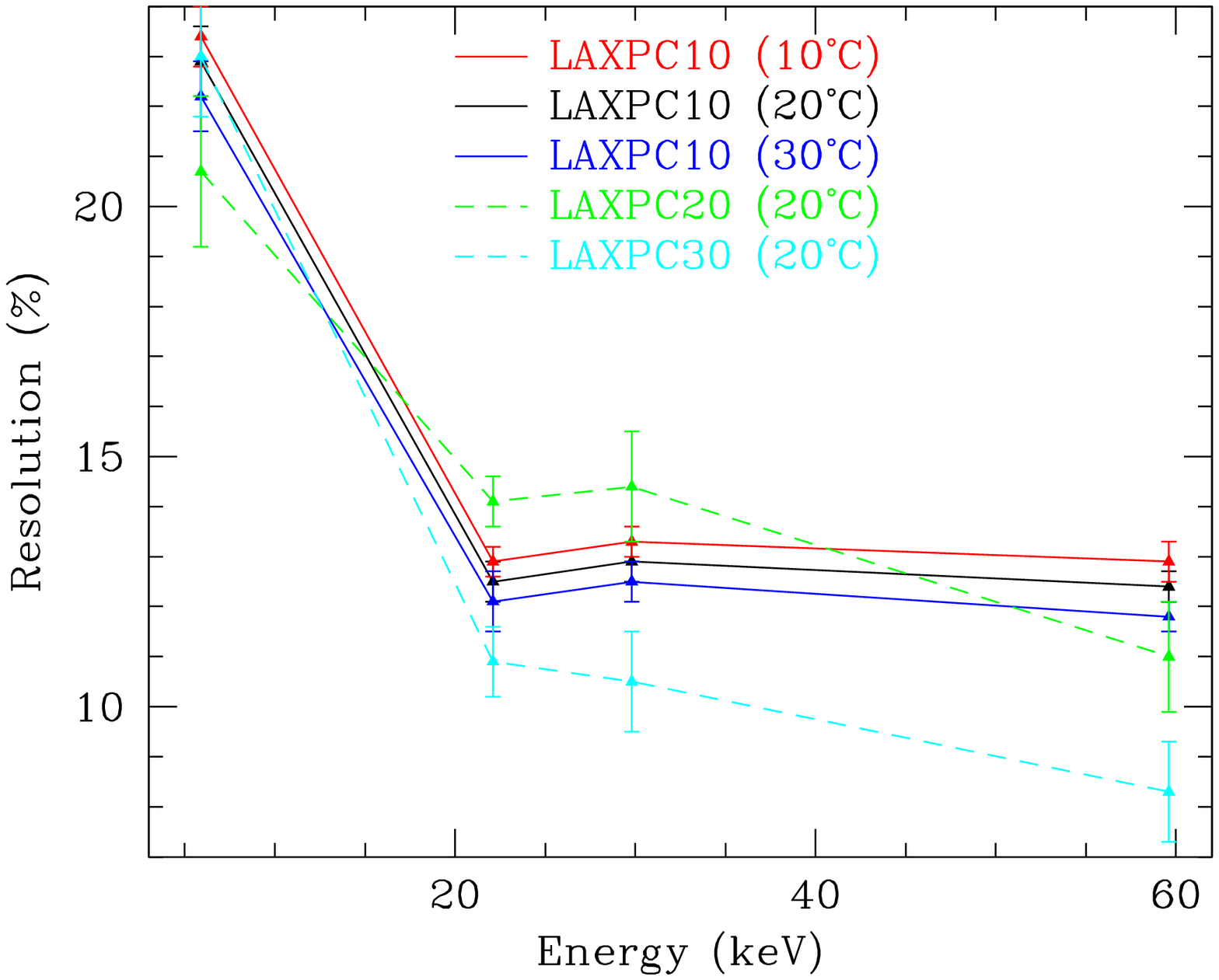}{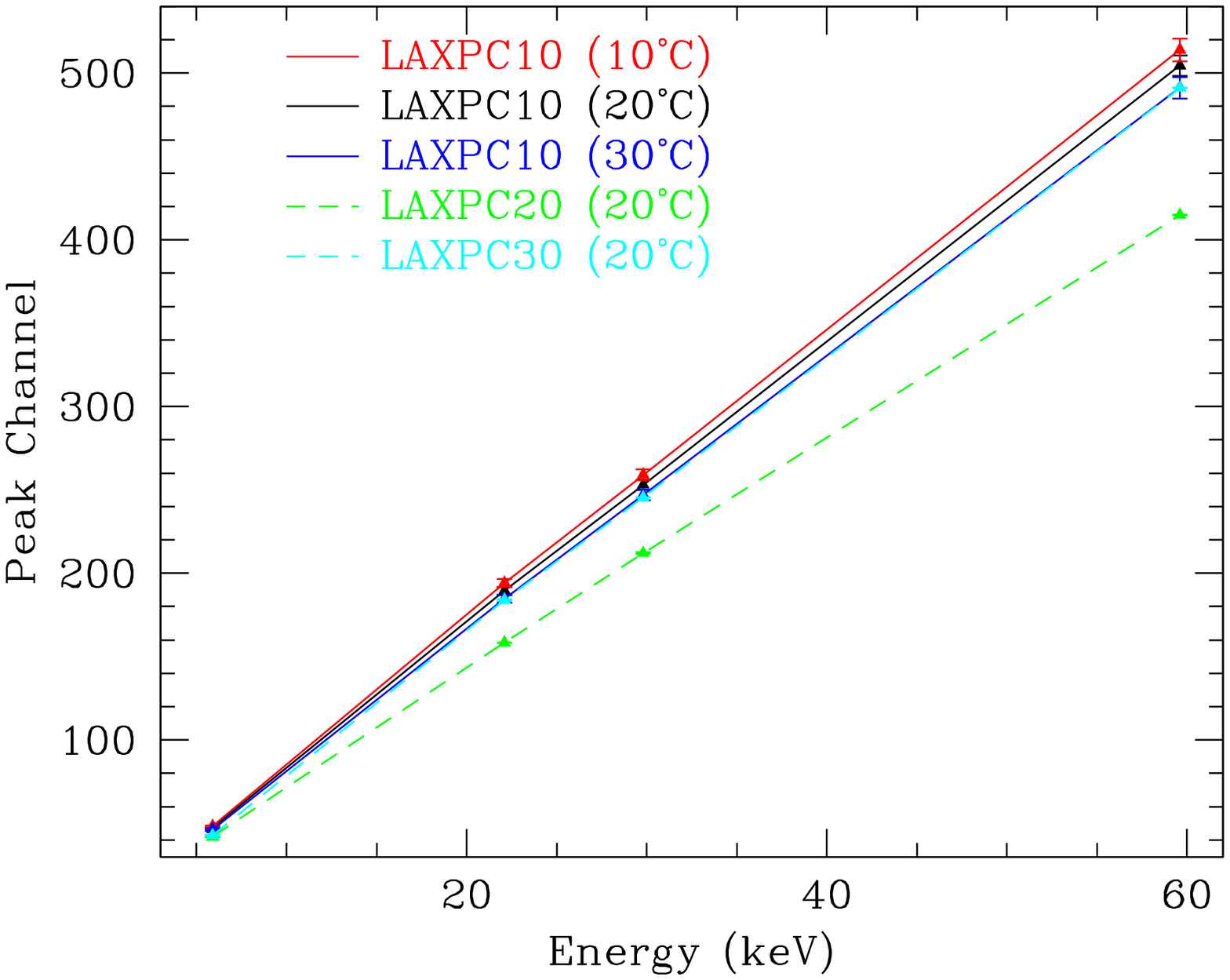}
\caption{The energy resolution (left panel) and peak position (right panel)
for LAXPC detectors as determined from observations on ground using
radioactive sources.}
\label{fig:reslab}
\end{figure}

For each of the X-ray lines, from the radioactive sources, the energy
resolution $R$ of the LAXPC detector, averaged over the entire area was
measured. The energy resolution is defined by
\begin{equation}
R=\frac{\mathrm{FWHM(channels)}}{\mathrm{Peak(channels)}}\times100\%.
\label{eq:resol}
\end{equation}
To estimate the resolution, the observed spectrum for the source is
corrected for the background and then fitted to a sum of Gaussian profiles to
obtain the peak position and resolution.
The observed spectra for the three sources and the background obtained
by adding the counts in all anodes (A1--A7) are shown in
Figure~\ref{fig:spec4}. These spectra were fitted to obtain the peak
position and resolution and the results are shown in Figure~\ref{fig:reslab}
for three different temperatures and for different detectors at $20^\circ$ C.

The background spectrum in Figure~\ref{fig:spec4} shows the lower and upper
energy threshold beyond which the count rate drops sharply. On the lower
side, it drops to zero below the lower energy threshold as the anodes are not
triggered. Beyond the ULD the count rate does not drop
to zero as the contribution from double events is still accepted. This is because
this threshold is applied separately to each anode and sum of energy in
two anodes can exceed the threshold. Thus in principle, LAXPC can detect
photons with energies up to about 110 keV, though the efficiency reduces
beyond 80 keV as only the double events contribute. From the magnitude of
drop around channel 640, it appears that double events account for
25--30\% of all events at these energies in the background spectrum.

It can be seen from Figure~\ref{fig:reslab} that the resolution improves
as temperature increases, while the peak channel shifts to lower value.
In orbit, the temperature of detectors is maintained through heaters and
its variation by less than $2^\circ$ C, may not be significant. This resolution is calculated
when the counts in all anodes are added. If we consider single anode or
a single cell in an anode the resolution is better by 1--2\%.
This is because the gains of all anodes may not be perfectly aligned and the
scatter between different anodes broadens the peak slightly.
Similarly,
the two end cells in each layer (C1 and C12) have a different gain as
compared to the middle cells. The difference being about 5\% and has to
be accounted for while constructing the response matrix. This variation
also adds to the width of peaks. Below 20 keV the energy resolution
varies as $1/\sqrt{E}$ as expected for a proportional counter \citep{kno00}.

From the peak channel figure,
it appears that the relation between energy and channel number is almost
linear. This is misleading, as the peak around 59.6 keV is actually a
combination of two peaks which are not resolved in the fit. Similarly,
the energy resolution appears to remain flat at high energies, which is
also misleading due to compound nature of the 59.6 keV peak.
It can be seen that the gains of LAXPC10 and LAXPC30 are similar, but
that for LAXPC20 is lower. Further, the gain in C1 cell of anode A4 in
LAXPC10 is about 75\% of that for other cells and this gives rise to
a small second peak on lower energy side (see Figure~\ref{fig:simfit}). This difference has to be
accounted for while generating the response matrix. 

The veto anode A10 in LAXPC10 failed during thermovac test and has been
disconnected. As a result, the background in this detector is higher as
compared to the other two. After launch, the gain of LAXPC30 has been shifting
steadily due to a suspected minor leak. As a result,
the high voltage of this detector is periodically adjusted to keep
the gain close to ground value.

\begin{figure}
\epsscale{.95}
\plottwo{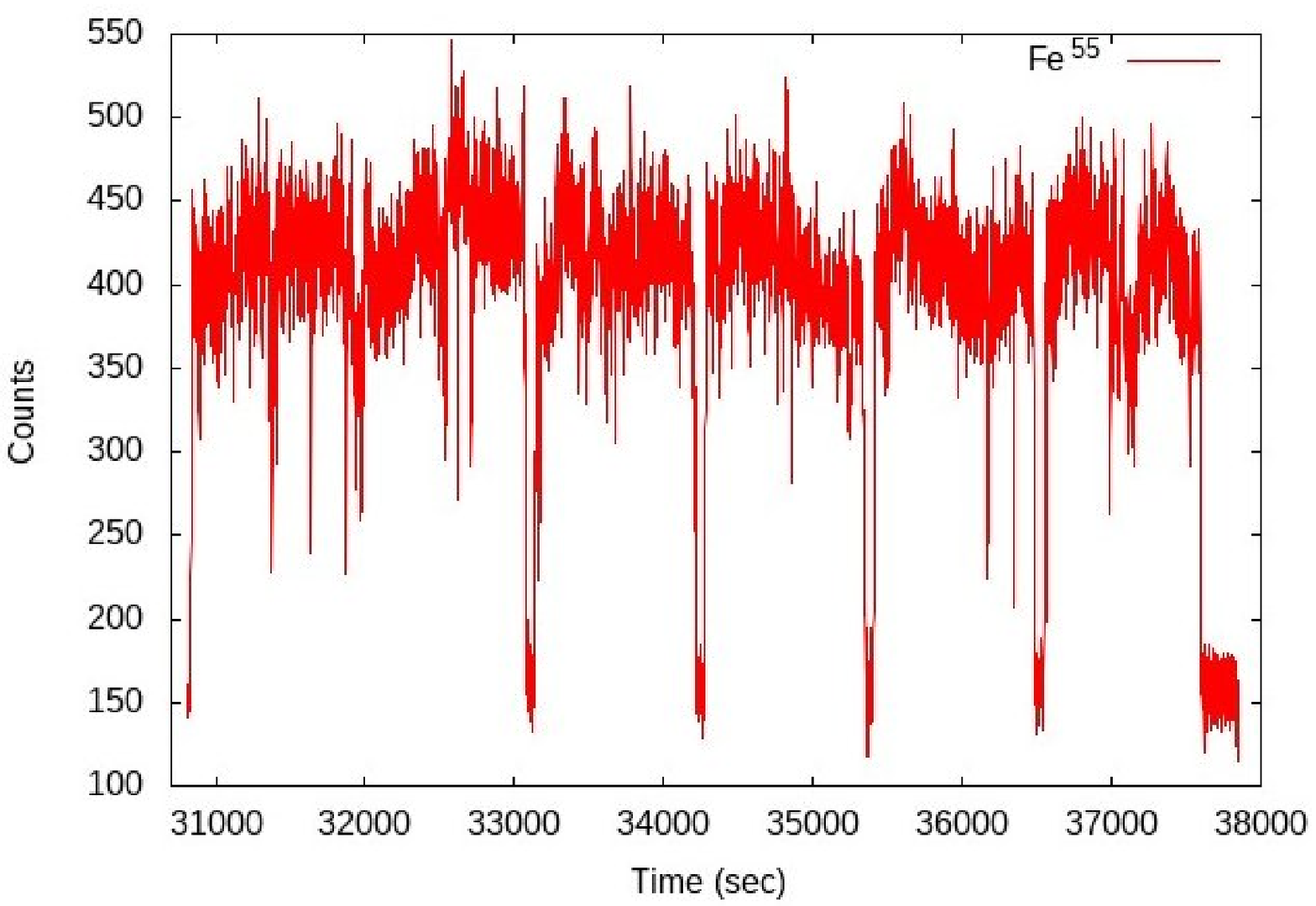}{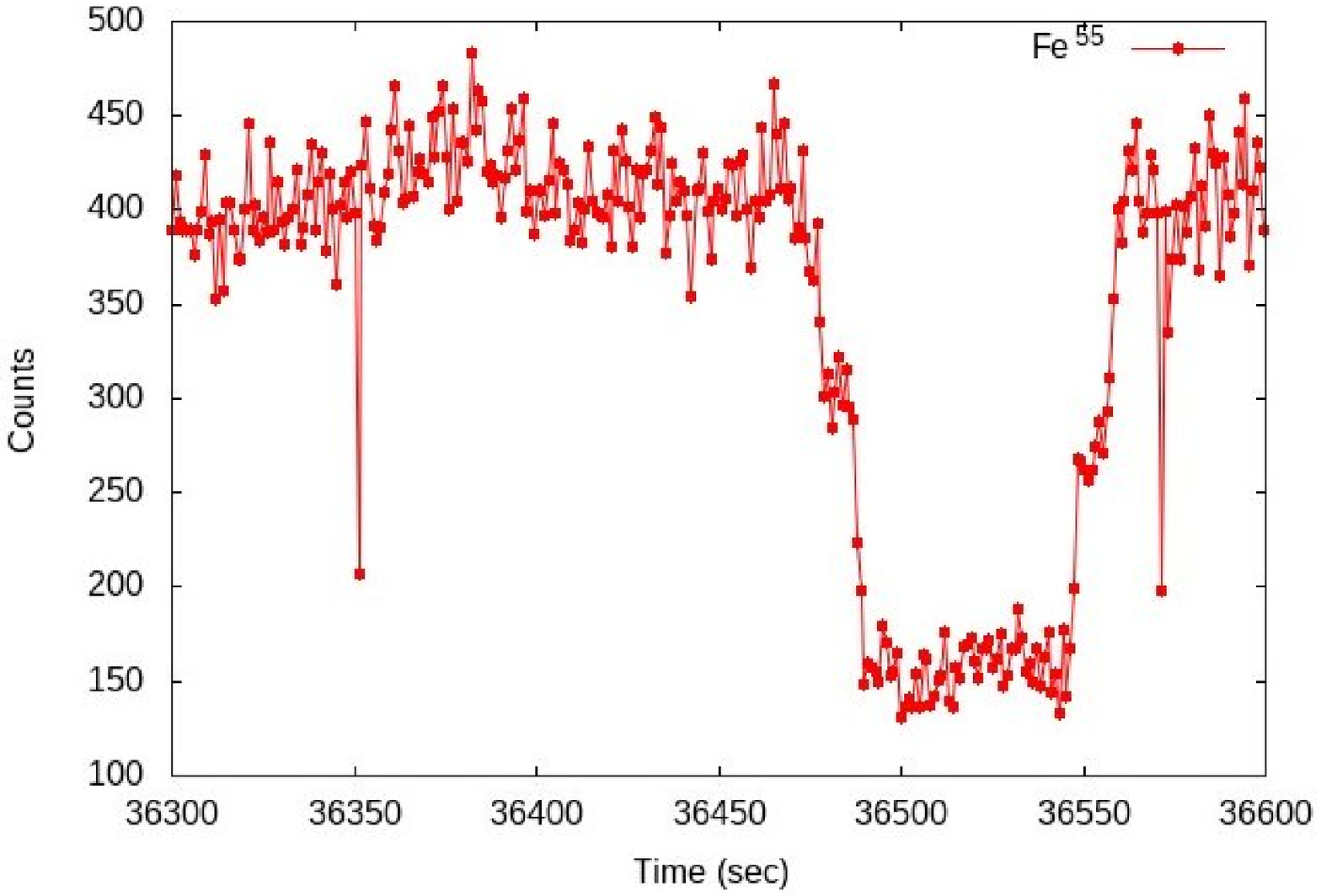}
\caption{Count rate during scan across the LAXPC20 detector using Fe$^{55}$
source. The right panel shows a part of the scan after magnification.
The big dips are due to the source going outside detector area during scan.}
\label{fig:fescan}
\end{figure}

The count rate was not uniform over the entire area of the detector
and variations of a few percent were observed, presumably because of
non-uniformity in the collimator. Figure~\ref{fig:fescan} shows the
scan over LAXPC10 detector using Fe$^{55}$ source. The prominent dips
at regular intervals are due to the source going outside the detector area
during the scan.  But there are other smaller dips which are likely to be due
to blockages in some collimator cells. The blocked area was found to be
less than 1\% of scanned area.  There were also some fluctuations
in the count rate over a time-scale of minutes.
These variations are most likely due to non-uniformity in the collimator.
Apart from these, there are fluctuations over
time scale of 10 s, due to the source crossing collimator cell boundary.
These show up clearly in the Fourier transform.
Since the radioactive sources are kept just above
the collimator, they are expected to illuminate only one or two collimator
cells at a time. If the source is moving almost exactly above the boundary of a cell,
it can illuminate up to 4 cells simultaneously. As it moves from one cell to the next, a part of the
beam will be blocked by the boundary of the cell, thus causing a dip in
the counts.
From these scans it is difficult to estimate the transmission efficiency
of the collimator. This was later estimated in orbit using scan over
standard Crab X-ray source.

\section{GEANT4 simulation}

In order to understand the characteristics of the LAXPC detectors and
to construct the response matrix, GEANT4 \citep{geant03} simulations 
were carried out. For this purpose the basic detector geometry, including
the aluminium box, shield, Mylar window and collimators were included
to define the detector. Other features like electronics, purification
system and the anode and cathode wires inside the detector were not
included for simplicity. The advantage of using GEANT4 simulation is that all
physical processes describing interaction of charged particles and photons
are included and absorption coefficients of all relevant materials are
incorporated. Similarly, all particles including secondaries, like
photons, electrons, protons, muons etc. are tracked.
Further, loss of efficiency due to escape of Xe K and L
X-rays is automatically included in the simulations once the anode boundaries
are specified. We did not include any dead zone between two anodes where the
interaction may not be recorded. The GEANT4 package can deal with
particles in energy range of 100 eV to 100 TeV. All secondary particles in this
energy range are followed till they interact or go outside the defined volume.
Apart from response matrix and efficiency of detector,
the simulations were also used
to estimate the field of view of the collimator as well as to estimate
the background from photons and charged particles.

The simulations typically use $10^6$--$10^7$ photons, which in most
cases were incident perpendicular to the Mylar window at the top. Further,
these particles are uniformly distributed over the active detector area.
These simulations were tested by comparing the results with observations
from radioactive sources described in Section 3.1.
For estimating the field of view of the collimator, the photons were
assumed to be coming at a small angle to the axis of the detector. To estimate the
background, the particles were assumed to be coming isotropically from
all directions and were uniformly distributed on all bounding surfaces.

\subsection{The detector response and efficiency}

In order to approximate the observations of radioactive sources, simulations
were carried out with the same distribution of energy as expected from these
sources. The simulations yield the energy deposited in each anode cell from
any event. The same logic as used in the detector electronics was
applied to reject or accept the event. The result was compared with the
observed spectrum. Since the absolute strength of the sources were not
known, as they were collimated through a small hole in the container
and placed over the field of view collimator, the number of events
accepted by the simulation in a broad energy band covering the peak
were normalized to match the number of counts observed in the same range.
Thus the normalization was not a free parameter in the simulation.
Once the normalization of the simulated spectrum was fixed,
the counts in each anode and channel were compared with observed spectrum.

The voltage pulse in a proportional counter is only approximately
proportional to the energy deposited in a given anode, instead the pulse is
proportional to the number of electrons produced in the corresponding
anode. The observed pulse height is essentially, the number of electrons
produced in the gas multiplied by the electronic gain from the proportional
counter and the amplifier. To generate the response matrix we need a
mapping from the energy deposited to pulse height channel that is recorded.
For this purpose, following \citet{jah06} we define an energy scale, $E_p$,
proportional to the number of electrons produced, but is normalized such
that it is approximately equal to the energy deposited $E$.
The average energy $w(E)$ in eV required to produce one ionization
electron in Xe is close to 22 eV \citep{dias91,dias93,dias97}. It is convenient
to define
\begin{equation}
E_p=\frac{22.0}{w(E)} E.
\label{eq;ep}
\end{equation}
The function $w(E)$ is shown in Figure~4 of \citet{jah06} and has discontinuities
at the Xenon K and L edges. As a result of this, the energy to channel mapping
may not be monotonic in this region. This function is
used in the simulation to calculate $E_p$, which is then used to calculate the corresponding channel.

In order to calculate the spectrum of simulated events we need a
mapping from energy to channel. This was provided by the observed
positions of the peaks in spectra for radioactive sources. We
used a quadratic function
\begin{equation}
n_c=e_0+e_1E_p(1+e_2E_p),
\label{eq:gain}
\end{equation}
to fix the energy to channel mapping. Here, $E_p$ is the effective energy in
keV as defined above, and
$n_c$ is the ADC channel where energy is mapped. The parameters $e_0$
and $e_1$ are determined by matching the 5.9 keV peak in Fe$^{55}$ spectrum,
the 22.1 keV peak in Cd$^{109}$ spectrum and the 29.8 keV peak in Am$^{241}$
spectrum. The parameter $e_2$ was determined by fitting the 59.6 keV peak
in the Am$^{241}$. One more iteration was performed by recalculating
$e_0$, $e_1$ with fitted value of $e_2$. Since the position of low energy
peaks is not sensitive to $e_2$ one iteration was found to be sufficient.

Near the absorption edges of Xe the mapping given by Eq.~\ref{eq:gain} is further
tuned to fit the observed spectrum of Crab.
Since the temperature in orbit is maintained constant to within $2^\circ$ C,
no temperature dependence of coefficients in Eq.~\ref{eq:gain} is required.
The same is true for energy resolution also.

\begin{figure}
\epsscale{.95}
\plotone{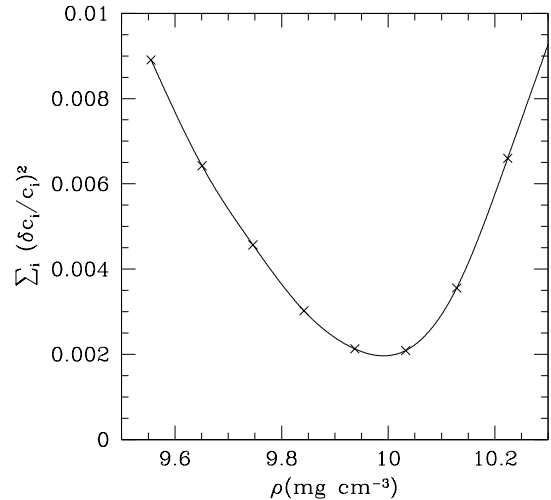}
\caption{The function $F(\rho)$ as defined in Eq.~\ref{eq:den} is shown
for LAXPC10.}
\label{fig:den}
\end{figure}

\begin{figure}
\epsscale{.55}
\plotone{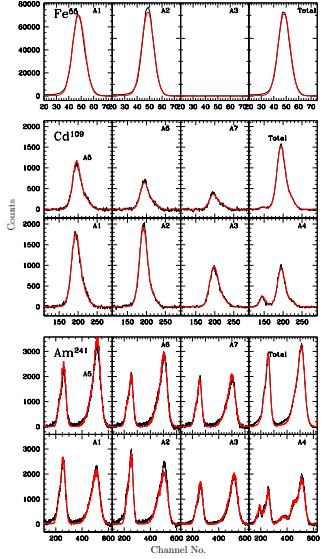}
\caption{The fits to observed spectra for radioactive sources by simulations
for LAXPC10.
The black lines show the observed spectrum while red lines show the simulated
spectrum. The counts in the panel for combined spectra are scaled down to
fit in the same axis.
The anode A4 shows a secondary peak
on the lower side, due to difference in gain in cell C1.}
\label{fig:simfit}
\end{figure}

The relative counts (total counts under a peak) in different anode
layers depends on the absorption coefficient and density (or pressure) of
the gas. Using this ratio for Cd$^{109}$ source, which shows systematic
variation in counts with layers, it is possible to estimate the density
of the gas. It was found that varying the density by a few percent can
get the relative count rates in better agreement. To estimate the density
of the gas, simulations were done with different values of density and the
relative difference in count rates
\begin{equation}
F(\rho)=\sum_{i=1}^5\left(\frac{O_i-S_i}{O_i}\right)^2,
\label{eq:den}
\end{equation}
where $O_i$ and $S_i$ are respectively, the counts in the observed and
simulated spectra in $i$th layer. The sum is over the 5 layers. Since the
total counts in observed spectrum are of order of $10^6$, the statistical
errors in $O_i$ are very small and any departure in $F(\rho)$ from zero is
due to systematic discrepancies in simulation. As a result, we have not tried to
define a $\chi^2$ function for this purpose. This function
shows a well-defined minimum as a function of density as shown in Figure~\ref{fig:den}. The difference in density could arise due to a small variation in
pressure or temperature. This value of density is used in all further
simulations. The density is found to be 10.0 mg cm$^{-3}$ in LAXPC10,
10.7 mg cm$^{-3}$ in LAXPC20 and 11.5 mg cm$^{-3}$ in LAXPC30.

Apart from the energy to channel mapping we also need the energy resolution
as a function of energy to calculate the detector response. In principle,
we can use the measured resolution shown in Fig.~\ref{fig:reslab} for this
purpose, but that is not satisfactory as that will not leave any free
parameter in simulations to match the observed spectrum for the radioactive
sources. Further, the peak at 59.6 keV in the Am$^{241}$ spectrum is a
multiple peak arising from single and double events with different
resolution and channel. Thus we keep the energy resolution as a free
parameter in simulations, which is determined by matching the observed
spectrum. Thus we have 4 free parameters which give the energy resolution
for the four peaks at 5.9, 22.0, 29.8 and 59.6 keV. The 5.9 keV peak is not
seen in lower layers and its resolution is determined by the top layer
only. In any case here we have assumed that all anodes have the same
resolution. The 22.0 keV peak in
Cd$^{109}$, has a small component around 25 keV, but we assume the same
relative resolution for the entire peak. Similarly, 29.8 keV peak in
Am$^{241}$ spectrum also has a contribution at lower energy from escape
of Xe K$_\beta$ X-rays, but we assume the same relative resolution for the entire
peak. The 59.6 keV resolution applies to the peak from single events where
all energy is absorbed in the same anode. For the double events the two
events have energy close to 29.8 keV and that resolution is used in
each anode, before the channel number in the two anodes are added to get
the final value. Thus we have only 8 free parameters,  $\sigma_1,\sigma_2,
\sigma_3,\sigma_4$ to specify the energy resolution, $e_0,e_1,e_2$ (Eq.~\ref{eq:gain}) and the density,
to match the 3 spectra in each anode. Only the regions near the peaks are 
used to fit the spectra.

To determine these parameters we first determine $e_0,e_1$ by matching
the peak positions and then $\sigma_2$ and density are determined
to match the Cd$^{109}$ spectrum for all anodes. This fixes the value of
density as explained above which is used in all further simulations.
Then Fe$^{55}$ spectrum is fitted to obtain the resolution $\sigma_1$.
Since the low energy photons are absorbed in the top layer we only need
to fit the spectrum in anodes A1, A2.
The other three parameters, $\sigma_3,\sigma_4,e_2$ are used to
match the spectrum for Am$^{241}$. The resulting fits are shown in
Figure~\ref{fig:simfit} for LAXPC10. The anode A4 shows a secondary peak
on the lower side, due to difference in gain in cell C1.

\begin{figure}
\epsscale{.95}
\plottwo{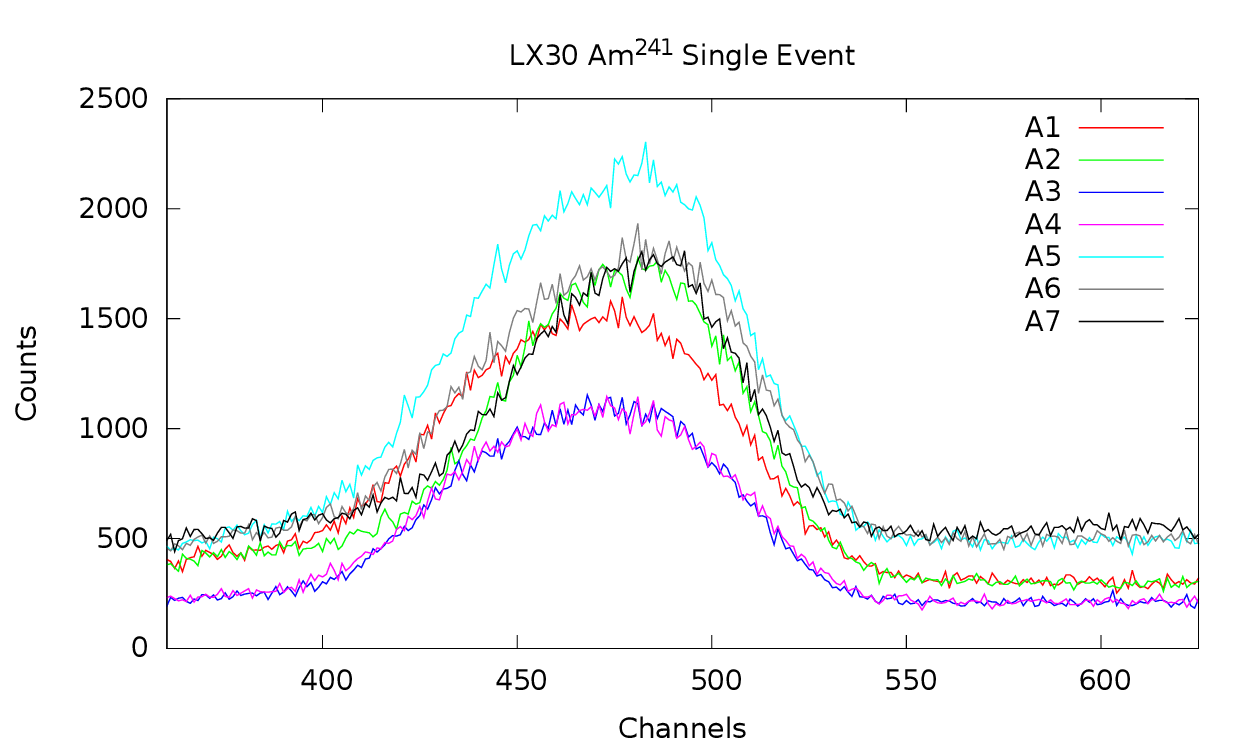}{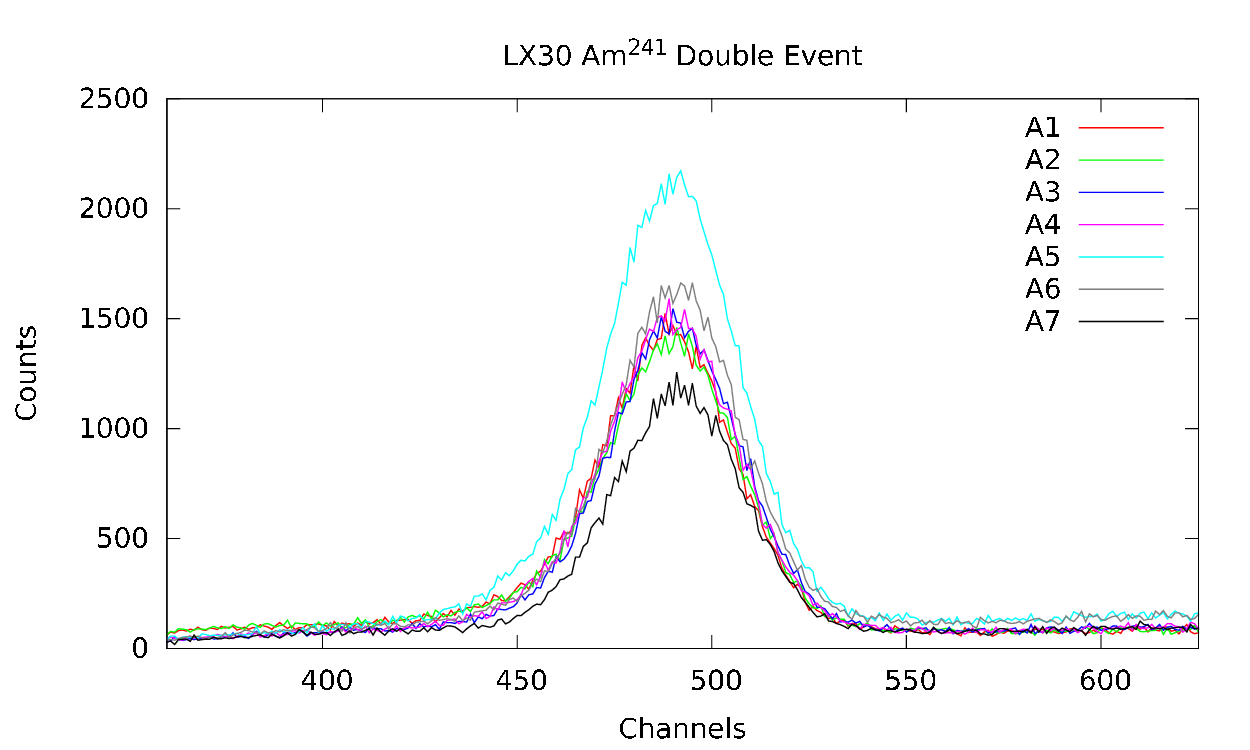}
\caption{The observed spectrum of Am$^{241}$ in LAXPC30.
The left
panel shows the spectrum when only single events are included, while the
right panel shows the spectrum when only the double events are included.}
\label{fig:am241}
\end{figure}

The resolution obtained by fitting the simulated spectrum for the $5.9,22.1,29.8$ keV lines are close to those determined by fitting the observed spectrum
described in Section 3.1. However, for 59.6 keV peak the resolution obtained
by fitting the simulated spectrum is much poorer than that from fitting the observed
spectrum directly. This is because the simulated spectra give the resolution
for the peak defined by single events, while the fit to observed spectra
did not attempt to resolve the two peaks. It is possible to fit the
two peaks but in that case the fit is not very stable. To illustrate the
difference we show in Figure~\ref{fig:am241} the observed spectrum
for Am$^{241}$ when only single or double events are included.
It can be seen that the peak in single events is broader and shifted
to lower channel as compared to that in double events. The resolution
obtained by fitting the simulated spectrum matches that for single event.
The shift to the negative side occurs because the quadratic term in
Eq.~\ref{eq:gain} is negative. The values of the parameters defining
the gain for the detectors are listed in Table~\ref{tab:ei}.

\begin{figure}
\epsscale{.95}
\plotone{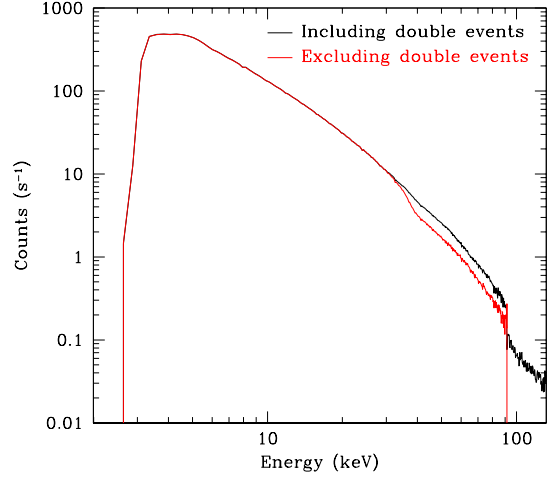}
\caption{The background subtracted spectrum of Crab observed during January
2017 in LAXPC10 by including and excluding the double events.}
\label{fig:crabsin}
\end{figure}

The inclusion of double events in the detector logic does introduce
some complication in the construction of response matrix, but this is needed
as above the K-edge of Xenon, typically 30\% of the events are double events
and their exclusion will reduce the efficiency significantly. The reduction
in efficiency would be about 40\% as in another 30\% of the events the
Xe K fluorescence X-ray escapes the detector and the event is recorded
at lower energy. The event mode data
from the detectors gives the channel information for each anode and it is
possible to reject these double events in software, though the current
pipeline software does not have that option. Figure~\ref{fig:crabsin} shows
the observed spectrum of Crab in LAXPC10 when the double events are excluded.
It can be seen that at high energies the efficiency is reduced by about 40\%
when double events are excluded. Further, excluding the double event will not
remove all complications in response matrix, as at energies above the K-edge of Xenon a significant
fraction of single events also correspond to cases where the Xenon K X-ray
has escaped the detector and this will anyway have to be considered while
generating the response, as there is no way this event can be distinguished
from an event produced by a lower energy photon.

\begin{figure}
\epsscale{.95}
\plottwo{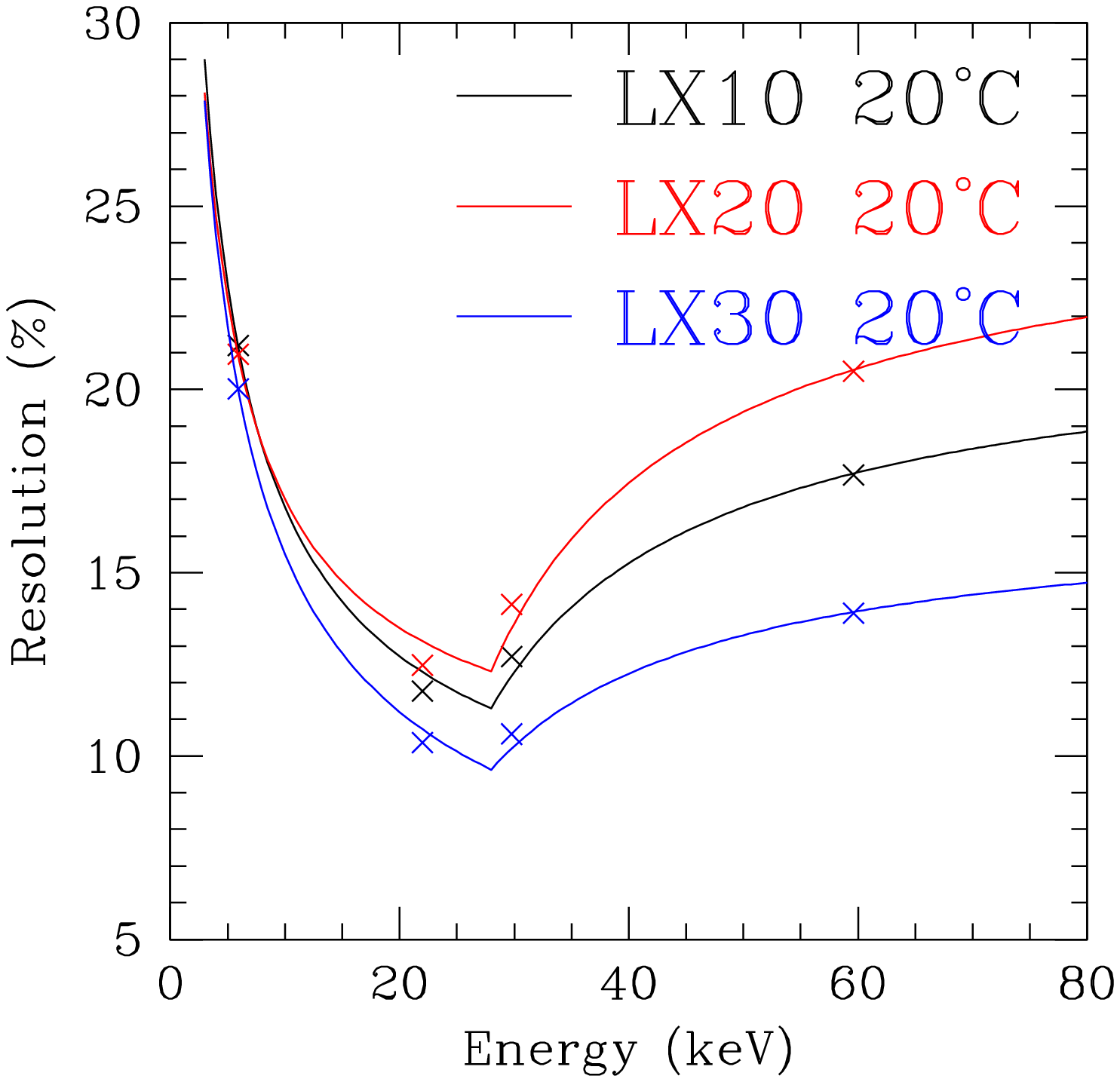}{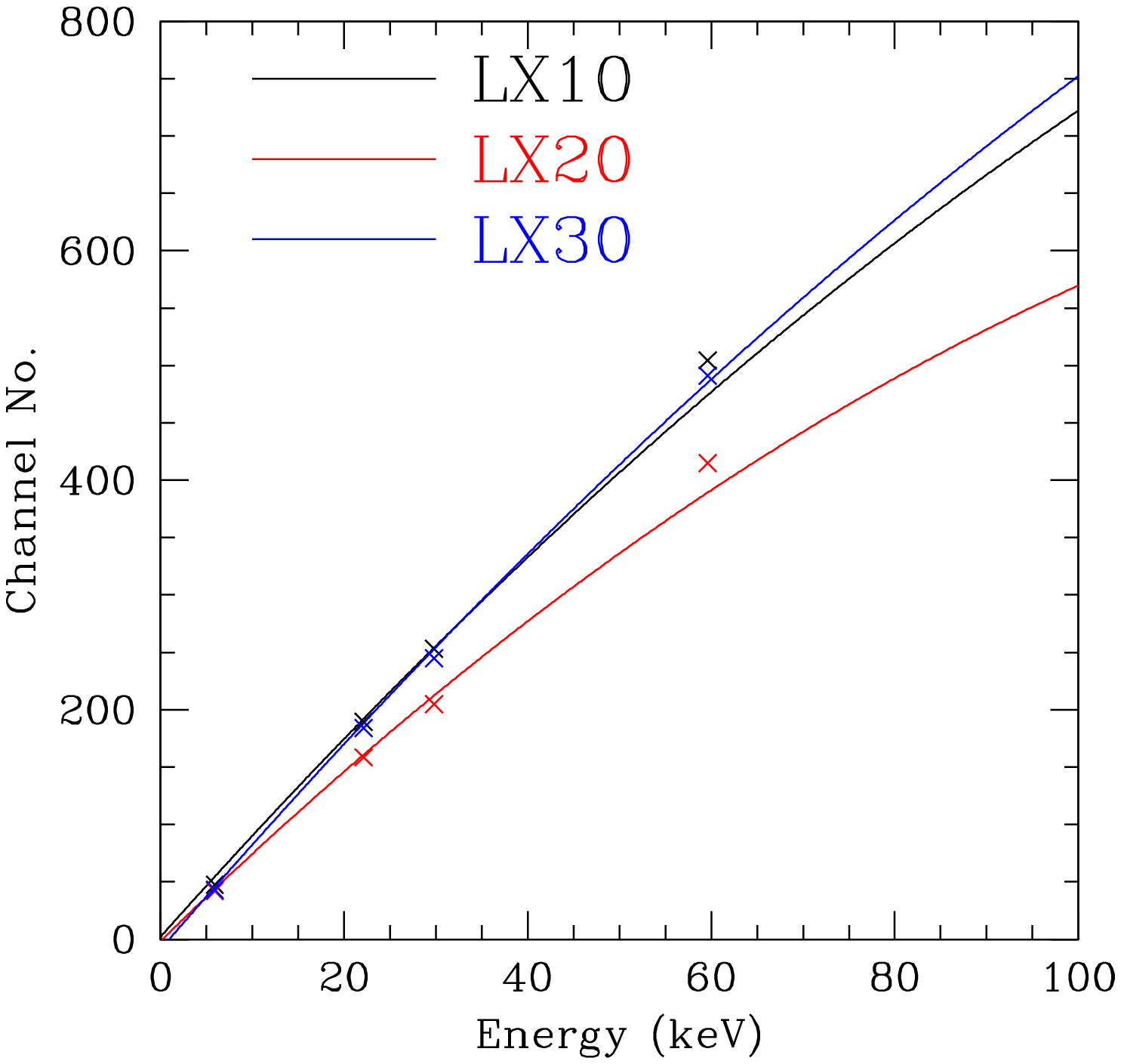}
\caption{The energy resolution (left panel) and the energy to channel
mapping (right panel) as determined by fitting the simulated spectrum to
observed spectrum for radioactive sources. The points in the left panel mark
the actual value for the 4 peaks, while those in right panel show the
values obtained by fitting the spectrum by a sum of Gaussian peaks.}
\label{fig:sig}
\end{figure}

The observations of radioactive sources give the resolution at only
four energy values, while for constructing the response matrix we need
the energy resolution for all energies in the range of LAXPC. To obtain this
we use a fit to $\sigma^2$ by a linear B-spline basis functions using 3 knots in
$1/E$. We fix the knots at $1/80,1/28,1/3$ keV$^{-1}$. This gives a piecewise
linear approximation in $1/E$. Figure~\ref{fig:sig} shows the energy resolution
and energy to channel mapping as determined from simulations. This can be
compared with Figure~\ref{fig:reslab} based on fitting the observed peaks in
spectra for radioactive sources. The main difference arises for energies
above 35 keV which is due to composite nature of the peak in this region.
The larger nonlinearity in LAXPC20 is clearly seen in Figure~\ref{fig:sig}.
Because of this nonlinearity the effective ULD is higher
in LAXPC20.

Using the energy to channel mapping and the energy resolution shown in Figure~\ref{fig:sig},
it is possible to simulate the detector response for any incident energy in
the relevant range. From these simulations it is possible to calculate
the response matrix, which provides the information about the probability
that an incident photon of any given energy will be observed in a particular
channel of a particular anode. By summing over the probability in all channels
and anodes we can get the
detector efficiency at that energy, which is the probability that a photon
of given energy will be observed in the detector. This efficiency multiplied by
the geometric area of the detector would give the effective area of the detector.

\subsection{The effective area of LAXPC detectors}

To calculate the effective area we carry out simulations with incident
photons at various energies and calculate the fraction that would be
detected by the detectors. The geometric area of each LAXPC detector is
$100\times36=3600$ cm$^2$. Assuming a perfectly aligned collimator in the
simulations, we find that at best 79\% of photons are detected as the
rest of the area is blocked by the collimator. This gives an area of about
2800 cm$^2$ for each detector. This would give a total effective area of
three detectors around 8400 cm$^2$. From the calibration on ground it was
impossible to estimate the effect of imperfections in collimator. In particular,
even the alignment of collimator may not accurately match that of the aluminium
box containing the collimator which was used to align the detectors on
satellite deck. Since all instruments on AstroSat are expected to be
co-aligned it would be necessary to find the offset of each LAXPC detector
with respect to the Satellite pointing axis. This offset as well as
field of view of the collimator was expected to be estimated by scanning
across the Crab. That should also give a better estimate of
collimator misalignment.

For a perfect collimator the dead-time corrected
count rate would vary linearly with pointing offset. The collimator of each
detector consists of about 7000 square cells of side 7 mm. The fabrication
and mounting of the collimator modules may introduce some level of
misalignment between different individual cells in the collimator assembly.
To produce a more realistic model of collimator, following \citet{jah06} we assume that the pointing
direction of each cell is randomly displaced from the detector axis. The random
offsets are assumed to have a Gaussian distribution centered at the detector axis
and with standard deviation, $\sigma_c$, which quantifies the imperfection
in the collimator.  The standard deviation would
need to be adjusted to approximate the observed scan profile. The actual
collimator may have a misalignment in the mean also which would again be determined by the scan.
Final effective area can only be
obtained by cross calibration with other instruments after launch  as
described in Section 5.2.

\subsection{The field of view}

\begin{figure}
\epsscale{.95}
\plottwo{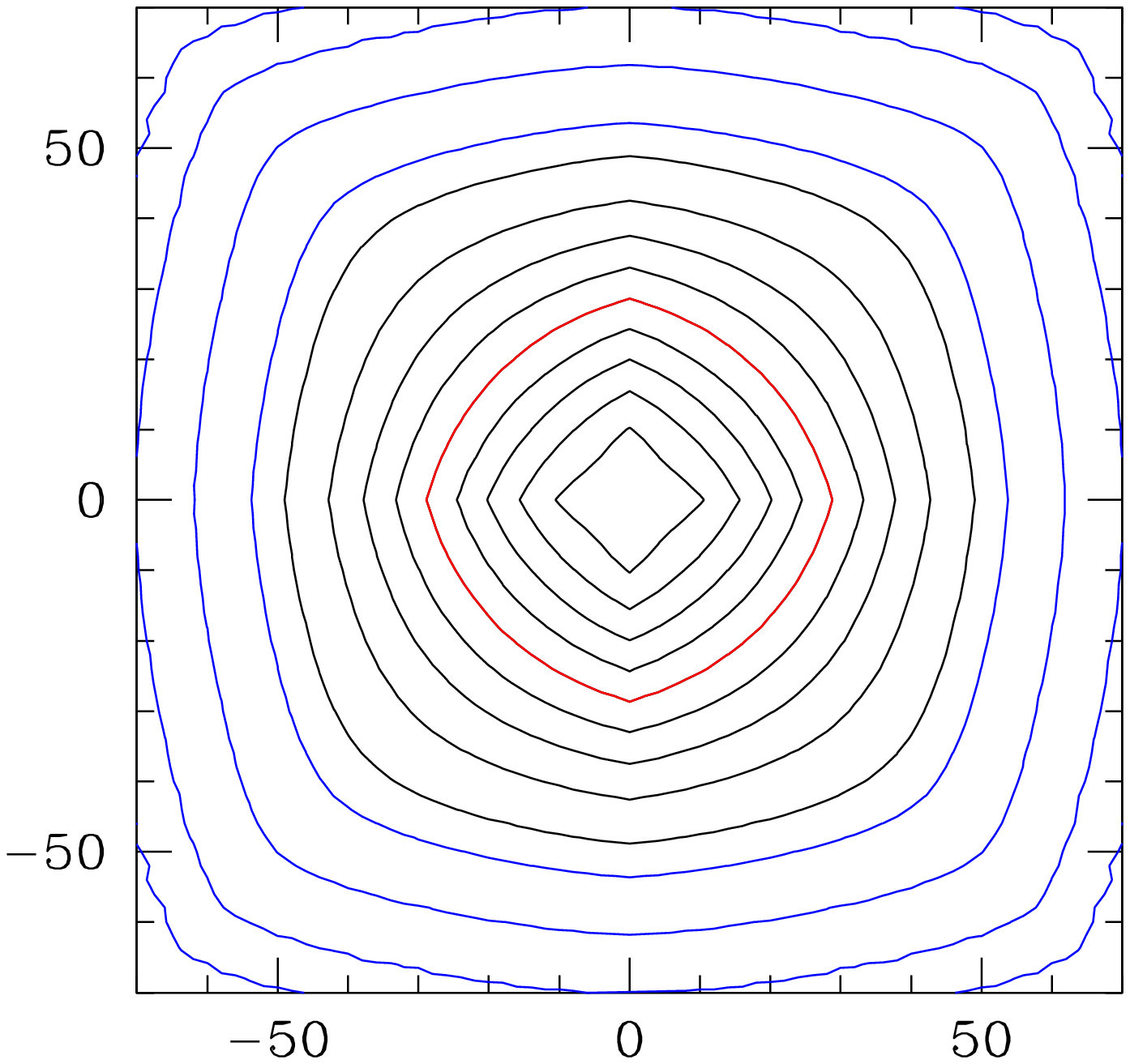}{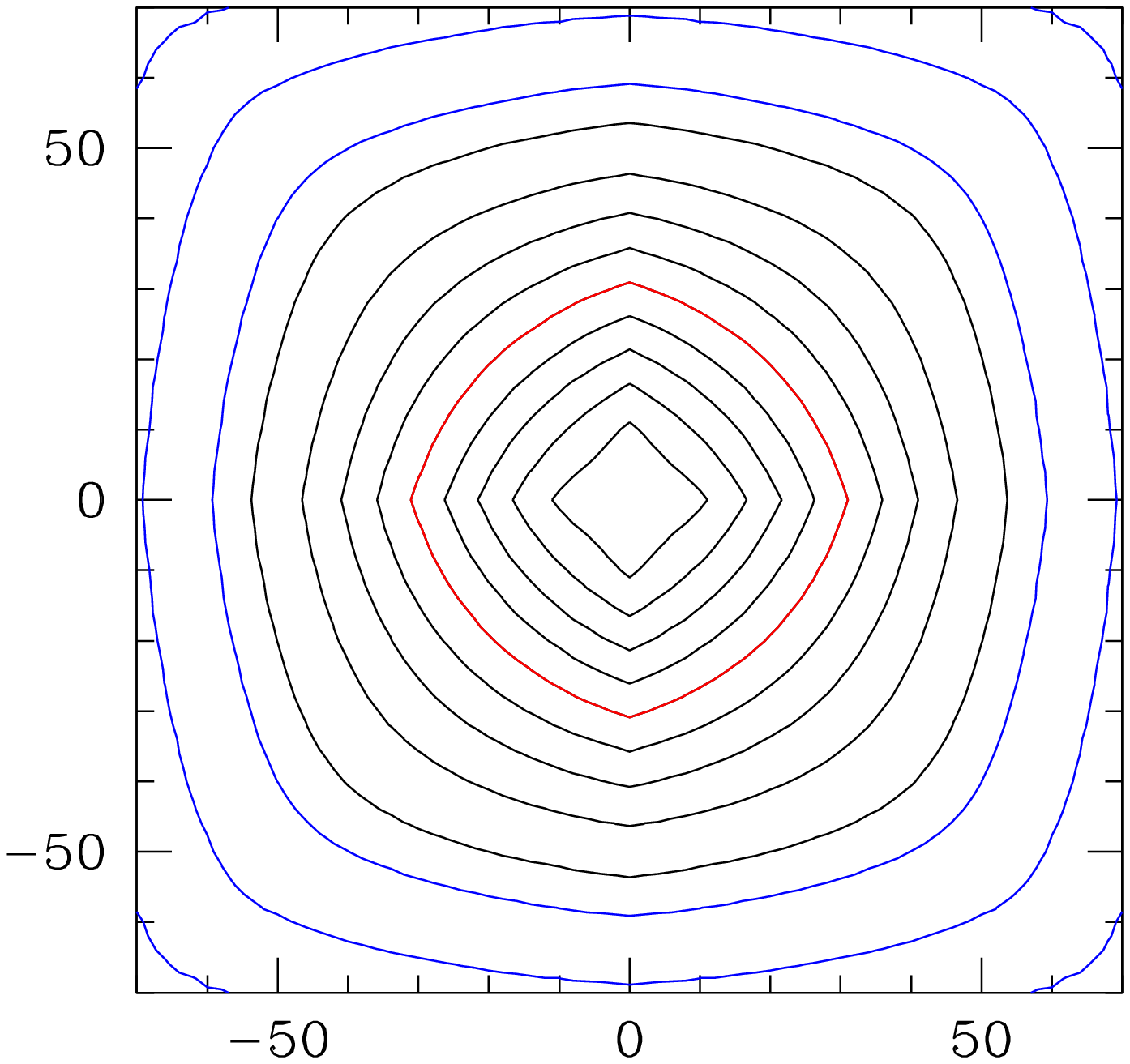}
\caption{Contours of constant efficiency (relative to the detector axis)
are shown as a function of angle from detector axis for photons of 15 keV (left
panel) and 50 keV (right panel). The axes are marked in arc minutes. The red
contour encloses the region where count rate is more than half of the peak
value and gives the FWHM of the field of view. The black contours are at
interval of 10\%, while the blue contours are at levels of 5\%, 1\%, 0.1\%
and 0.01\%.}
\label{fig:fov}
\end{figure}

The field of view of the collimator could not be determined using radioactive
sources on ground. Hence GEANT4 simulations were used for determining the field of view.
Each simulation included $10^6$ photons of fixed energy which were uniformly
distributed over the detector top and incident at a fixed angle to the detector
axis. The number of
events registered in the detector were noted for each angle. The secondary
particles produced in the collimator are also tracked and all events which
satisfy the selection criteria for the processing electronics were counted.
The maximum counts were obtained when the photons were incident normal to
the detector surface. The counts at other angles were divided by this
maximum number to get the relative efficiency of the detector in different
directions. There is some energy dependence in the field of view, as the
window support collimator which is made of aluminium is almost transparent
to high energy photons. As a result, at high energy the field of view
increases marginally. For a perfect collimator the full-width at half maximum
(FWHM) of the collimator is $43'=0.72^\circ$ at 15 keV and $47'=0.78^\circ$
at 50 keV when measured along diagonal of the detector.
Along the sides of detector the FWHM turns out to be $50'=0.83^\circ$ at
15 keV and  $54'=0.90^\circ$ at 50 keV.
However, all cells in the collimator are unlikely to be perfectly
aligned in the same direction and it is necessary to include this effect in
simulations. On ground it was not possible to estimate the magnitude of
dispersion in alignment of different cells and hence this was achieved by
a scan across Crab after launch, by comparing the scan profile
with simulations with different amount of dispersion in collimator alignment.
The dispersion was assumed to have a Gaussian distribution and it was estimated
that for LAXPC10 the standard deviation of the alignment is about $12'=0.2^\circ$. This reduces the detector efficiency by about 14\% which affects the
effective area also. This estimate of 14\% applies when photons are incident
perpendicular to the top window, there is a mild dependence on the angle of
incidence.
Figure~\ref{fig:fov} shows the contours of constant
efficiency for photons of 15 keV and 50 keV. The red contours which marks the
region of half the maximum counts has a FWHM when scan is taken along the sides of detectors of $57'=0.95^\circ$ at 15 keV
and $62'=1.03^\circ$ at 50 keV.
These contours are not circular in shape because the cells in collimator are
square.

For other detectors the dispersion in collimator cells is even larger giving
even lower efficiencies. For LAXPC20 the dispersion $\sigma$ is estimated to
be about $19'$ giving a loss of efficiency by about 22\%. For LAXPC30
$\sigma$ is estimated to be about $17'$ giving a loss of efficiency by 18\%.
It is possible that this dispersion is overestimated as there would also be
some contribution from misalignment between the field of view collimator and
the window support collimator.

\subsection{The detector background}

The detector background arises from the photons and charged particles in
space, apart from the X-rays coming from the source.
The detector background can be estimated from simulation, if the flux of
particles (photons and charged particles) responsible for the background is known
at different energies. The high energy particles interact with shield and other
surrounding material and may produce secondary X-rays or other particles of varying
energies. The GEANT4 simulations keep track of all these secondaries.
Although, the detector is not designed to detect high energy particles, many
of these particles may deposit only a part of their energy inside the detector,
thus triggering a valid event in the detector. Simulations include all secondaries
produced in the shield, collimators and aluminium box enclosing the gas and
collimator. But simulations do not include other materials surrounding the
detector in the satellite and as a result, some differences may be expected.
The main purpose of these simulations is to estimate the efficiency with which
background events are rejected.

There are the following mechanism to reject the background:
\begin{enumerate}
\item The shield: The shield including the tin shield, aluminium box and
collimators may not allow the particles to pass through. At low energies
this is very effective and practically all background events are suppressed.

\item The detector efficiency: Since the low energy particles are shielded
from the detector, only high energy particles will enter the detector volume.
At these energies, the detector is fairly inefficient to photons and a large
fraction of these photons will simply pass through the detector without
registering an event.

\item Coincidence: The logic in processing electronics rejects all events
which trigger more than one anode, except for the Xenon K X-ray peak as
explained in Section 2. This can reject events due to charged particles.

\item Veto layers: Any event which deposits energy in a veto anode is
rejected. This would also be effective for rejecting charged particles
entering from the sides.

\item Excess energy: If the energy deposited in an anode exceeds the
ULD of about 80 keV, it will be rejected.

\end{enumerate}
The relative contributions of all these factors were estimated in the
simulations. There can be multiple reasons why a given event is rejected.
In particular, the last three options can have significant overlap. In the
simulations, any event which registers in more than two anodes or two main
anodes is accounted for in contribution from coincidence, while those
events which are recorded in two anodes which include a veto anode are accounted for in
contribution of veto layers.
The excess energy
refers to only those cases where only at most two main anodes are triggered. The overlap
between different options was not estimated in these simulations.

The background flux was assumed to be isotropic and uniform. Only gamma,
electrons and protons were considered separately as the primary particles.
Each simulation run consisted of $10^6$ events for a chosen particle with
energy uniformly distributed in a specified interval. The particles were
incident on the virtual surface of a box of size $120\times60\times80$ cm
with uniform probability of landing at any point on the surface. The detector
was inside this virtual box. The initial direction of incident particles
was also considered to be uniformly distributed in the solid angle of
$2\pi$ constituting the particles directed inside the volume. With this
size of virtual box, it was found that if the shield, aluminium box and
collimator were removed, then on an average about 23\% of the particles
reached the active detector volume ($100\times39\times16.5$ cm). This is
purely a geometric effect because the active detector is only a fraction of
total volume considered. When the shielding material is included, the
number of particles reaching the detector were much smaller at low energies,
but at high energies they approached or even exceeded 23\%. The latter is
because of secondaries produced in the shield. Nevertheless, because of
various rejection criterion, the number of events actually registered is
much smaller. Thus for consistency all rates were normalized with this number.

For each event the total energy deposited in each anode volume was calculated
and the rejection criteria outlined above were applied. In order to identify
the contribution to background rejection from various measures, all contributions
were separately counted. In addition, count of events rejected by each of the
veto layers were also kept separately. These include only those events which
also triggered one of the main anodes. These are counts which will be added
if one of these veto anodes is not functioning, as is the case with LAXPC10.

\begin{figure}
\epsscale{.45}
\plotone{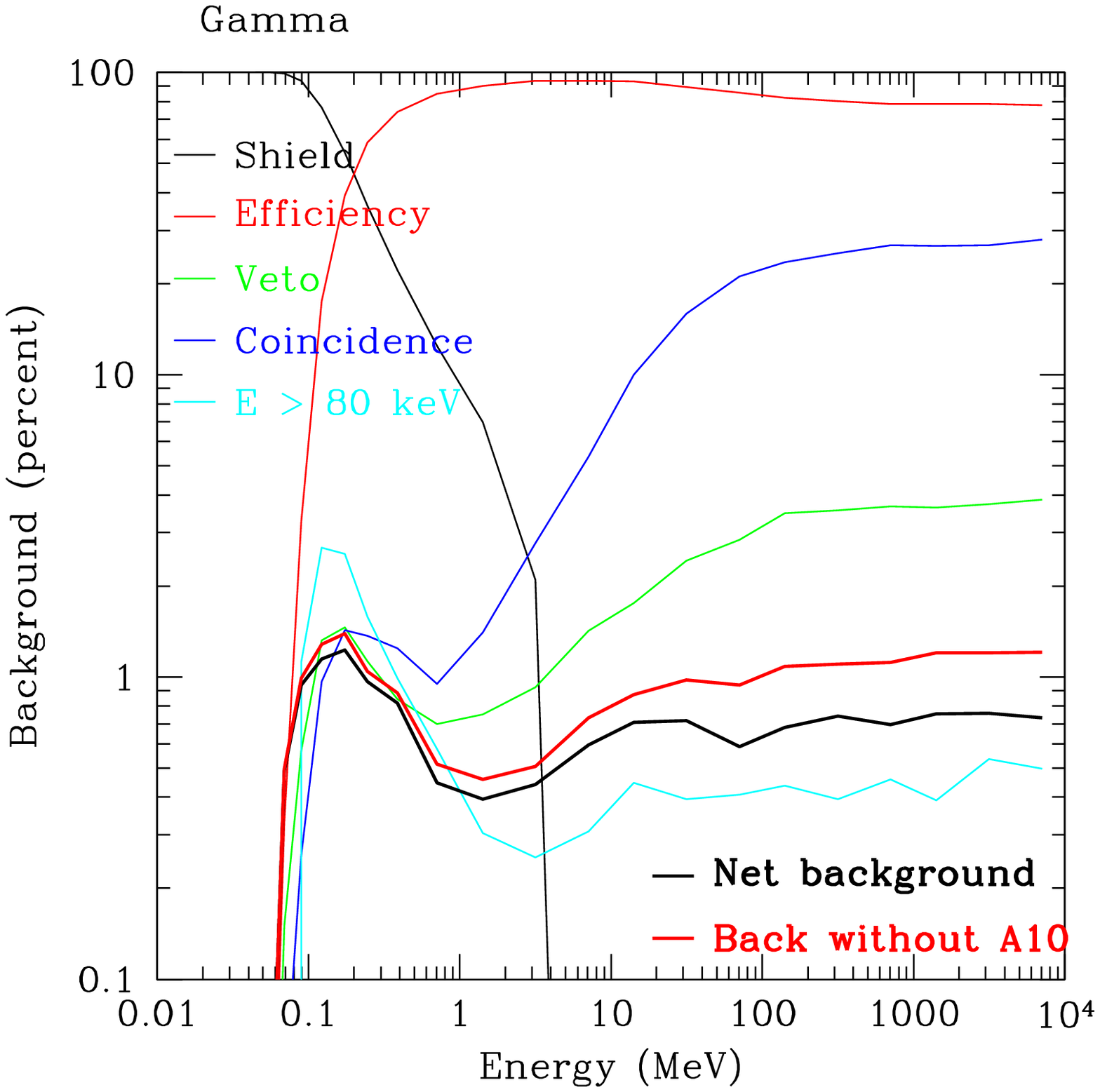}
\plotone{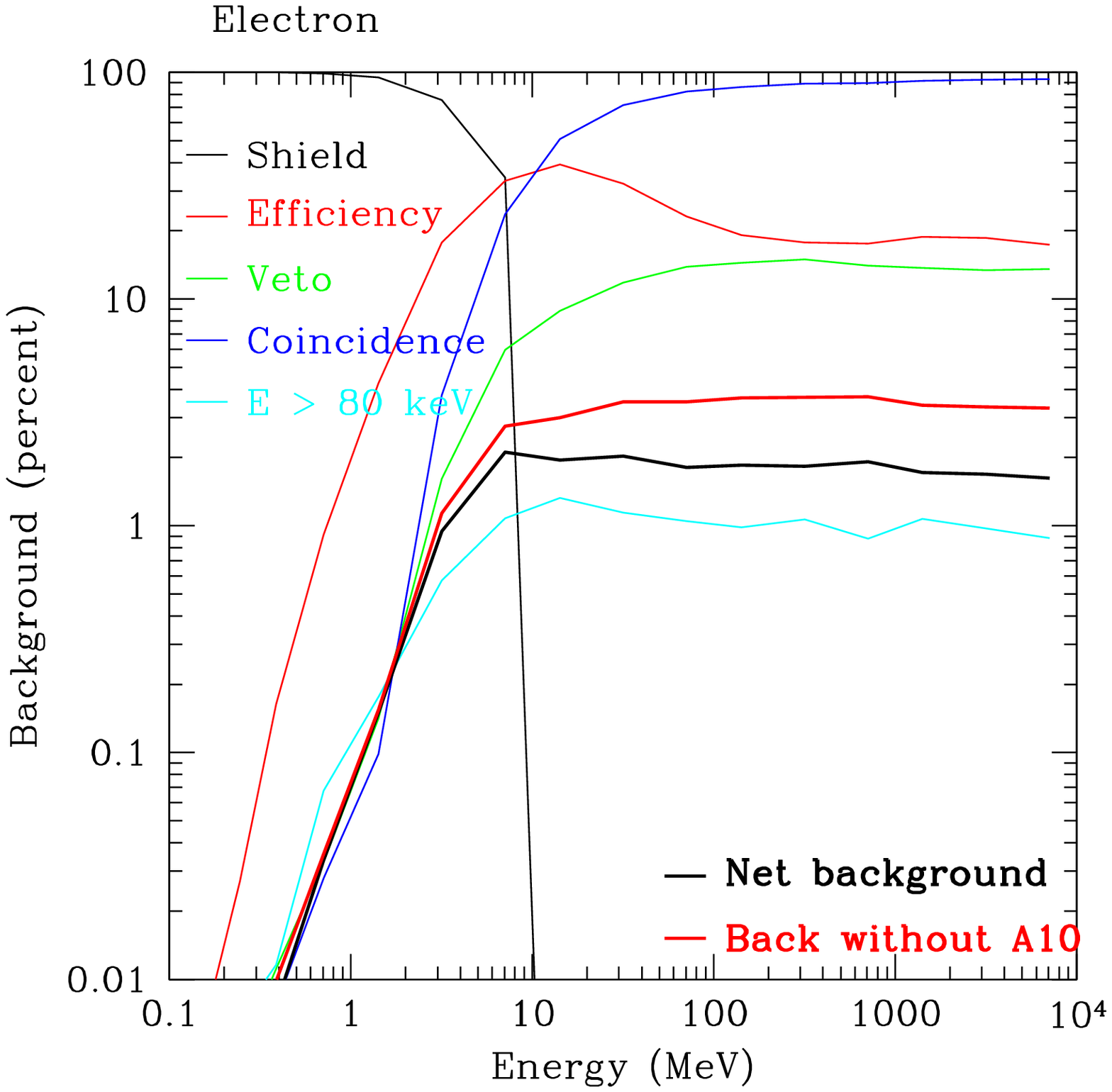}
\plotone{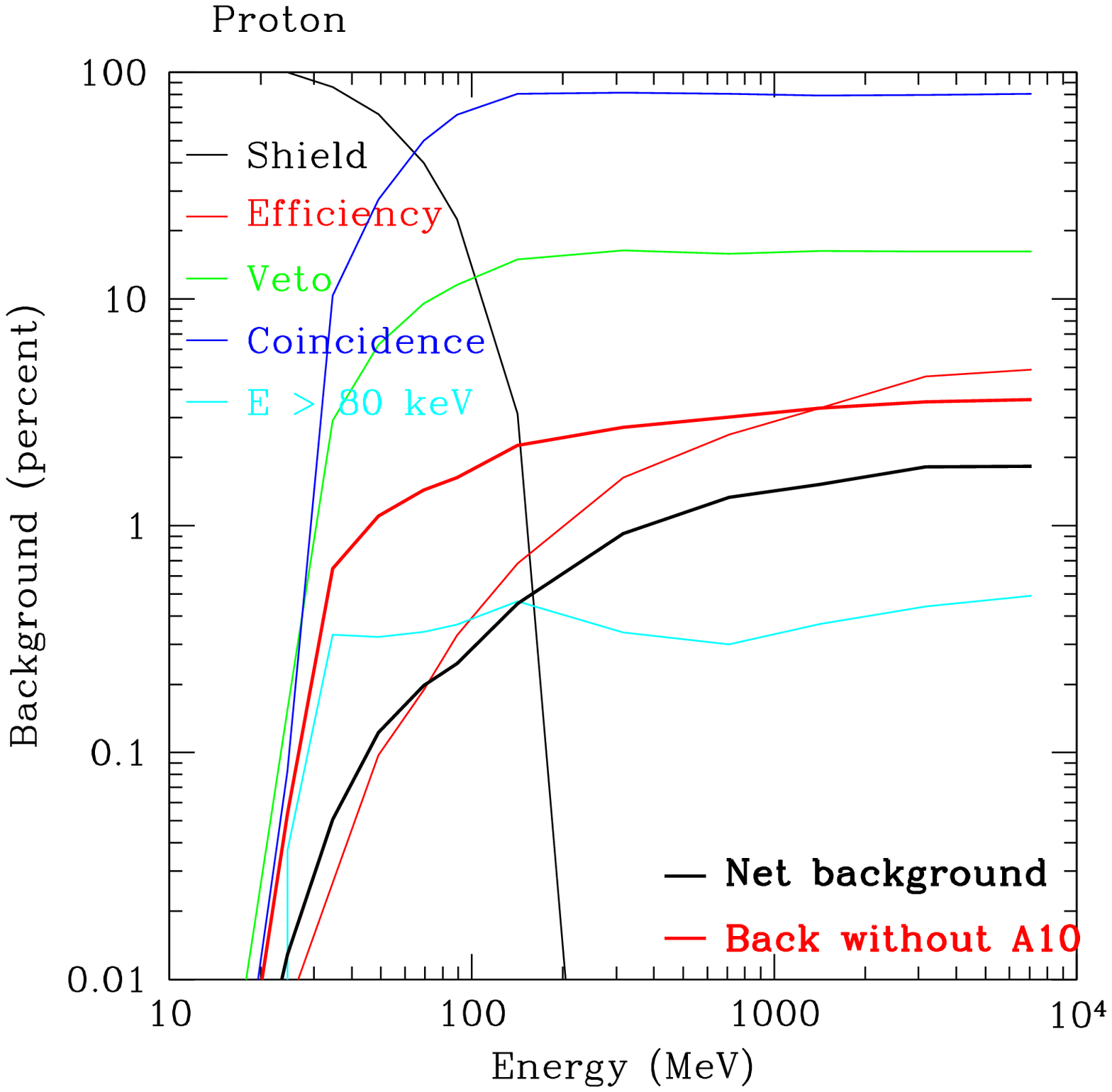}
\caption{Contribution to background rejection from Shield, Efficiency,
Coincidence, Veto layers and excess energy for gamma, electrons and protons
is shown as a function of energy. The net background in the detector with
and without veto anode A10 is also shown.}
\label{fig:backsim3}
\end{figure}

The results are shown in Figure~\ref{fig:backsim3}.
Photons at energies below about 50 keV are almost completely absorbed by the shield.
At higher energies an increasing fraction penetrates through the shield to reach the
detector, but at these energies the efficiency of detector is rather low and a large
fraction of these pass through the detector without interacting. For energies above 1 MeV the efficiency of
the detector is very low and 70\%--90\% of the events are rejected because of this.
The anti-coincidence logic accounts for rejection of up to 30\% events at high energies,
while veto layers reject up to 4\% events at high energies. For energies above 10 MeV
the anti-coincidence is about a factor of 4--7 more effective in rejecting background
as compared to veto layers.
The fraction of events which pass through all rejection criterion is about 1\% or less
at all energies. Also shown in the figure is the result if veto anode A10 is not functioning.
It can be seen that the increase is typically about 60\% at highest energies and drops
down to 10\%--20\% at low energies, where the flux of gamma rays is likely to be larger.

Similarly, electrons up to energy of 1 MeV are almost completely absorbed in the shield,
but the shield is totally ineffective beyond an energy of 10 MeV. For electrons the
detector is fairly efficient and only about 20\%--30\% of events are rejected on that count.
However, in this case anti-coincidence is very effective and about 90\% of the events
at highest energies are rejected by that. The fraction of events rejected by
detector efficiency decreases to 10\% at 5 MeV.
The veto layers account for about 10\% of the rejections at high energies and is typically
less effective by a factor of 6 as compared to anti-coincidence at energies above 10 MeV.
The net background recorded by the detector is about 2\% of the incident flux at high energies and increases
by about a factor of 2 when veto anode A10 is not functioning. At lower energies of
1-10 MeV, where the electron flux is likely to be large, the increase due to A10
not functioning is much less.

For protons the shield is effective up to an energy of 20 MeV, beyond that anti-coincidence
is the dominant factor in suppressing the background, accounting for over 80\% at high
energies. The veto layers account for about 20\% events at high energies. The net
background recorded by the detector reaches 2\% at high energies, but at these energies
the flux of protons should be negligible. At energies below 500 MeV the background is
less than 1\%. If A10 is not functioning the background could be significantly higher
and drops below 1\% at about 50 MeV. The large contribution from A10 at intermediate
energies comes because at these energies the protons get significantly attenuated in
the shield and veto layer and after traveling through these they do not have much energy
left and hence the event will be registered if the veto anode is not functioning.
At higher energies the particles pass through multiple layers and can get rejected
through anti-coincidence even when one veto anode is not functioning.

\begin{figure}
\epsscale{.95}
\plottwo{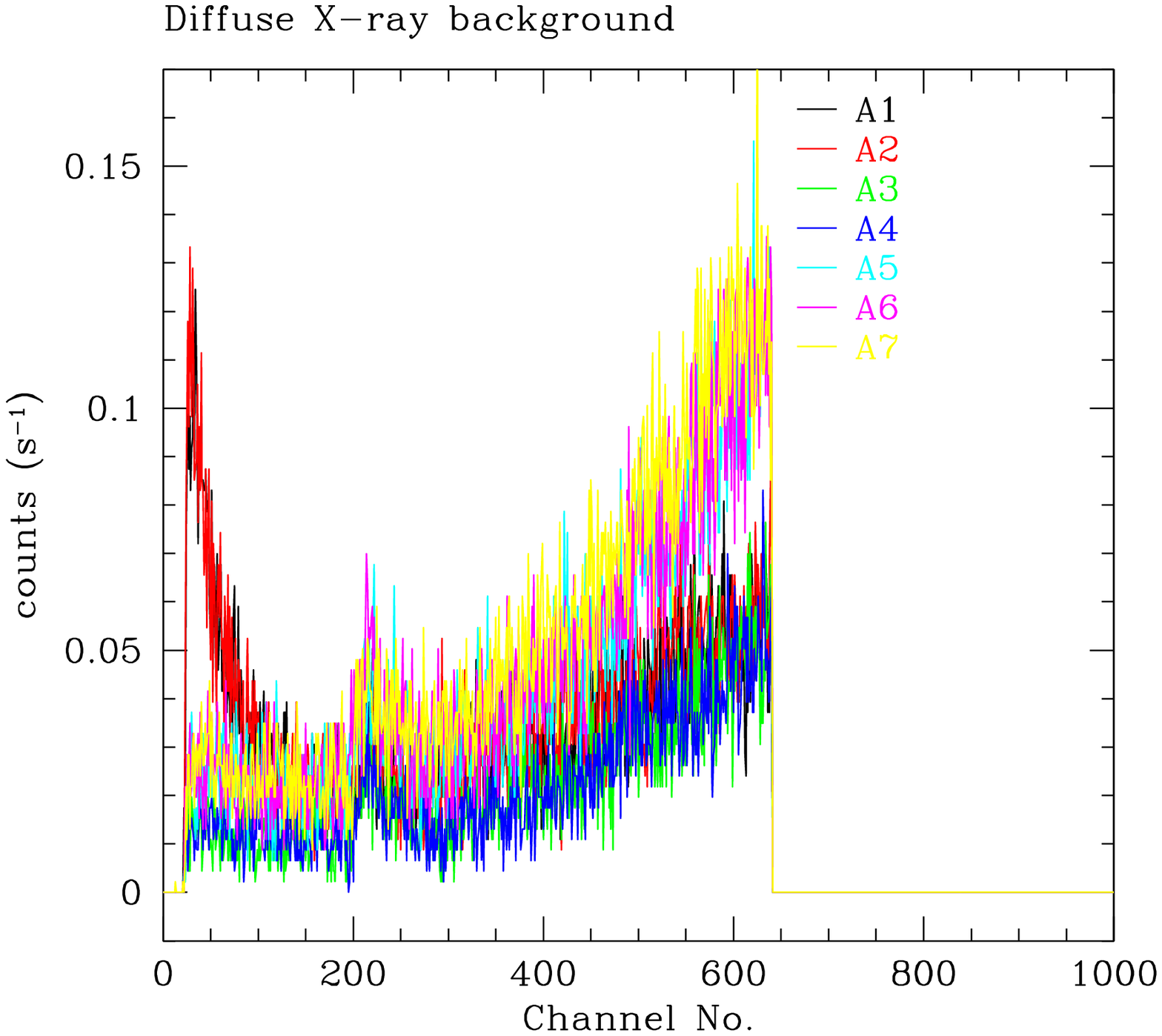}{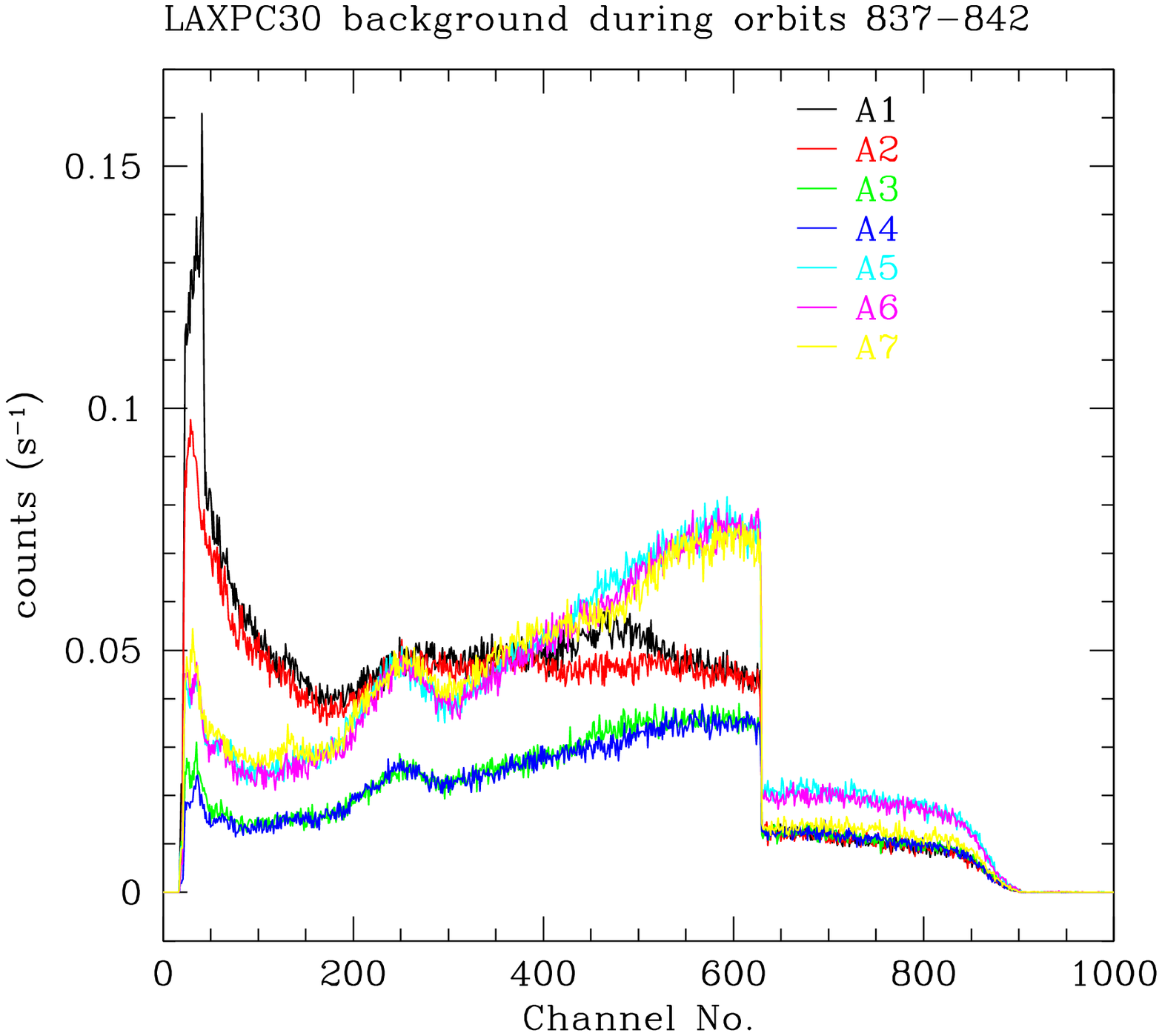}
\caption{The simulated background from cosmic X-ray background (left panel)
is compared with the observed background in detector LAXPC30 in orbit
(right panel).}
\label{fig:back}
\end{figure}

The charged particle flux in orbit outside of SAA region, where the detector
is switched off, is rather small and most of the contribution is expected
from cosmic X-ray background. The spectral form of this radiation flux
can be approximated by \citep{man79,sch80,dean91}
\begin{equation}
\frac{\mathrm{d}N}{\mathrm{d}E}(E_\gamma)=87.4E_\gamma^{-2.3}\;\mathrm{cm}^{-2}
\mathrm{s}^{-1} \mathrm{keV}^{-1} \mathrm{sr}^{-1}.
\end{equation}
The resulting background spectrum is compared with actual background observed
in the orbit (Figure~\ref{fig:back}). The simulations suppressed all events with
energy exceeding 80 keV and as a result there are no counts beyond the ULD
in the simulated spectrum, while in the observed spectrum the double events
contribute to counts in this energy range.
It can be seen that considering
the approximations made the agreement is reasonably good. More comparisons
are discussed in Section 6. It may be noted that a smoothed version of this
background spectrum was only used for simulating the LAXPC observations
before launch. After
launch the observed background is available.

\section{Calibration in orbit}

\begin{figure}
\epsscale{.65}
\plotone{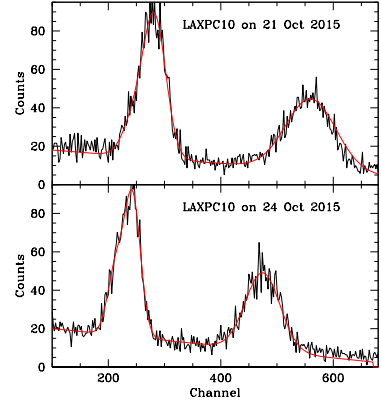}
\caption{The spectrum in veto anode A8 from on-board calibration source
before (top panel) and after (bottom panel) purification for LAXPC10.}
\label{fig:pure}
\end{figure}

After the launch of AstroSat on September 28, 2015, LAXPC detectors were
switched on in phases and on October 19, 2015 the high-voltage was raised
to the value set during the ground calibration.
The first round of gas purification was carried out during October 20--22, 2015.
The energy resolution of the detector was estimated using the on-board
calibration source in veto anode A8. The results are shown in Figure~\ref{fig:pure} which compares the spectrum before and after purification. Since the
detectors were kept in air for several days before launch some impurities
had accumulated in the detectors thus degrading the energy resolution.
Before purification the energy resolution at 29.8 keV of LAXPC10, LAXPC20 and LAXPC30 was
respectively, 17\%, 12\%, 13\%, while after purification it improved to
14\%, 12\% and 10\%, respectively.
After purification, energy resolution was close to that on ground.
As expected the gain also reduced after purification.
Nevertheless, a second round of purification was performed on November 22,
2015. After this for LAXPC10 and LAXPC20 purification was done on
August 18, 2016.

\begin{figure}
\epsscale{.95}
\plotone{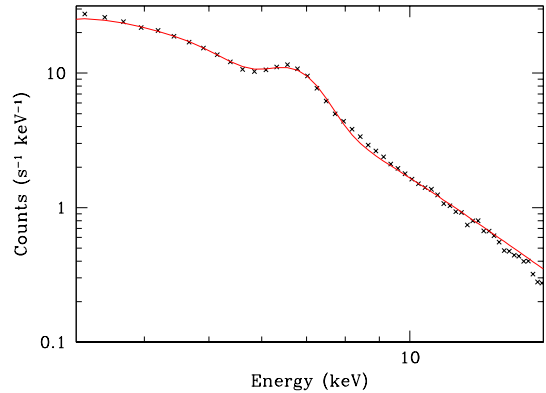}
\caption{The spectrum of Cas A in LAXPC10 is fitted by a power law and Gaussian line
to get the energy scale. The black points show the observed
spectrum, while red line shows the fitted spectrum.}
\label{fig:cas}
\end{figure}

During the performance verification phase lasting till March 2016, various
sources were observed to calibrate the instruments. Among them was Cas A,
which shows an Iron line at 6.62 keV \citep{yam14}. The spectrum from observation
on January 4, 2016, was fitted to check
the peak channel mapping at low energies and the result is shown in Figure~\ref{fig:cas}.
The peak was found at energies of $6.7\pm0.1$, $6.8\pm0.1$ and $6.7\pm0.1$ keV for LAXPC10,
LAXPC20 and LAXPC30, respectively.
The width of the peak was found to be consistent with the expected
energy resolution of the detectors. Since the detector response was used
in the fit, the fitted width was very small as compared to the expected energy
resolution of about 20\% at this energy.
This confirmed that the used channel to energy mapping at this
energy is within 0.2 keV of actual value.

All data received after launch have been analyzed using LaxpcSoft
software\footnote{http://www.tifr.res.in/\~{}astrosat\_laxpc/LaxpcSoft.html}
for analyzing LAXPC data.
All results reported in this work have been obtained using this software.

\subsection{The field of view and alignment}

\begin{figure}
\epsscale{.95}
\plotone{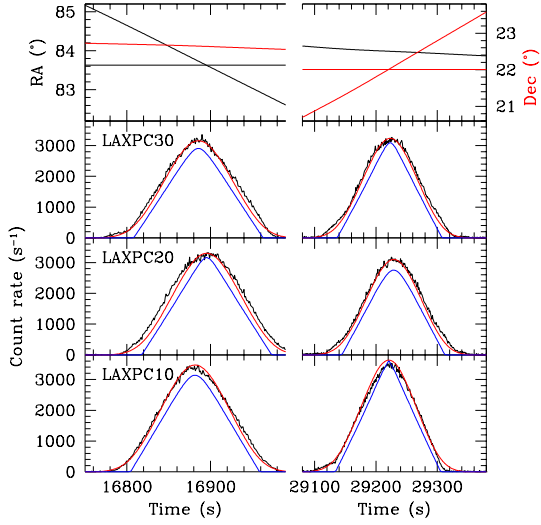}
\caption{The count rate during scan across the Crab in right
ascension
(left panel) and declination (right panel) for all three LAXPC detectors.
The black lines show the observed count rate corrected for dead-time and
background, the blue lines show the count rate expected for an ideal
collimator scaled to the observed maximum, while the red line shows the
same when a random misalignment by $12'$ in collimator is included.
The top panel shows RA (black lines) on scale shown on left side and
declination (red lines) on scale shown on the right side. The horizontal
lines in the top panel show the coordinates of the Crab.}
\label{fig:scan}
\end{figure}

To find the alignment of LAXPC detectors with respect to the satellite
pointing direction we performed a scan across the Crab.
The scan was performed along both right ascension and declination at a
rate of $0.01^\circ$ s$^{-1}$ covering a range of $\pm3^\circ$ from the
nominal position of the source.
This exercise was repeated three times and during each run three scans
were done along right ascension and declination. The result obtained during
the last run in February 2016 is shown in figure~\ref{fig:scan}.
It can be seen that all detectors do not show the peak counts at the
same time and also not when the pointing direction matches Crab position.
Analysis of these data gives the offset for each detector as listed in
Table~\ref{tab:crab}. It is possible to choose a pointing direction
to maximize the total counts from LAXPC detectors and some observations
have been made with such a pointing. However, other instruments on board
AstroSat have a smaller field of view and placing the source at the
LAXPC efficiency peak would place it far off the axis of some of these.

Apart from the misalignment, the scan profile can also
give some estimate of the quality of collimator, from the rate at which counts
fall off with offset from the source. For an ideal collimator the counts
should decrease linearly with offset. There is a significant deviation from
this linear profile in the region where the count rate is
close to maximum as well as where the count rate approaches the background.
This departure can be modelled, if we assume that all cells in the collimator
are not perfectly aligned with each other and that there is a scatter with
Gaussian distribution. For LAXPC10 scan profile, the width of the distribution is
found to be about $12'$, while for other detectors it is somewhat larger.
Figure~\ref{fig:scan} compares the observed profile with that expected for
an ideal collimator as well as one with a scatter of $12'$. Both the simulated
profiles are scaled to give the maximum count rate that is observed during
the scan. It can be seen that the profile for ideal collimator shown by
blue lines in the figure does not match the observed profile and gives a
somewhat smaller field of view. In some cases the peak counts in the profile
for ideal collimator are below the observed counts. This is because in
these scans the offset angle was never close to zero and at the minimum
offset the ideal collimator gives significant reduction in count rate.
For LAXPC10 the profile with a scatter of
$12'$ is close to the observed profile considering the assumptions made.
The simulations only try to account for misalignment between different
collimator cells, but do not account for other imperfections in the collimator
and hence we do not expect a perfect agreement. Nevertheless, for LAXPC20
and LAXPC30 the deviations are somewhat larger and it appears that these
detectors have larger scatter or other imperfections in their collimator.
All collimators have the FWHM of field of view of about $55'=0.92^\circ$
at low energies when measured along the sides of the detectors.
These scans show the total count rate integrated over the entire energy
range, but the count rate is dominated by low energy photons and hence the
resulting field of view is applicable at low energies.

\begin{figure}
\epsscale{.95}
\plotone{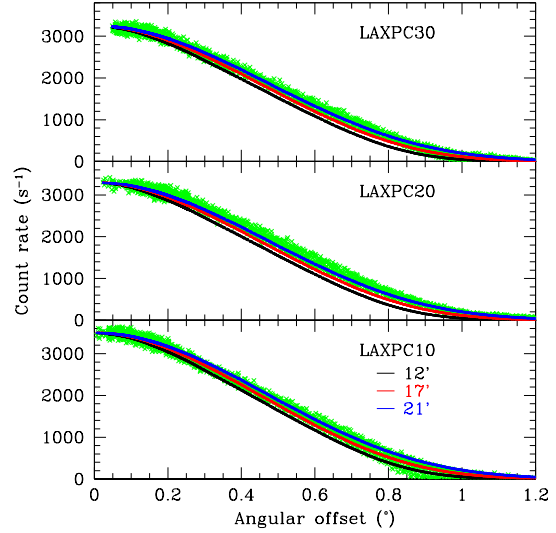}
\caption{The count rate during scan across the Crab as
a function of the calculated angular offset for the three LAXPC
detectors. The green points are the observed counts and other lines
are the simulated profiles for various levels of collimator misalignment
as marked in the bottom panel.}
\label{fig:fovsig}
\end{figure}

Figure~\ref{fig:fovsig} compares the observed profile during scan by
combining all six scans as a function of calculated angular offset. It can
be seen that the profiles of all scans merge into one which shows that the
calculated offset is correct. Further, this figure compares the observed
profile with simulated profiles with a few different levels of scatter in
the alignment of collimator. The shape of the simulated profile does
not perfectly match the observed profile, and it is likely that there are
other imperfections in the collimator which contribute to the differences.
Nevertheless, it is clear that LAXPC10 appears
to be consistent with lower scatter in collimator,
while for other detectors the scatter is larger, with LAXPC20 having
the largest scatter. This is consistent with the estimated effective areas of the
three detectors as discussed in the next subsection.
It can be seen that for LAXPC10 the scatter is between $12'$ and $17'$.
It may be noted that this result is only used to estimate the loss of efficiency
due to collimator imperfection. The final effective area is estimated by
simultaneous fit with NuSTAR observation and does not use this estimate.

\begin{figure}
\epsscale{.95}
\plotone{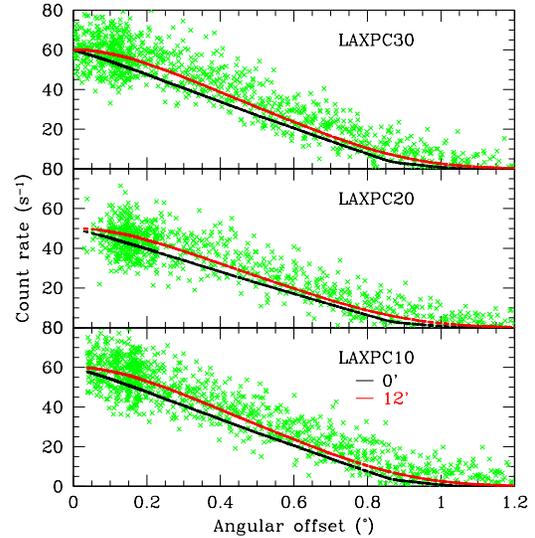}
\caption{The count rate in energy range of 40--60 keV during scan across the Crab as
a function of the calculated angular offset for the three LAXPC
detectors. The green points are the observed counts and other lines
are the simulated profiles for two levels of collimator misalignment
as marked in the bottom panel.}
\label{fig:fovsig50}
\end{figure}

We can estimate the field of view at higher energies by considering
the counts in different energy bands. If we consider counts in the
energy range of 40--60 keV the FWHM of field of view is about
$63'=1.05^\circ$. Figure~\ref{fig:fovsig50} shows the results for this
energy range and compares them with simulations with perfect collimator
and one with dispersion of 12'.

The offset angle for the detectors range from
$0.07^\circ$ to $0.15^\circ$. For an ideal collimator an offset of
$0.15^\circ$ would reduce the efficiency by about 15\%, but considering the
scatter in alignment of individual cells at the level that matches observations
the decrease is about 7\%. The scatter in the collimator alignment
reduces the efficiency by about 15\%, which gives a total reduction of
about 20\% in efficiency for LAXPC10. For other detectors the reduction
is comparable as the smaller offset compensates for larger scatter in collimator
pointing.  Table~\ref{tab:crab} also lists the estimated loss of efficiency
for all detectors due to pointing offset and collimator quality.
This reflects in the effective area considered in the
next subsection.
It may be noted that these offsets have been obtained with respect to the
AstroSat satellite axis. However, all
instruments are not perfectly aligned to this axis and as a result different
pointings are used while observing with different instruments. Hence there
could be an offset of about $0.1^\circ$ with respect to the pointing
obtained here. As a result, it may be better to keep the normalization of effective
area as a free parameter while simultaneously fitting spectra from different
detectors or instruments.

\subsection{Effective area}

As mentioned in Section 4.2, the effective area of the combined LAXPC detector
was expected to be about 8400 cm$^2$ assuming nearly perfect collimator and
alignment. From the scan across the Crab we have estimated that
the misalignment with respect to the satellite pointing direction and
imperfections in the collimator reduce the efficiency and hence the
effective area by about 21\% for LAXPC10.

Apart from the collimator imperfection the count rate would also be affected
by the dead-time associated with rejected events. This count rate is more or
less independent of the source being observed, though it varies during an orbit
by about 25\%. This rate may be estimated using appropriate counters in BBC mode.
However, this rate in orbit is very different from that on ground and hence
it could only be estimated after launch. From the BBC mode data the total
counts in rejected events is estimated to be 2000--2500 s$^{-1}$.
With an estimated dead-time (from electronics design)
of about 35 $\mu$s for rejected events, it gives a loss of efficiency by
7--9\%. Combining both these factors gives a reduction in effective area
by a little over 27\% to about 6000 cm$^2$.

Figure~\ref{fig:eff} shows the effective area of the three LAXPC detectors
with correcting factor as explained below.
The efficiency or the effective area is defined by the probability that a
photon of given energy will register a valid event in the detector, the
energy deposited may be less than the incident energy, for example, due
to Xe K fluorescence X-ray escaping from the detector volume. Thus all these photons
may not be detected at the correct energy.
The difference between the effective areas of the three detectors is mainly from the collimator
quality. For a perfect collimator all detectors show similar
effective areas. The total effective area of the three detector is
shown in the right panel. The sharp dip at low energies is due to absorption
in the Mylar window at the top of the detector. The dip around 34 keV is
due to reduced absorption coefficient just below the K-edge of Xenon. At high
energies the efficiency decreases because of decrease in the absorption
coefficient of the gas.

To get a better estimate of effective areas for each LAXPC detector,
simultaneous observations of the Crab with NuSTAR \citep{har13}
were carried out on 31 March 2016. Simultaneous fit of NuSTAR and each of
the LAXPC detectors to a power law spectrum gave the relative normalization
of each detector assuming the response matrix for NuSTAR.
The resulting fit for LAXPC10 is shown in Figure~\ref{fig:nustar}.
From these fits the relative normalization with respect to NuSTAR of the three LAXPC detectors were
calculated to be 0.92, 0.84 and 0.89, respectively for LAXPC10, LAXPC20 and
LAXPC30. The simultaneous fit with NuSTAR including 1\% systematics in LAXPC
detector response gave a $\chi^2$ of 1.0 per degree of freedom.
The effective area of each detector using the normalization as determined
above and the combined effective
area is shown in Figure~\ref{fig:eff}. The maximum effective area of combined
LAXPC detectors is about 6000 cm$^2$ around 15 keV and reduces to about
5600 cm$^2$, 4100 cm$^2$ and 2200 cm$^2$, respectively at 40 keV, 60 keV and
80 keV. These effective area estimates are for pointing as determined
by the satellite axis. Since the pointing directions are not the same for
all observations, there could be an uncertainty in effective area by 5\%.

\subsection{The response matrix}

To calculate the response matrix we use GEANT4 simulations with energy
resolution and energy to channel mapping as obtained in Section 4.1, for
photons of fixed energy. We have
used 338 energy bins in the range 2.05 to 145 keV to calculate the response.
For each energy we use $10^7$ photons incident normally at the top of detector
and uniformly distributed over the area. The efficiency of detection is
multiplied by appropriate factor to take care of factors mentioned in
the previous section. As explained in the previous section these
factors are determined by a simultaneous fit to Crab spectrum with NuSTAR
observations. Figure~\ref{fig:resp} shows the response of LAXPC30
for a few selected energies. All these are combined into a response file
which incorporates the effective area and can be used with XSPEC package
for fitting observed spectra.

For energies above the K-edge of Xe we get
multiple peaks. The peaks at lower energies are due to escape of Xe K fluorescence X-rays
outside the detector. For energies of 40.5 and 50.5 the two peaks due to
escape of K$_\alpha$ and K$_\beta$ are resolved. For energies of around
80 keV or above, the main peak may not be complete as due to finite
resolution a part of the peak may go beyond the ULD.
However, the counts do not go to zero beyond this limit as the double
events are still recorded. At high energies even the main peak due to
single and double events may be resolved as the difference increases
due to nonlinearity of response. This has not been verified from any
source with known energy.

The calculated response matrix was used to fit the spectrum for Crab
taken during January 2016. Although, the overall fit was good there
were some differences around the Xe K- and L-edges. To correct for this
the energy to channel mapping was locally modified in these regions.
Another correction which was required to correct for excess events around
30 keV, which are coming from a gap in the shield. The aluminium box
containing the collimator has some ribs of about 1 cm on all sides and the
shield is put on this ribs. This leaves a small gap between the shield
and collimator box (see Figure~\ref{fig:detec}), through which high energy photons can pass through as
the ribs made of aluminium are not enough to block them. Only photons coming in
the direction close to the direction of pointing can pass through this gap.
Although, this region is outside the volume covering the main anodes, the
chamber enclosing the gas is larger and these photons can still interact
with Xenon in this extra volume. Any Xe K fluorescence X-rays emitted in this region can
enter the detector volume and trigger the detector. This contribution has
been adjusted by matching the Crab spectrum. Typical contribution is about
1\% of high energy photons above the Xe K-edge. This gap was not included
in GEANT4 simulations.

This tuned response has been used for all calculations. The fits to Crab spectrum
by a power-law for all 3 detectors with 1\% systematics is shown in Figure~\ref{fig:crabfit}.
It can be seen that for all detectors the fit is good to within 2\%.
Further, for LAXPC10, LAXPC20 the fit is good even beyond 80 keV, while
for LAXPC30 it deteriorates beyond 80 keV.

The drift in gain of detectors need to be accounted for while fitting the
spectrum using a response matrix as well as while subtracting the background.
Since the background spectrum is observed at a different time the gain
may not be the same as that during source observation and it may be necessary
to shift the background to the same gain as that during the source observations.
The extent of shift required can be obtained by fitting the peak of
calibration source in Anode A8. The gain shift is applied by assuming that
the observed shift is due to change in coefficient $e_1$ in Eq.~\ref{eq:gain}.
This assumption may not be correct as other coefficients also could have
changed, but it is not possible to determine the change in all coefficients
from the shift in 30 keV peak.

A simple shift in gain will not
give the correct spectrum at energies beyond about 35 keV as the logic for
adding the double event checks for the energy to be in the interval of Xe
K photons. The shift in gain will cause the double events to be missed
and it is not possible to account for this while shifting the gain by a
linear transformation in energy scale. This limits the ability to correct
for gain shift between source and background and can introduce features in
the resulting spectrum.

Another possibility is to use the spectrum obtained
during the Earth occultation during the same orbits when the source was
observed. This ensures that there is no gain shift between the source and
background, but the observed spectrum during Earth occultation is not
the same as real background as the Earth albedo would also contribute, but
this may be the only option if no background observation with nearby gain
is available. This issue is discussed in the next section. Once the
background corrected source spectrum is available a response which has
nearby gain can be used to fit the spectrum. The LaxpcSoft software also
has option to shift background spectrum to align with source spectrum and
it also identifies the response to be used. It may help to make finer
adjustment in gain by applying a linear transformation to the energy
scale, particularly the constant term, $e_0$ in Eq.~\ref{eq:gain}.

\subsection{Timing characteristics}

Since the prime use of LAXPC instrument is for timing studies, the timing
characteristics were investigated during ground calibration and the
Fourier transform of the time series did not show any unexpected features.
After launch this was checked in more detail with additional data for bright
sources \citep{yad16b}.

There are some known problems in time
tagging of events. When the lowest time byte in time, T1, rolls over from
FF to 00, the next higher byte, T2, does not increment at the same time.
There is a delay of about 20 $\mu$s between updating the two bytes of time.
If an event occurs in between the time when two time bytes are updated
and if the count rate is sufficiently high, apparent time ordering of events
may also get disturbed as apparently this event will precede the previous event.
Similar problem is also present in roll over of higher time bytes, though
these occur less often and are easier to correct.
This problem has been corrected in software and after that all events are
correctly ordered in time.
However, the distribution of
lowest time byte is not uniform and shows some features as shown in Figure~\ref{fig:hist}.
The origin of this effect is not known, but it can result in spurious
periodicities if a period of exactly 2.56 ms (frequency of 390.625 Hz) or its
multiples is used. The resulting pulse profile will match that shown in
Figure~\ref{fig:hist}. The reason for this distribution is not
understood but it is independent of time. Since the profile is strongly peaked at one value, most of the
power would be in higher harmonics of this frequency and this feature is
not seen in Fourier transform, but it
can show up if period folding is applied.
This effect is seen if the instrument time as recorded through STBG is used.
If the arrival times are converted to UTC (red line in Figure~\ref{fig:hist})
the effect goes away because of a slow drift in the STBG clock.
We do not believe that this reduces the time resolution as we have
obtained power density spectrum up to a frequency of 50 kHz and it agrees
with that expected from a Poisson distribution with dead-time \citep{yad16b}.

After launch a peak around 50 Hz was found in the power density
spectrum of bright sources in LAXPC20.
This feature is present only in anode A1 and at energy of
less than about 5 keV. It is also found in weak sources with a smaller
significance. This has been traced to noise in the amplifier for anode A1.
This anomaly was also present during ground tests.

Another source of
possible instrumental feature in the power density spectrum is the calibration
source in veto anode A8. To allow this spectrum to be observed, $1/128$ of
the events registered in veto anodes are not rejected and get entered in
the event mode file. Since the count rate in veto anodes is roughly constant,
this gives a peak around 8-11 Hz in count rate from veto anodes. The lower
frequency is seen in LAXPC10 where veto anode A10 is disabled and hence the
count rate is lower. During observation of bright sources veto anodes count rate
can be reduced due to effect of dead-time, thus reducing the frequency at
which the peak occurs. For bright sources it is possible that this frequency
will show up in the source counts also and in some cases a weak feature
has been seen around this frequency in the power density spectra.
This has to be accounted for while interpreting a power density spectrum.

\section{The detector background}

Although, the background in the laboratory was very stable, observations
in orbit shows some variation with time and hence it is necessary to model
the background, so that source contribution can be extracted from any
observation. This is particularly important for faint sources. The total
background counts averaged over an orbit in LAXPC10 is about 260 s$^{-1}$, while in LAXPC20 and
LAXPC30 it is about 200 s$^{-1}$. The higher count rate in LAXPC10 is
because one of the veto anode is switched off. The variation in background
count rate is about 20\% around the mean value.

The observed spectrum of background with counts in all anodes added is
shown in Figure~\ref{fig:back3} for all LAXPC detectors during three
different observations. One of these observations was when the satellite
was pointing towards the Earth. For comparison simulated background spectrum
calculated before launch is shown in one of the panels and it is clear that
the agreement is reasonably good. Since the simulations did not include contribution
from charged particles, it shows that this contribution is small in the orbit.
There is some difference between the three spectra for all detectors as shown
in the right panel. It can be seen that there is some energy dependence in
the difference, hence scaling the spectrum by the total counts may not be
enough. It is known that the cosmic X-ray background which is the main
contributor to the LAXPC background, has spatial fluctuation of about
7\% ($1\sigma$) \citep{rev03} in the 2--10 keV band. These fluctuations
cannot be modelled and set the flux limit below which faint sources may
not be observable. It is not clear how these spatial fluctuations will
contribute to LAXPC background, as the background is contributed by particles
coming from almost all directions, much of these fluctuations will be
averaged out.
Even a bright X-ray source with hard spectrum which is off-axis can also
contribute to the background as the typical rejection efficiency for hard photons
above 100 keV is about 99\% (Section 4.4), so 1\% of these photons can be detected. For
example, for Crab, we would expect a few counts per second
when the source is shining on the side of the detector.

To cover some variation LAXPC has observed the blank sky coordinates listed in
Table~\ref{tab:back}. The table also gives the total count rate observed in
each detector. Considering the exposure time, the statistical error in
these counts is less than 0.1 s$^{-1}$ and all the variation can
be considered as systematic error.
Some of the variation seen
in LAXPC30 counts is due to shift in gain and decrease in density as discussed in the next section.
The gain of the other detectors has also been drifting slowly and that could
account for some of the variations. The background in LAXPC10 appears to
be increasing with time, particularly, after August 2016 the count rate is
higher than that during earlier observations.
Table~\ref{tab:backlay} shows the background counts in different layers
and different energy bins as observed during March 2016.
Some of the variation in background could
be due to variation in satellite position and we attempt to model this.
We have tried two models one based on observed correlation with the count
rate of events which exceed ULD, and another based on satellite position.
We also tried to model the background variation by using induced
radioactivity during SAA passage, but did not find any improvement in the
model with this addition and hence we have not included this effect.

\subsection{Background model based on ULD counts}

Figure~\ref{fig:back1} shows variation in background during one day in March 2016. All
the three detectors show increase in counts as the spacecraft approaches
the SAA region. In the SAA region the high voltage is switched off and hence no
counts are recorded. The figure also shows ULD count rate which measures the
number of events which exceed the upper energy threshold. This count rate
also shows similar behavior. Further, the increase in the count rate as
the satellite approaches SAA region is not the same in every orbit. It
depends on the latitude at that point. AstroSat is placed in a nearly circular
orbit with inclination of $6^\circ$ to the equator. If AstroSat is close
to $6^\circ$ S while passing the SAA region, the effect is maximum. On the
other hand when it is close to $6^\circ$ N, the SAA passage is short
and the charged particle flux is also much smaller.
The background counts are correlated to ULD counts as shown in the right panel
of Figure~\ref{fig:back1}.
The correlation coefficient
between the two is found to be 0.98, 0.93, 0.96 respectively for LAXPC10, LAXPC20,
LAXPC30. Since the ULD counts do not vary too much with source counts, this correlation
can provide a simple model for background. For high count rate sources we
need to apply correction for dead-time in the ULD count rate also.
To get the resulting model
background spectrum the range of ULD counts is divided into 4 equal bins
covering the full range for each detector. The boundary of these bins is
marked by vertical lines in Figure~\ref{fig:back1}. For LAXPC30 the background
as well as ULD counts are shifted to lower values due to a shift in gain.
The background count rate in each
of these 4 bins is shown in Table~~\ref{tab:uld}.

The spectra obtained for each ULD bin is shown in Figure~\ref{fig:uld}.
For a given source observation, the time is divided into these ULD bins and
the background spectra for these bins is weighted with time spent in the bin
to get resulting background spectrum. The ULD counts need to be corrected for
dead-time using the count rate of genuine events. This method is sensitive
to shift in gain which affects the ULD counts. If the gain shifts on positive
side some of events at high energy end will go beyond the ULD and the observed count
rate would reduce while the ULD count rate will increase thus affecting the
observed correlation. For LAXPC30 the ULD counts
have reduced significantly due to leak and it is not possible to use this
model.
This background model does not give the background count rate as a function
of time, but only corrects for the average background spectrum during an
observation. Hence, it cannot be used to obtain the light curve for source
counts.

\subsection{Background Model Based on Satellite position}

Since the SAA region is located mainly just south of the equator, the background
counts depend on both longitude and latitude. A background observation over
1 day is expected to cover most of the region where the detector is not
switched off. Thus using such an observation it is possible to fit the
2D dependence using product B-spline basis functions and this model was
used. There is also some variation with altitude. Although, the AstroSat
orbit is nearly circular there is a variation by about 15 km in altitude
during the orbit, which may be enough to give some variation in background
counts. Hence we also attempted to fit in 3-dimensions using product B-splines.
Some regularization was also applied while calculating the fits
\citep{ant12}. However, the fitted background did not show any significant
improvement over the 2-dimensional fit.
The fitted background for all
three detectors are shown in Figure~\ref{fig:backfig} for observations
during March 2016. Since the detectors are off during SAA
passage that region is excluded from the figure.
The residuals obtained after subtracting the fitted background from the
observed value is shown in Figure~\ref{fig:backfit}.

Until  August 4, 2016 the SAA region was defined to be between the longitudes
of $-110^\circ$ to $0^\circ$ (or $250^\circ$ to $360^\circ$). It was felt that this includes some region
in the north where the charged particle flux may be low enough for the
detectors to be operated. Hence, a modified definition based on a SAA model
was implemented which accounted for the latitude dependence, and the entry
to SAA was defined to be the point where longitude equals $-110^\circ+4(\hbox{latitude}+6^\circ)$.
With this change in definition of SAA region, a revised background model is
needed and this was obtained from observations of background on August 16, 2016
and the results are shown in Figure~\ref{fig:backfigsaa}. There appears to be
some region in the north where background counts are quite large and it is
necessary to eliminate this region from good time interval to get reliable
model for background. There is also some variation in the average background
count rate between the two background observations, mostly because of shift
in the gain of detectors between the two observations (see Table~\ref{tab:back}
and Section 7).

This model gives the total background counts at any position of the spacecraft.
For source observation, we take the mean counts over the period when the source
was observed and scale the background spectrum to this mean count. Apart
from the spectrum this model also gives the background counts at any time
and this can be subtracted from the total light curve to get the source
contribution. In most regions the background model agrees with actual counts
to within 2\%. But there is some unexplained variation with a period of
slightly less than a day and an amplitude of a few counts per second, which is not modelled. The origin of this periodic
variation is not known.
The background spectrum obtained from this model does not account for
variation in spectrum over the observed period. It only accounts for the
total count rate, which is used to scale the spectrum.

In order to obtain the fits to background we need observations
covering the entire range in longitude and latitude, which requires observations
spanning at least one day. Apart from this we also require that the entire
range of longitude and latitude are covered during the time when the
satellite is not pointing towards the Earth. Some pointing directions are
excluded by this criterion at a given time. For example, the observation
of Sky-5 during August 30, 2016, did not include any region with latitude
greater than $4^\circ$ and hence cannot be used to obtain the background
fit.

\subsection{Background from observation during Earth occultation}

During observations of all sources with declination less than about $60^\circ$
there are periods during orbit when the source is occulted by the Earth
and the satellite is facing the Earth. The source will not be visible
during this period but the observed spectrum would not be identical to
normal background as there would be some contribution from the Earth albedo and
the Earth would block the diffuse X-ray background which directly goes
through the collimator. The advantage of using the spectrum during the
Earth occultation as a background
is that a separate background observation is not required and more
importantly it ensures that there is no shift in the gain between the
source and background spectrum. Hence we have tried to study the difference
between the true background, as obtained when the satellite is pointing in
a direction where there are no X-ray source detectable by LAXPC, and the Earth
occultation spectrum obtained when the satellite is pointing towards the Earth.
This can be conveniently done during the background observations as most
of the background regions observed are at latitudes low enough to have
Earth occultation.

Figure~\ref{fig:eo} shows the ratio of counts during Earth occultation and
background during five different background observations for all three
LAXPC detectors. It can be seen that particularly, at low energies the
ratio is far from 1 for LAXPC20 and LAXPC30. LAXPC10 shows a different
behaviour because one of the veto anode (A10) has been disabled in that.
It is clear that use of Earth occultation as background can introduce
significant error at low energies, particularly for faint sources.
However, in some cases it may be preferable to use this, particularly for
LAXPC30 as its gain shifts significantly during a day and it is difficult
to account for that from independent background observation.

\section{Long term performance of LAXPC in orbit}

The health parameters of the detectors, like the temperature, high
voltage and various energy thresholds are monitored regularly. The temperature
has been maintained at $18\pm2^\circ$ C for LAXPC10 and LAXPC20, while that
for LAXPC30 is maintained at $22\pm2^\circ$ C. The energy thresholds have also
remained steady, except for one instance when on 19 April 2016, the LLD
of LAXPC10 was accidentally changed. This was corrected within a day. The
high voltage has held steady, but as explained below it is being
adjusted from time to time to maintain the gain of detectors.
To monitor the long term stability of detectors the peak position and
energy resolution of 30 and 60 keV peaks in the veto anode A8 from the
on-board radioactive source is monitored regularly. Figure~\ref{fig:a8}
shows the peak positions and energy resolution of the two peaks as
a function of days after launch of AstroSat. Some of these variations are
due to purification of gas and adjustment of high voltage. Table~\ref{tab:hv}
shows the times when the gain was adjusted.
The gain in LAXPC30 has been constantly shifting due to
suspected leak. As a result, the high voltage is regularly adjusted for this
detector. The 30 keV peak had shifted by up to 85 channels (out of 1024 channels) before
the high voltage was adjusted downward for the first time on 17 March 2016
after the leak was verified. During October to December 2015 the shift
in gain was small and similar to that in LAXPC10. During the first week
of January 2016, the rate of gain shift in LAXPC30 started increasing.
This gain shift has to be accounted
for in the response matrix and these have been generated for different
amounts of shift. Fig.~\ref{fig:a8z} shows the peak position of 30 keV
peak on a magnified scale.

Unfortunately, the on-board pressure gauge does not have adequate sensitivity
to get accurate measure of the leak. Hence, we attempted
to estimate the density of the gas using ratio of count rates in different
layers as was done on ground (Section 4.1). For this purpose we use counts
in the range of 20--24 keV. Since the background spectra is essentially
flat we have to subtract the background from the source spectra for this
purpose and we also need a bright source with soft spectrum to get sufficient counts in all
layers. We found Cyg X-1 as a suitable source for this purpose.
Following a procedure
similar to that on ground as described in Section 4.1, we take the ratio
of counts in energy interval 20--24 keV for different layers with respect
to the top layer and minimize the function
\begin{equation}
F_x(\rho)=\sum_{i=2}^5 (r_i-s_i)^2,
\label{eq:denx}
\end{equation}
where $r_i$ is the ratio of counts in $i$th layer to the top layer
in the observed spectrum and $s_i$
is the same ratio in simulations with prescribed density.
Figure~\ref{fig:denorb} shows
the function defined above for 5 different observations
of Cyg X-1 on January 8, April 29, June 1, July 1 and October 9, 2016.
The density is shown in
the units of initial density estimated on ground to estimate the loss
due of leakage. This appears to indicate that density in LAXPC30 is decreasing by
about 5\% of original value every month and has now reached about 30\% of the
original value. During the first observation on January 8, 2016, which was the time when
the leak became significant, the density is comparable to that estimated on ground.
This gives us some confidence on this estimate. All observations before launch
also did not show any significant gain shift.

Since the detector response depends on density, the response matrix has been
generated for different values of density differing by steps of 5\% of the
original value. These responses have been used to fit the observed spectra
for some of the sources to find the density which gives the best fit.
That also gives an independent estimate of density which agrees with the
previous estimate. By fitting the Crab observations during different times
between February 2016 and February 2017 an estimate of the density in
LAXPC30 has been obtained.
These estimates are comparable to those obtained using Cyg X-1
observations described above and also with pressure estimated from the
pressure gauge as shown in Figure~\ref{fig:lx30den}. Since the absolute
calibration of pressure gauge is not reliable, we have divided all values
by the maximum value observed and we assume that pressure is proportional
to the density.
Figure~\ref{fig:eff30} shows the effective area of
the detector with different densities. It can be seen that at high energies
the effective area has been steadily decreasing with density. But at energies
of less than 10 keV, there is little difference even when the density is
10\% of the original value.

The Figure~\ref{fig:a8z} shows that the gain in LAXPC10 is also shifting
steadily upwards, though at a much smaller rate as compared to LAXPC30.
The rate was about 0.15 channels per day, which is more than 20 times
smaller than that for LAXPC30. This rate has also increased to about
0.25 channels per day since September 2016. This could also be due to a very fine leak,
which has been there at least, since launch. If this rate is held, the
detector should function for several years. No change in density or pressure
has been detected in LAXPC10 till now. Considering the rate of change of
gain and density in LAXPC30, it can be estimated that the observed gain
shift in LAXPC10 could be due to 2--3\% reduction in density. This is at the
limit of the sensitivities of techniques employed here.
Fit to Crab spectrum observed during January 2017, with response using
original density
was very good and there is no indication of any reduction in density.
For LAXPC20 the gain has been
slowly shifting downwards, which is the expected behavior due to impurities
accumulating in the detector. The resolution of this detector is also
deteriorating with time. The gas purification was attempted on August 18,
2016, but it did not change the gain or resolution. Further, attempt
needs to be made to purify the gas in LAXPC20. The gain appears to be
drifting at a rate of about 0.06 channels per day. For LAXPC10 purification
cycle on August 18, 2016 shifted the gain as expected, but the energy
resolution did not improve. Nevertheless, the resolution of LAXPC10
appears to be stable for the last one year. The resolution of LAXPC30
is improving with time, presumably due to lower pressure.

A large part of the gain shift can be adjusted by changing the linear term,
the coefficient $e_1$ in Eq.~\ref{eq:gain}, but other coefficients may also
be changing. Another measure of the change can be obtained by taking the
ratio of channel for the 60 keV and 30 keV peaks. The energy ratio of the
two peaks is close to 2 and for a linear gain the ratio in channels should
be 2 and any departure from this value could be due to the other two terms
in Eq.~\ref{eq:gain}. Figure~\ref{fig:a8n} shows the difference
$2-p_2/p_1$ where $p_1$ and $p_2$ are peak positions for the 30 keV and 60
keV peaks. It can be seen that for both LAXPC20 and LAXPC30 this ratio has
been decreasing in magnitude. Though for LAXPC20 it appears to have stabilized
after May 2016. Thus it is clear that the coefficients, $e_0$
and $e_2$ are also changing with time. It is unlikely that all the observed
variation can be accounted for variation in $e_0$ as that will require
a large variation. Thus, it appears that $e_2$ is decreasing with time
for these detectors.

To show the effect of long term variation in response, we have fitted the
observed spectrum for the Crab during different observations
using response which is adjusted for the shift in gain and for LAXPC30,
decrease in density and the results are shown in Figure~\ref{fig:crabres}.
The fit for LAXPC20 and LAXPC30 during August 2016 is poor due to shift in
gain and decreasing density, respectively.
There is also a general reduction in normalization for fit during this
observation which is most likely due to a difference in pointing.
For LAXPC30 the fitted power-law index, $\Gamma$, has also reduced. This is because the
fit with lowest $\chi^2$ with different densities has been selected. This
probably underestimates the density and $\Gamma$. The use of slightly higher
density increases $\Gamma$, but also increases the $\chi^2$.

Because of the leak, the efficiency of LAXPC30 is also changing as
shown in Figure~\ref{fig:eff30}. As a result of this, the background count
rate in LAXPC30 is decreasing with time. Figure~\ref{fig:back30} compares
the observed background during different times
in the three detectors. It can be seen that for LAXPC30 the count rate
is decreasing with time and the difference is more at high energies.
There is also some shift in the gain of LAXPC10 and LAXPC30
which also affected the background.
The shift in the gain for all detectors can be seen from
the position of the hump near 30 keV due to Xe K fluorescence X-rays.
The background in LAXPC10 has increased significantly during the last 6 months.
The increase is mainly seen at low energies in all anodes that are adjacent
to veto anode A10, which is disabled.
The reason for this increase is not clear.

Apart from observing the scheduled sources, LAXPC has also detected a number
of Gamma ray bursts.
The probability of occurrence of a GRB in the $1^\circ\times 1^\circ$ FOV of
LAXPC is very low (about 1/40000) and hence almost all the registered GRBs
are those that occur on the sides of the detectors. Gamma-rays from GRBs
have hard spectra extending to MeV and even beyond. The gamma-rays of
hundreds of keV and MeV strike the side walls of the detector and produce
secondary particles either in the walls of detector housing or in gas
volume. Some of these may get detected as genuine event.
As pointed out in Section 4.4, there is about 1\% probability of
high energy photons coming from random direction to register a valid event in the detector.
From time tagged events accurate time profile of the GRBs and be derived.
One of the events detected is shown in
Figure~\ref{fig:grb}. All detectors do not record the same count rate,
as depending on the direction of the burst with respect to satellite,
some detectors may be partially shielded by other material on the
satellite.
The bursts occurring during the SAA transit are obviously
missed. Similarly, those bursts which are occulted by the Earth are
also not seen.  The timing of these GRBs match that recorded in LAXPC
detectors. This gives us confidence in absolute timing accuracy at the
level of 1 s.
In principle, it is possible to obtain the spectrum
of the recorded events in LAXPC detectors during GRB but
the spectrum may not give
much information as most of these events are from secondaries produced
in the satellite. Since these photons are not coming from the top window,
the response matrix for the detector is not applicable to study these
events. Although LAXPC may not be useful for studying GRBs, the fact
that it can detect these bursts should be kept in mind as some of these
may be mistaken for bursts in the source being observed.

\section{Summary}

\begin{figure}
\epsscale{.95}
\plottwo{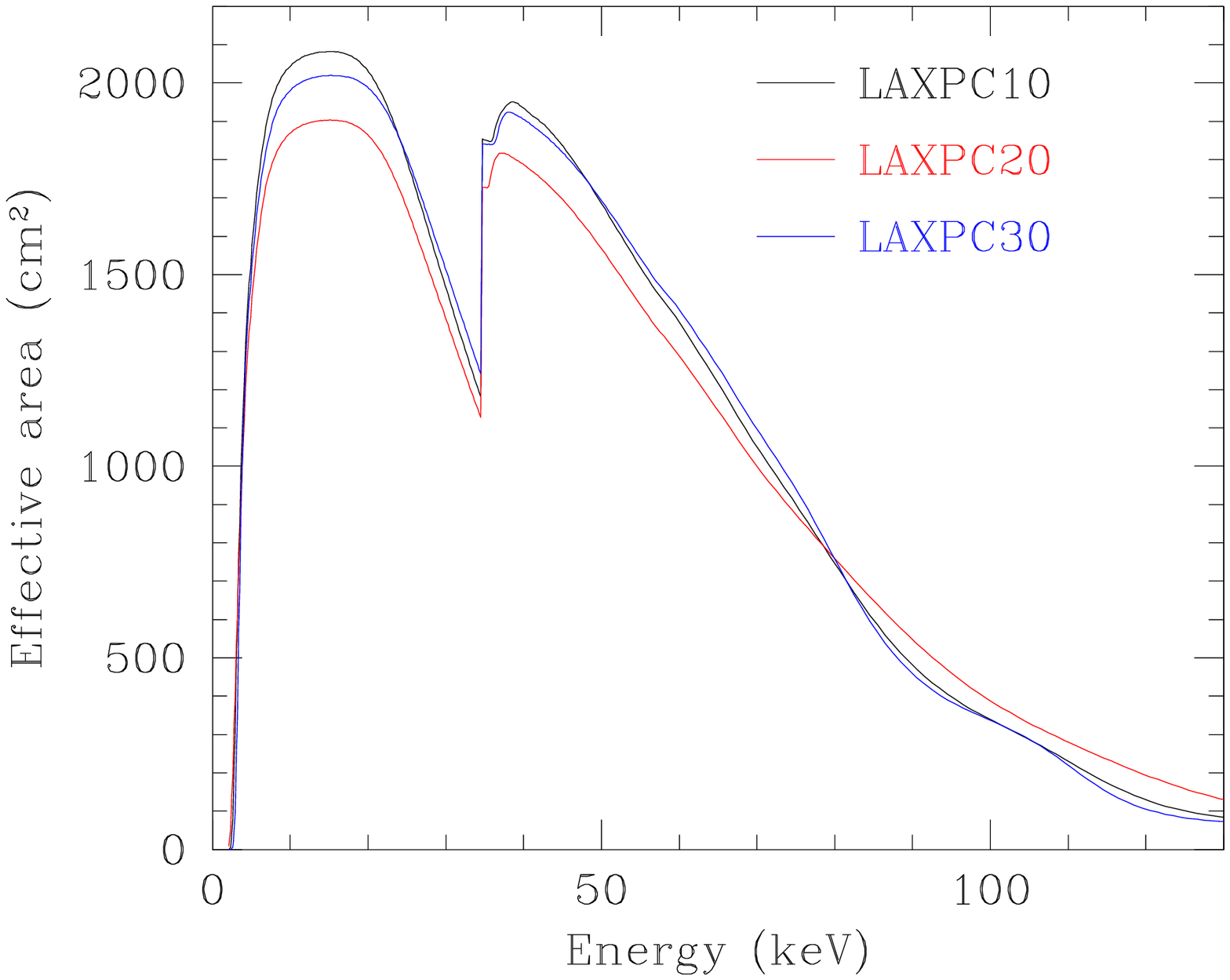}{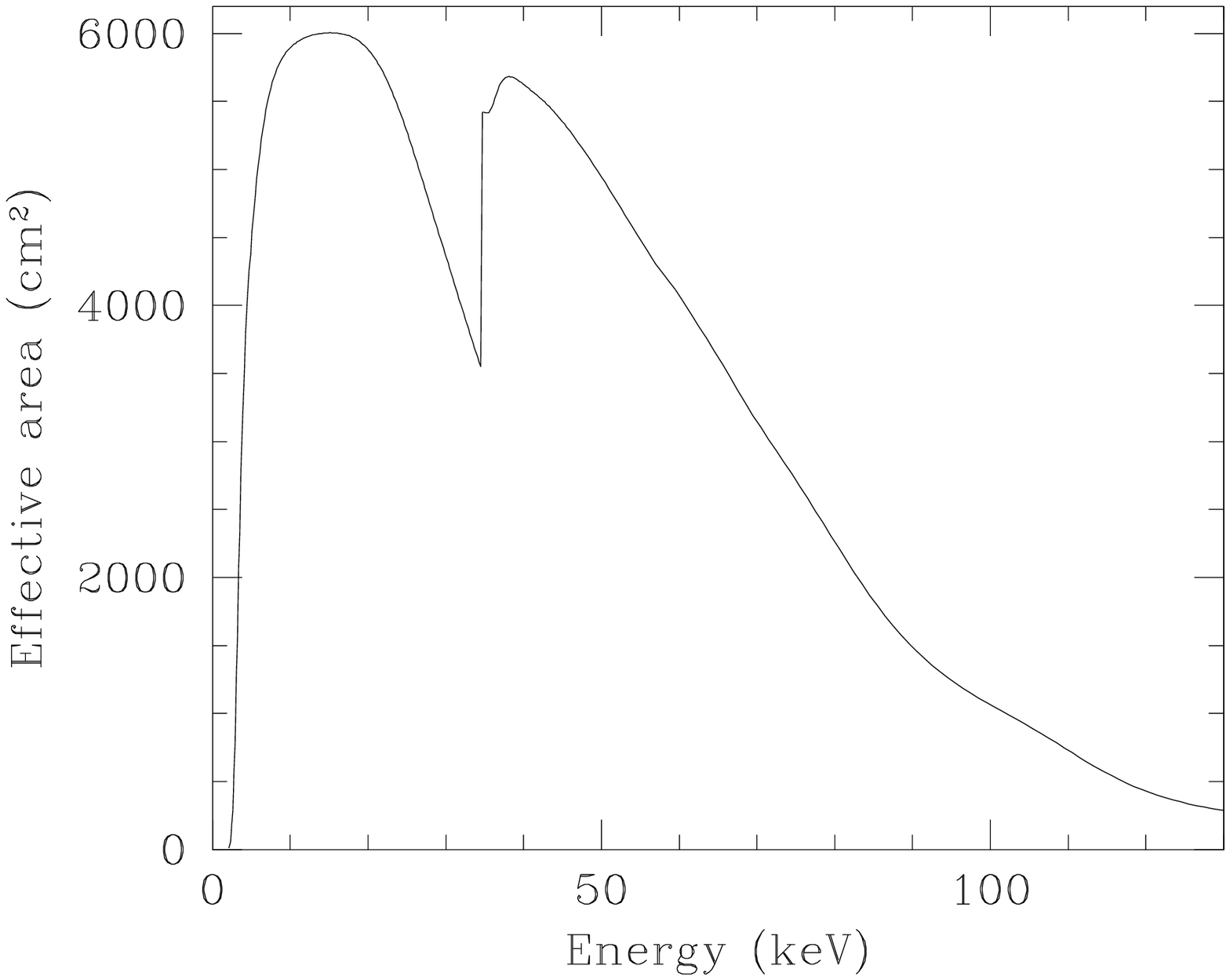}
\caption{The effective area of the three LAXPC detectors (left panel)
as estimated from
simulation with scaling determined by Crab observations after launch.
The right panel shows the effective area when all three detectors are
combined.}
\label{fig:eff}
\end{figure}

\begin{figure*}
\epsscale{.95}
\includegraphics[angle=-90,width=15 cm]{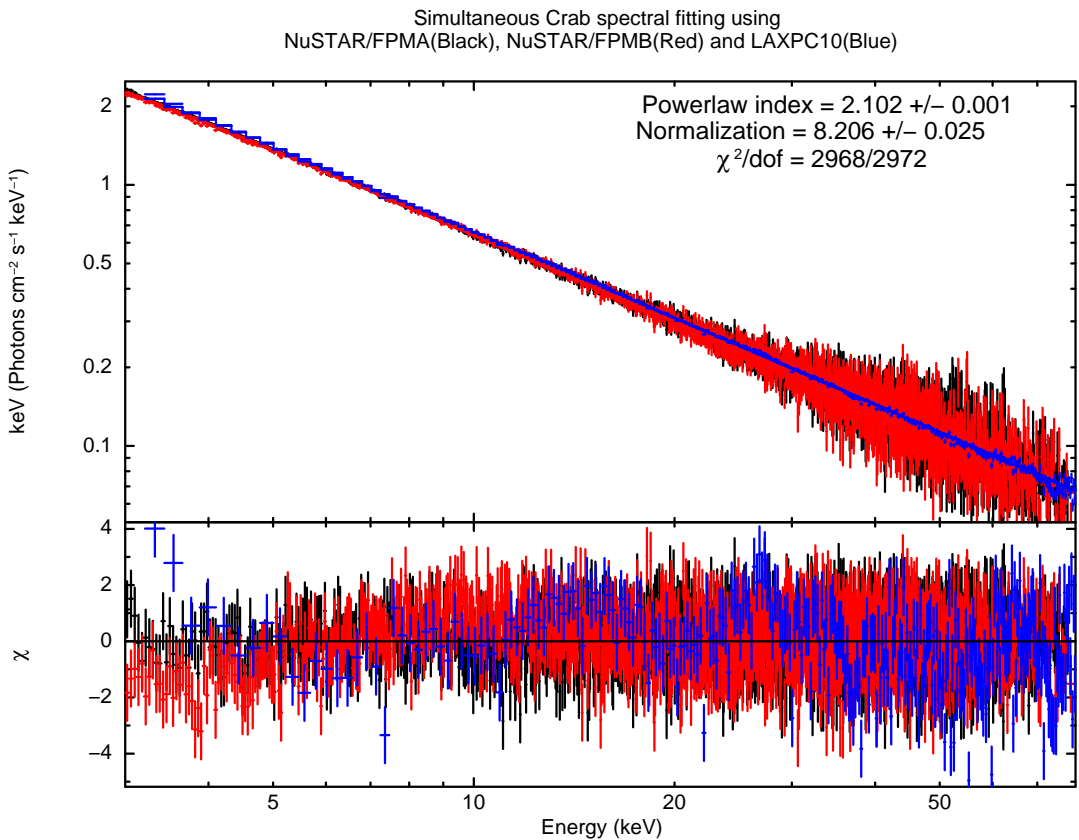}
\caption{The simultaneous fit to NuSTAR and LAXPC10 data for Crab.}
\label{fig:nustar}
\end{figure*}

\begin{figure}
\epsscale{.95}
\plottwo{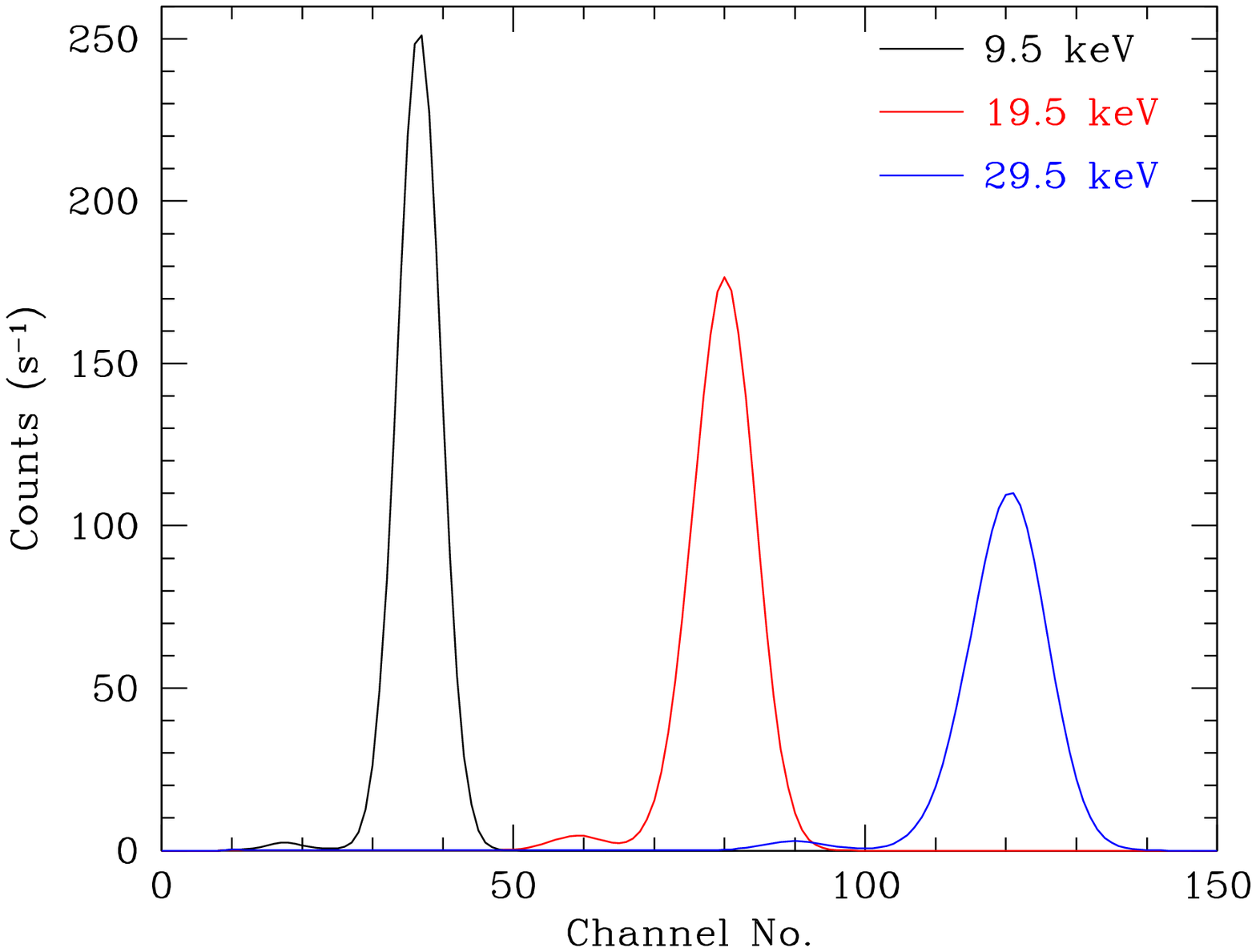}{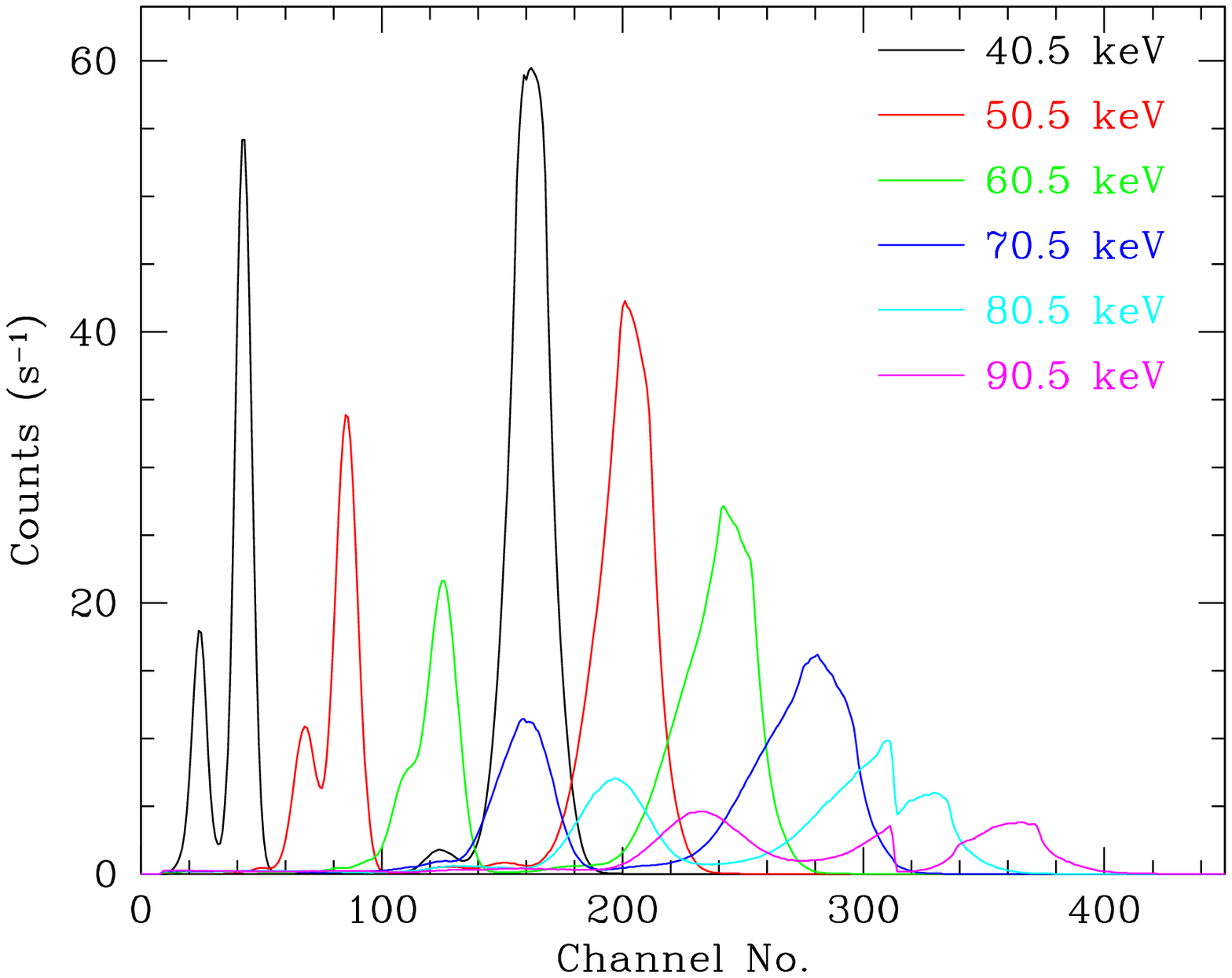}
\caption{The response of LAXPC30 for a few energies.}
\label{fig:resp}
\end{figure}

\begin{figure}
\epsscale{.45}
\plotone{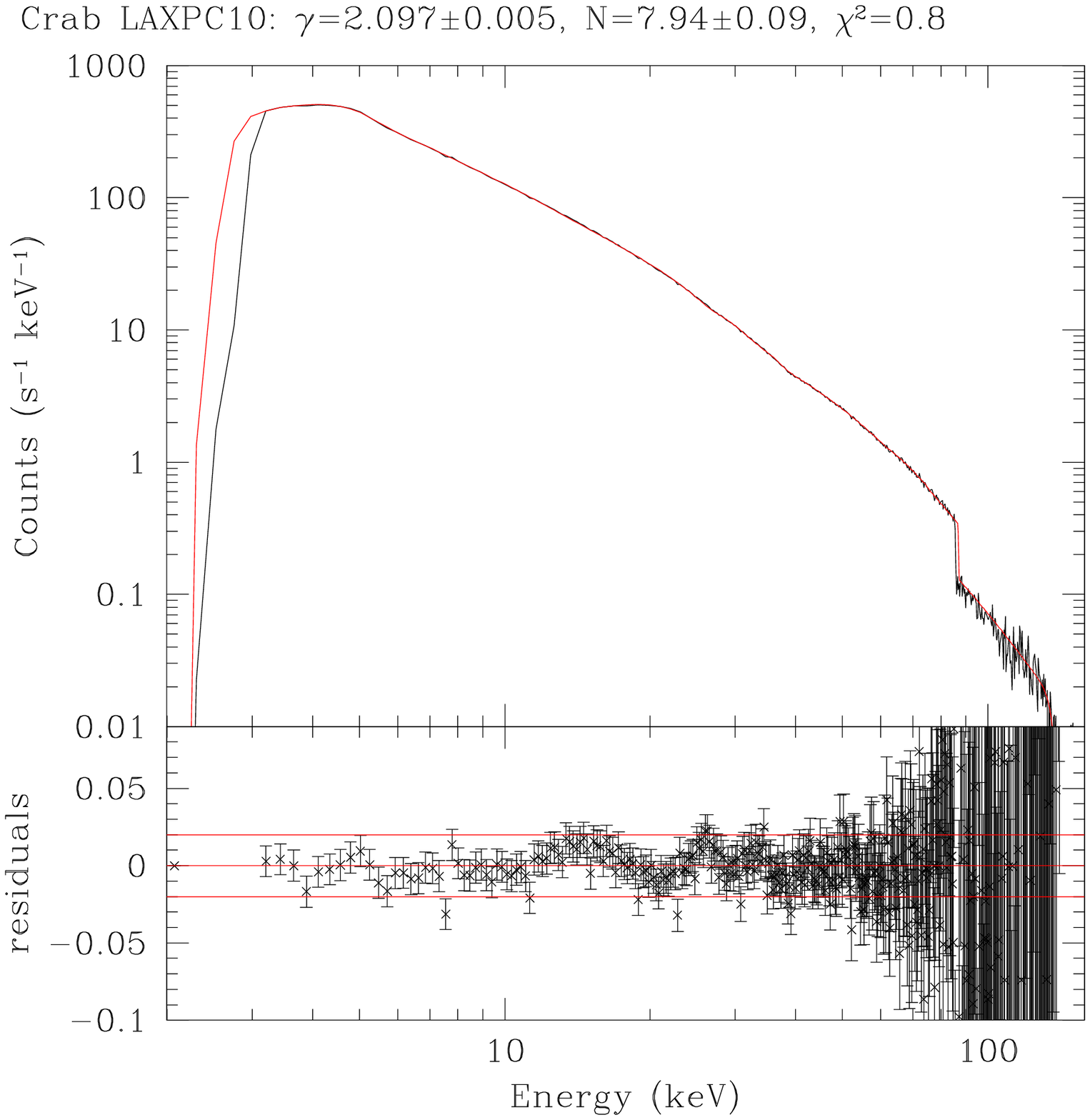}
\plotone{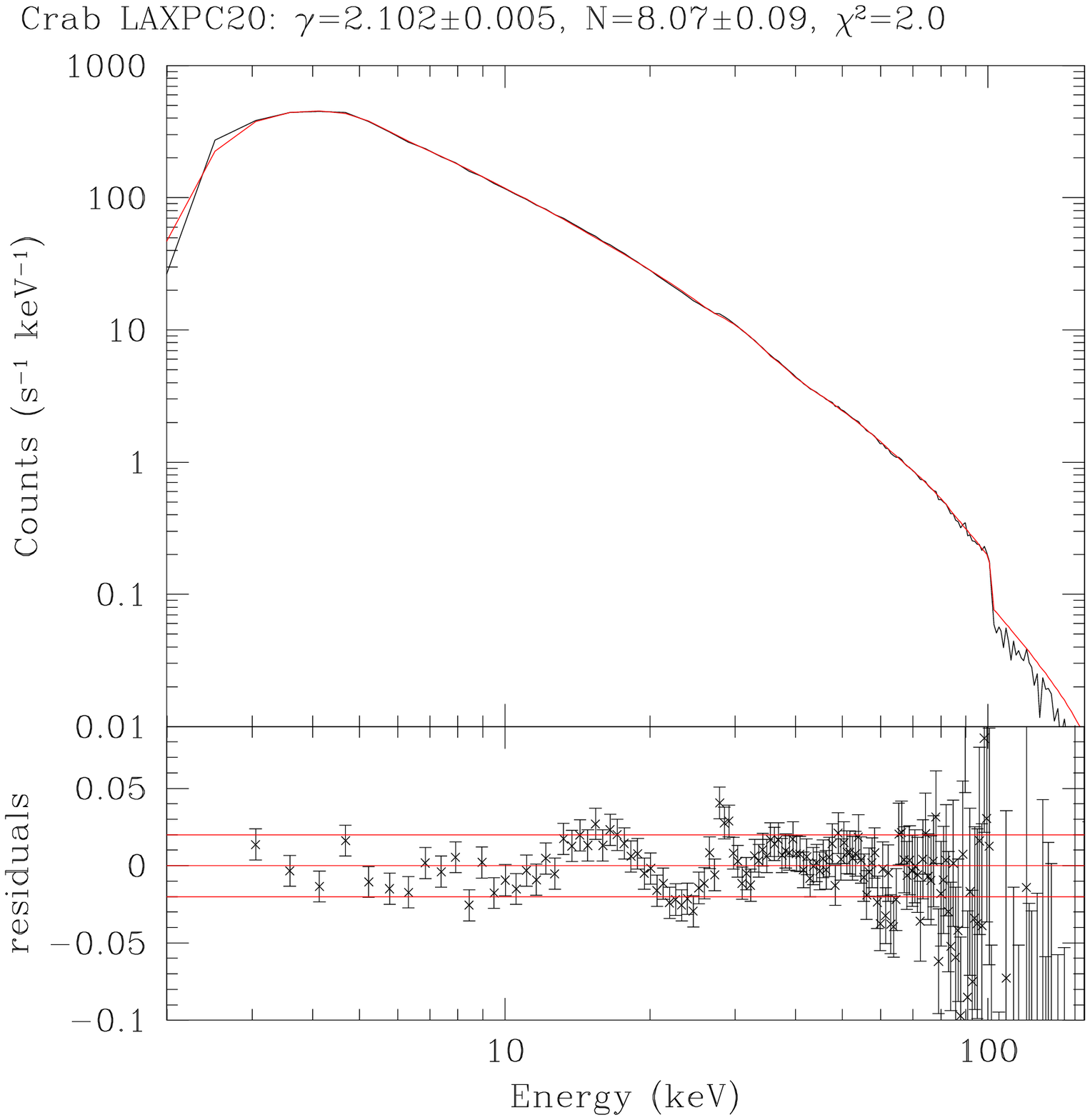}
\plotone{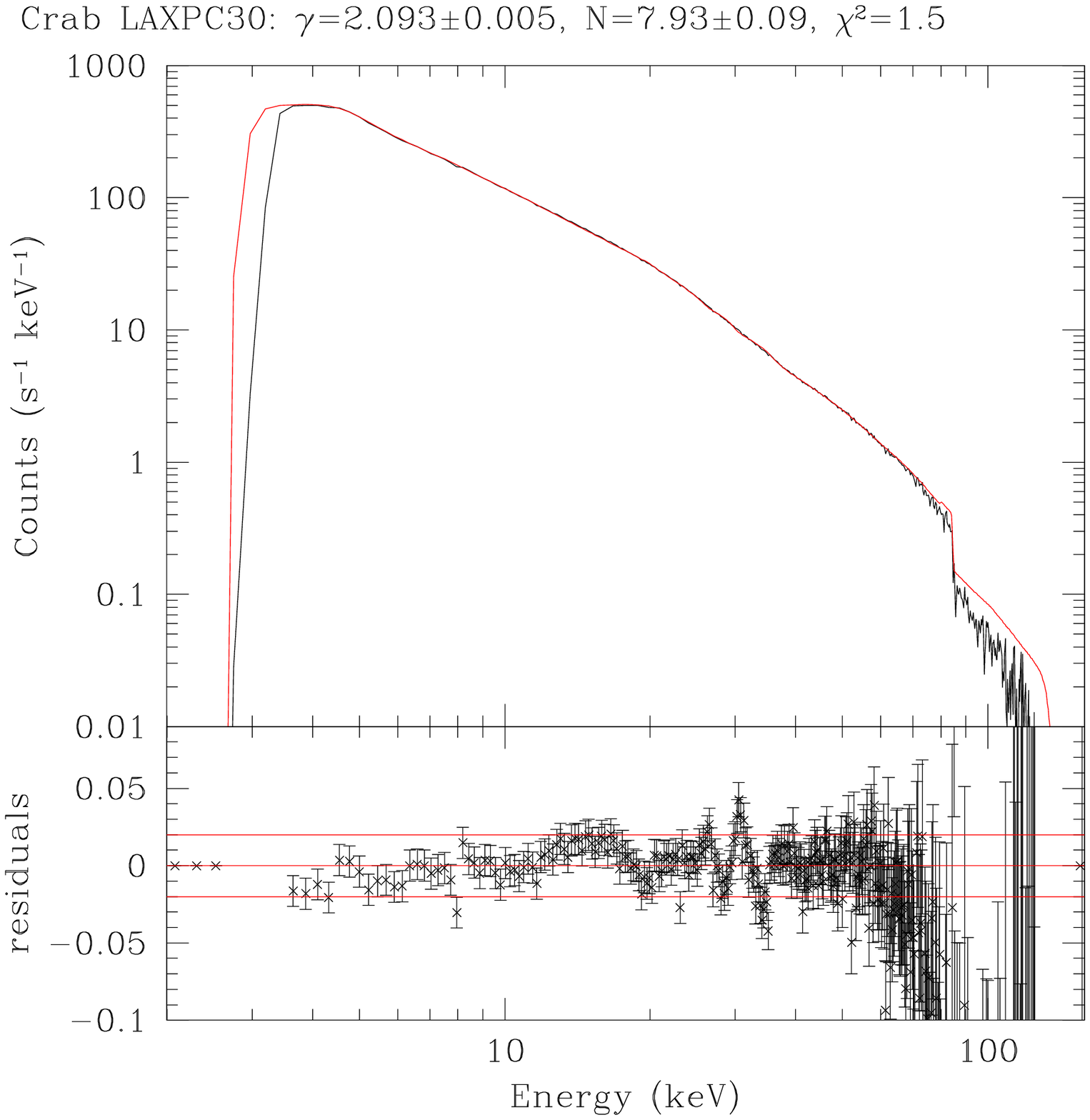}
\caption{The fit to Crab spectra for each LAXPC detector using a power-law
model with 1\% systematics.
The black lines show the observed spectra and red lines show the fitted
spectra.
The lower panel shows the relative difference between the observed and model
spectrum and the red lines mark the
region with differences of 2\%.}
\label{fig:crabfit}
\end{figure}

\begin{figure}
\epsscale{.95}
\plotone{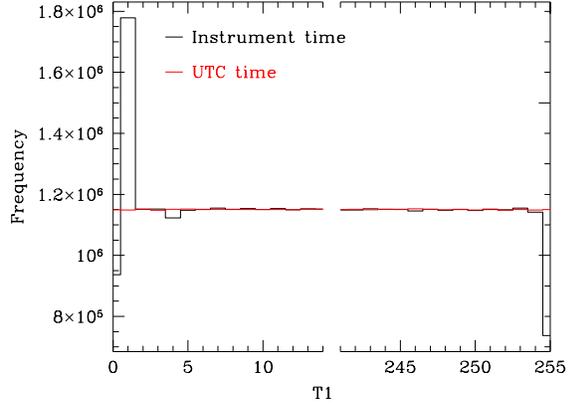}
\caption{Histogram of frequency of occurrence of different values of the
lowest time byte T1 (in units of 10 $\mu$s) during observation of Crab
by LAXPC10 in January 2016. The black line shows the histogram
using instrument time and red line shows the one when UTC time is used.}
\label{fig:hist}
\end{figure}

\begin{figure}
\epsscale{.95}
\plottwo{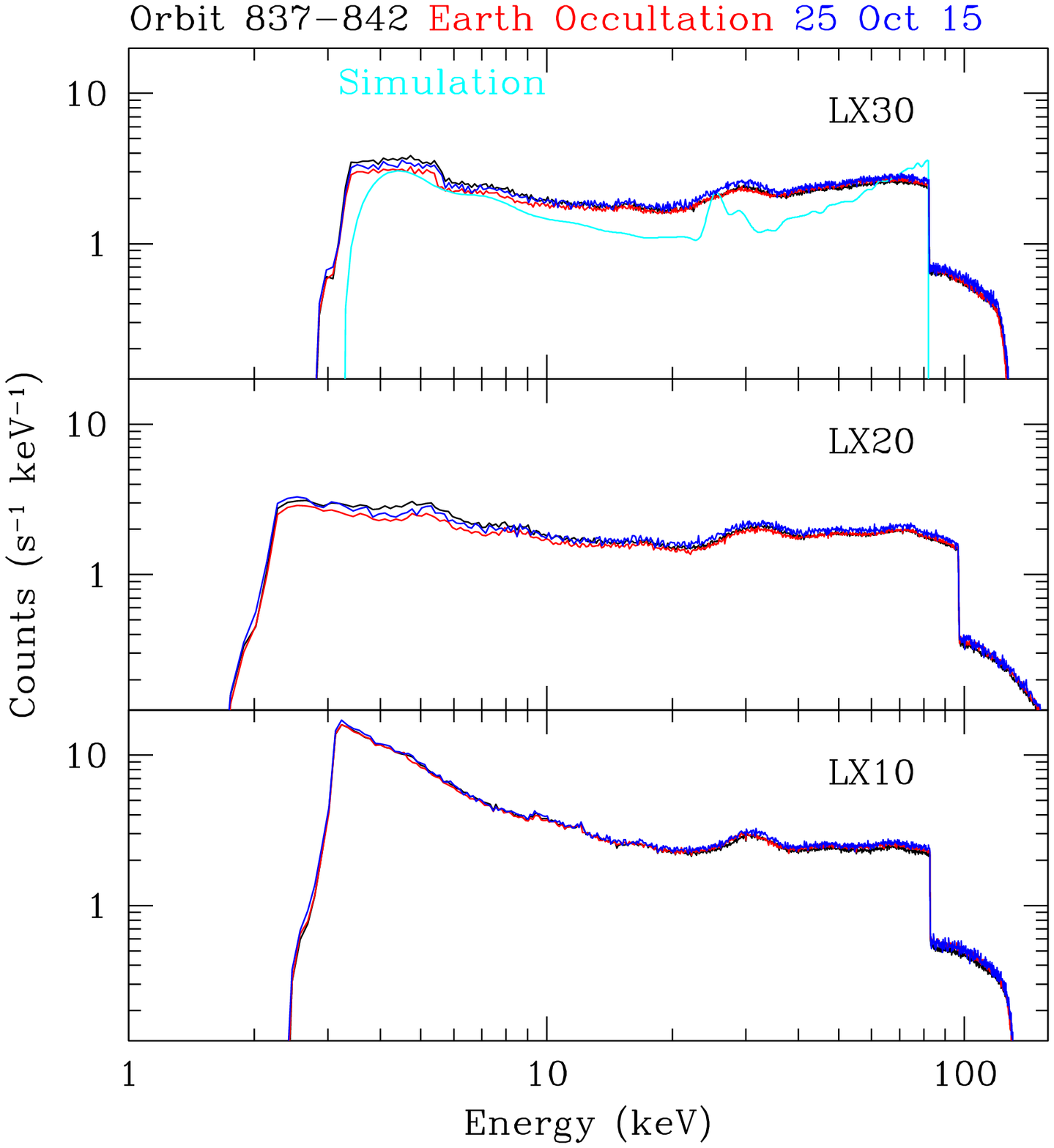}{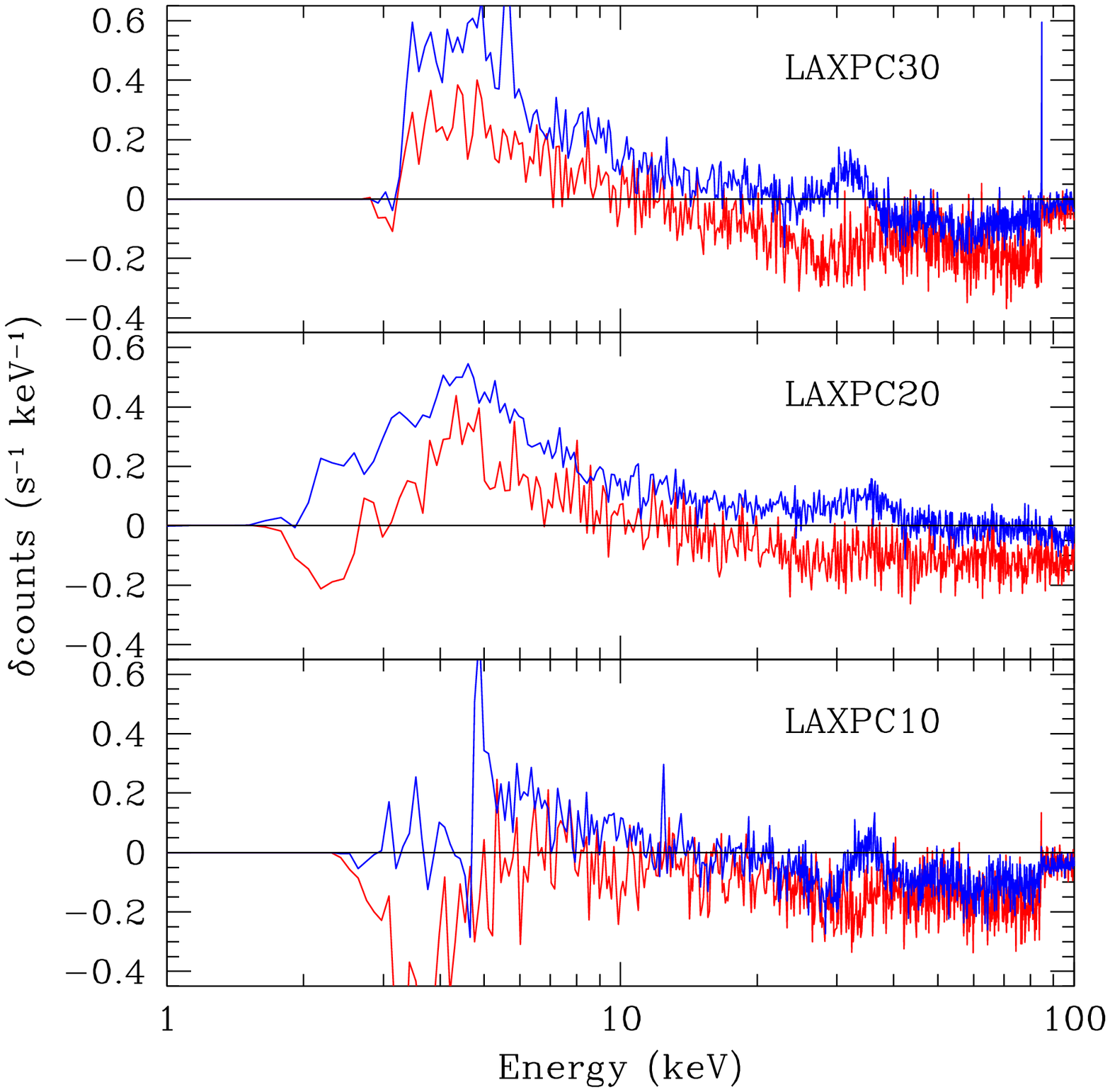}
\caption{The background spectra for all LAXPC detectors taken at three
different times in orbit is shown in the left panel. The black lines show the spectrum during
orbits 837 to 842 (November 23, 2015), the red lines show the spectrum during
the same period when the detectors were pointing towards the Earth, while
the blue lines show the spectrum during the first observation on October 25,
2015. The Cyan line in the top panel shows the simulated background obtained
before launch. The right panel shows the difference between the spectra. The red
lines show the difference between black and red lines in left panel (difference
between background and Earth occultation), while
the blue lines show the difference between black and blue lines (difference between
two backgrounds).}
\label{fig:back3}
\end{figure}

\begin{figure}
\epsscale{.95}
\plottwo{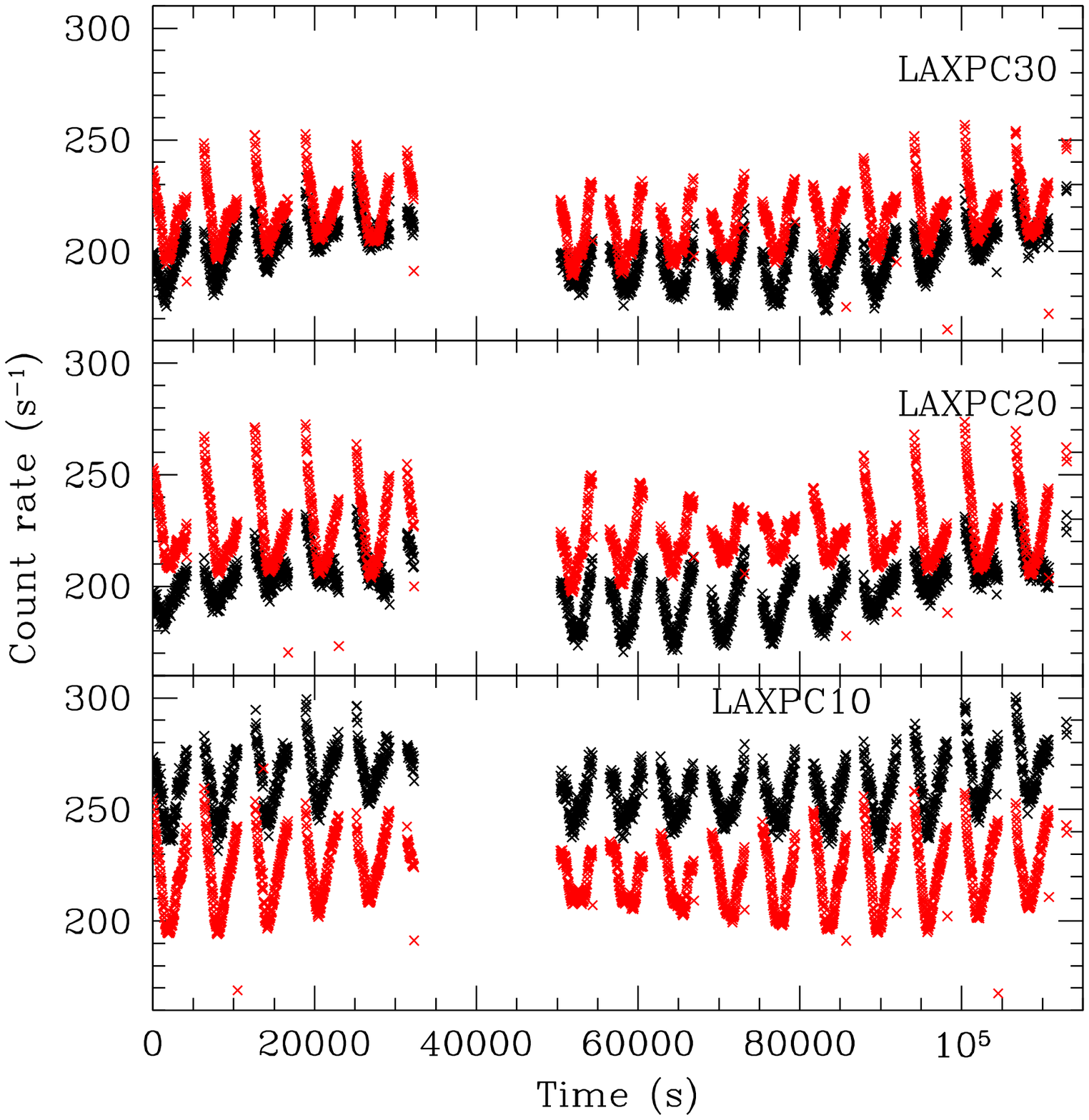}{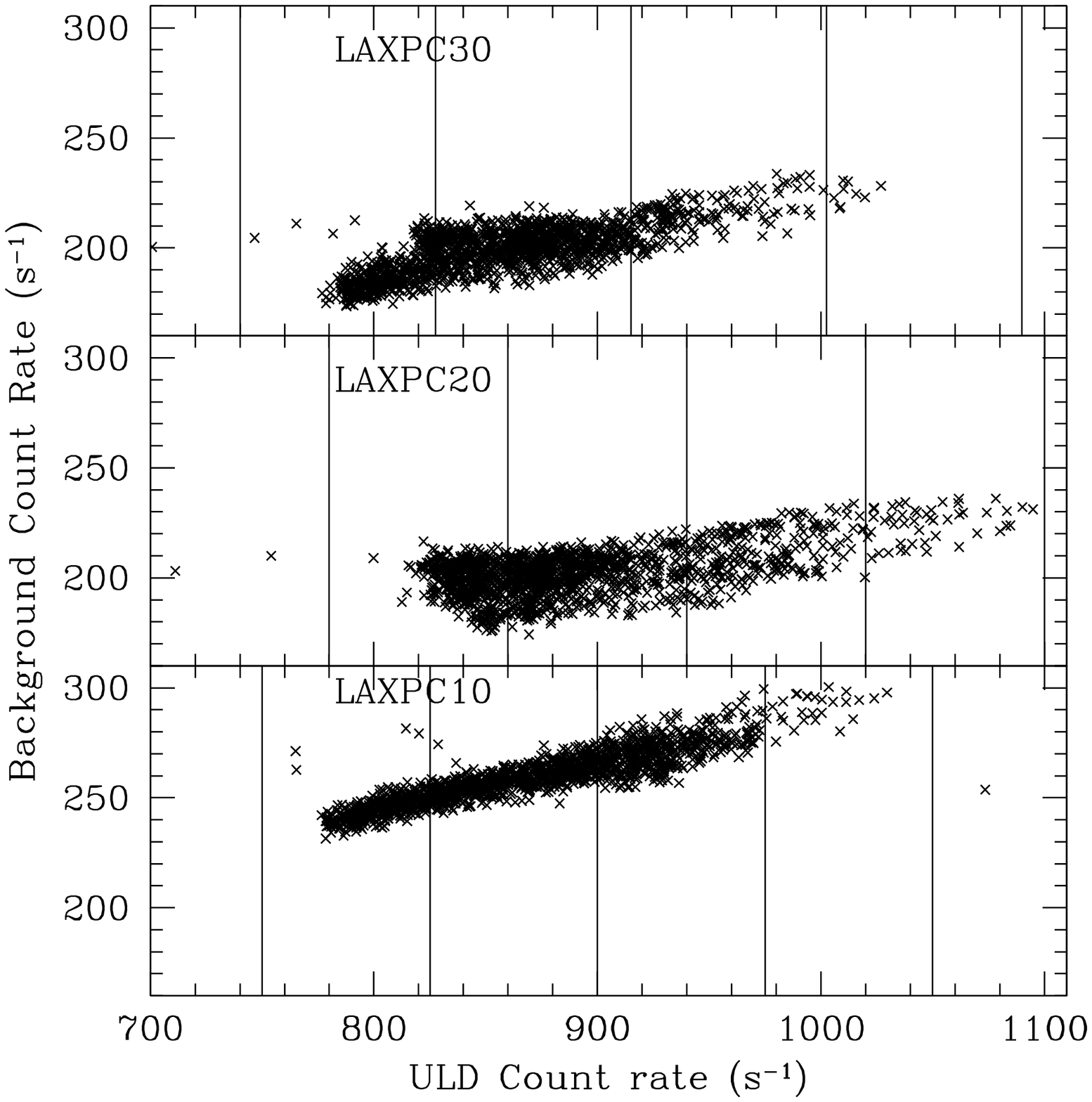}
\caption{Variation in count rate over about 1 day of background (black points) observation
in March 2016 is shown in the left panel. The ULD count rate (red points) is also shown.
The ULD rate has been divided by a factor of 4 to fit in the same scale.
The right panel shows the two count rates plotted against each other. The vertical
lines mark the 4 bins used to generate background spectra.}
\label{fig:back1}
\end{figure}

\begin{figure}
\epsscale{.95}
\plotone{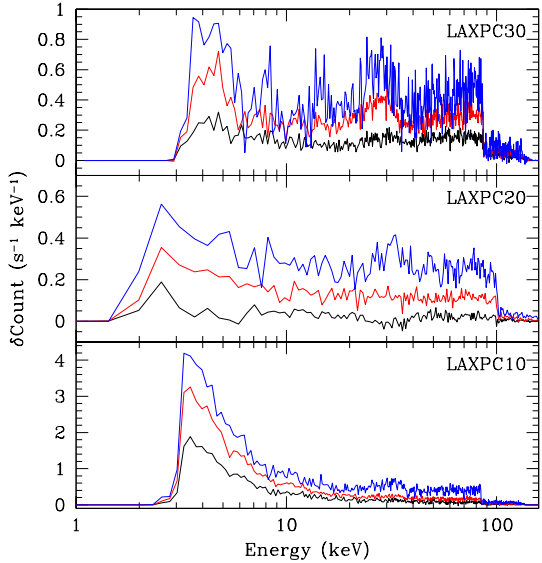}
\caption{The background spectra during different ULD bins for the three
detectors as observed during March 2016. For each detector the figure
shows the difference in count rate with respect to the lowest ULD bin.
The black lines show the difference $\mathrm{ULD2}-\mathrm{ULD1}$,
the red lines show $\mathrm{ULD3}-\mathrm{ULD1}$, while the blue lines
show $\mathrm{ULD4}-\mathrm{ULD1}$.}
\label{fig:uld}
\end{figure}

\begin{figure}
\epsscale{.95}
\plotone{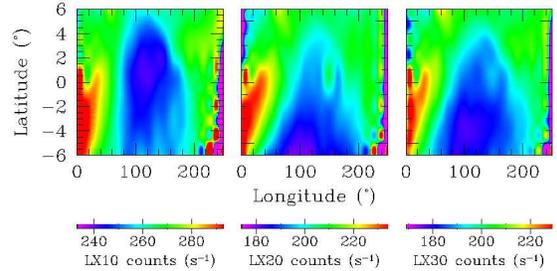}
\caption{The background count rate as a function of latitude and longitude
in all
detectors as observed during March 2016.}
\label{fig:backfig}
\end{figure}

\begin{figure}
\epsscale{.95}
\plotone{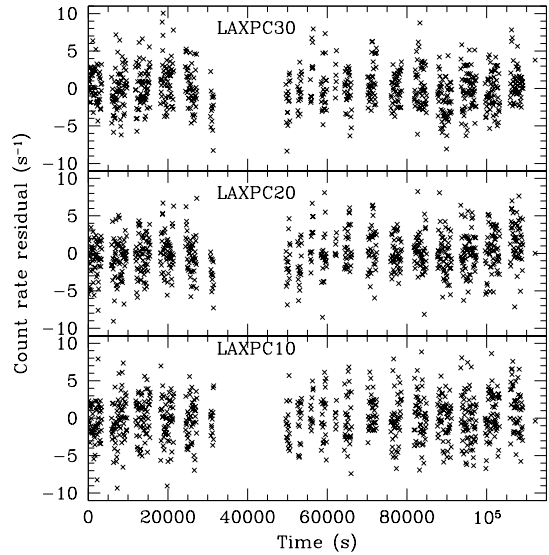}
\caption{The residual in the background after subtracting the fit for
observation during March 2016.}
\label{fig:backfit}
\end{figure}

\begin{figure}
\epsscale{.95}
\plotone{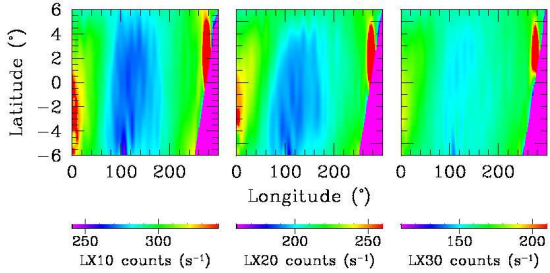}
\caption{The background count rate as a function of latitude and longitude
in all
detectors as observed during August 2016 after SAA criterion was changed.}
\label{fig:backfigsaa}
\end{figure}

\begin{figure}
\epsscale{.95}
\plotone{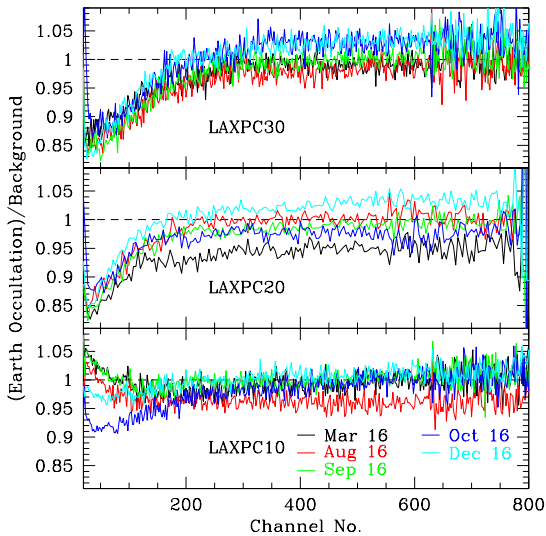}
\caption{The ratio of counts during Earth occultation and background
observations during different times.}
\label{fig:eo}
\end{figure}

\begin{figure}
\epsscale{.95}
\plotone{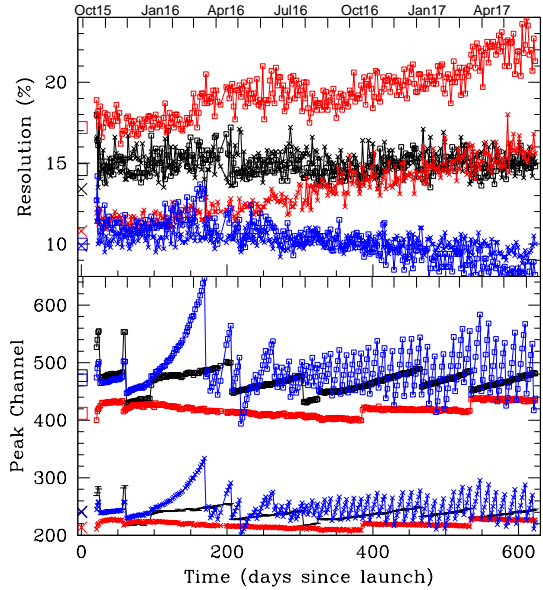}
\caption{The peak position and energy resolution of 30 and 60 keV peaks
as a function of time for all detectors as obtained from veto anode which
has the calibration source. The black, red and blue lines
respectively,  show the results
for LAXPC10, LAXPC20 and LAXPC30. The time is measured in days
from the launch of AstroSat on September 28, 2015.
The crosses
show the results for 30 keV peak and the open squares show those for 60 keV peak.
The point at $t=0$ shows the observed value during ground calibration.}
\label{fig:a8}
\end{figure}

\begin{figure}
\epsscale{.95}
\plotone{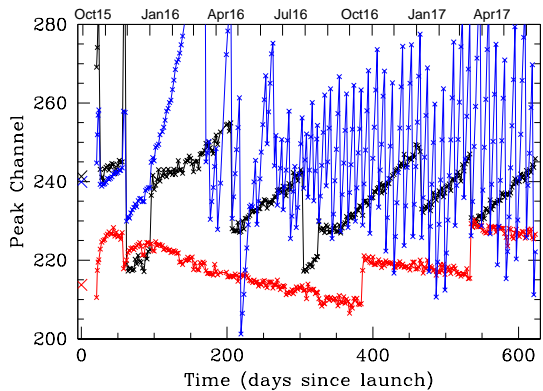}
\caption{The peak position of 30 keV peak
as a function of time for all detectors. The black, red and blue lines
respectively,  show the results
for LAXPC10, LAXPC20 and LAXPC30. The time is measured in days
from the launch of AstroSat on September 28, 2015.}
\label{fig:a8z}
\end{figure}

\begin{figure}
\epsscale{.95}
\plotone{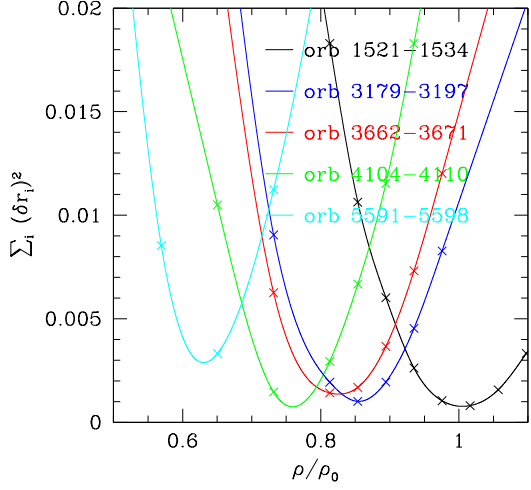}
\caption{The function $F_x(\rho)$ as defined in Eq.~\ref{eq:denx}, from five different observations of Cyg X-1
as a function of gas density in LAXPC30.}
\label{fig:denorb}
\end{figure}

\begin{figure}
\epsscale{.95}
\plotone{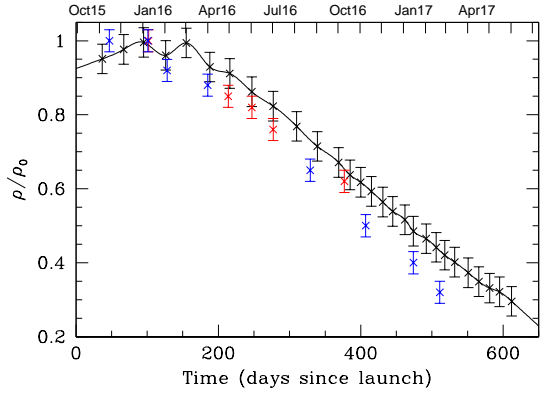}
\caption{The density of gas in LAXPC30 detector as a function of time using
different techniques. The black points show the result using on-board
pressure gauge which is normalized with respect to its maximum value,
the red points are those using observations for Cyg X-1 (Fig.~\ref{fig:denorb})
and blue points are using observations for Crab.}
\label{fig:lx30den}
\end{figure}

\begin{figure}
\epsscale{.95}
\plotone{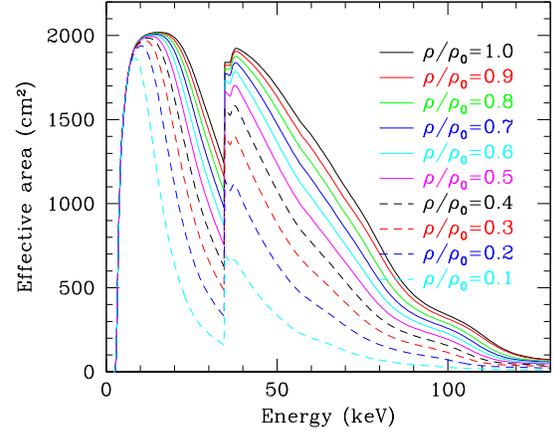}
\caption{The effective area of LAXPC30 with different values of gas density.
The values are in units of the original density.}
\label{fig:eff30}
\end{figure}

\begin{figure}
\epsscale{.95}
\plotone{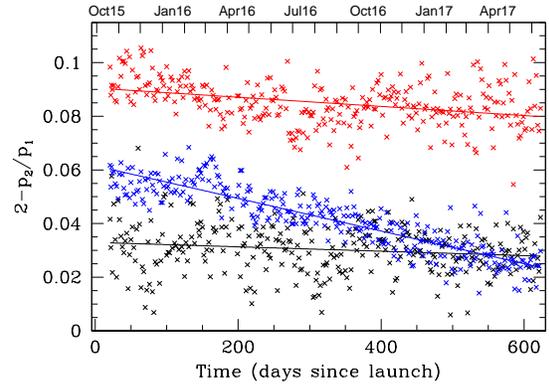}
\caption{The difference $2-p_2/p_1$, where $p_1$ and $p_2$ are positions
of 30 keV and 60 keV peaks, for the three LAXPC detectors
as a function of time for all detectors. The black, red and blue points
respectively,  show the results
for LAXPC10, LAXPC20 and LAXPC30. The time is measured in days
from the launch of AstroSat on September 28, 2015. The solid line shows the
linear fit to the points.}
\label{fig:a8n}
\end{figure}

\begin{figure}
\epsscale{.95}
\plotone{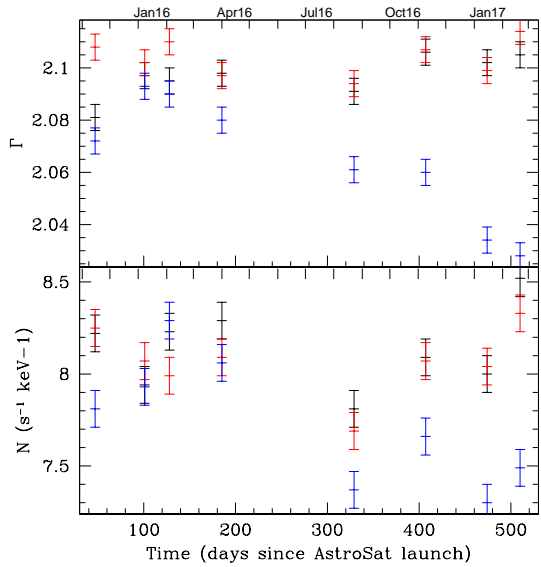}
\caption{Fit to Crab spectra observed at different times to a power-law model.
The black, red and blue points show the results for LAXPC10, LAXPC20 and
LAXPC30 respectively.}
\label{fig:crabres}
\end{figure}

\begin{figure}
\epsscale{.95}
\plotone{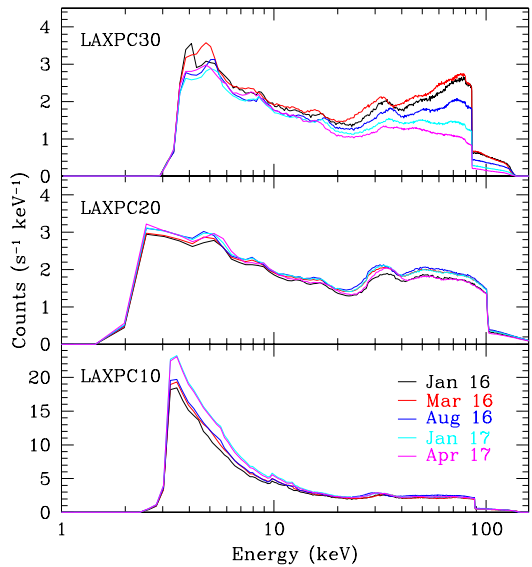}
\caption{The background spectra for all LAXPC detectors taken at five
different times is shown.}
\label{fig:back30}
\end{figure}

\begin{figure}
\epsscale{.85}
\plotone{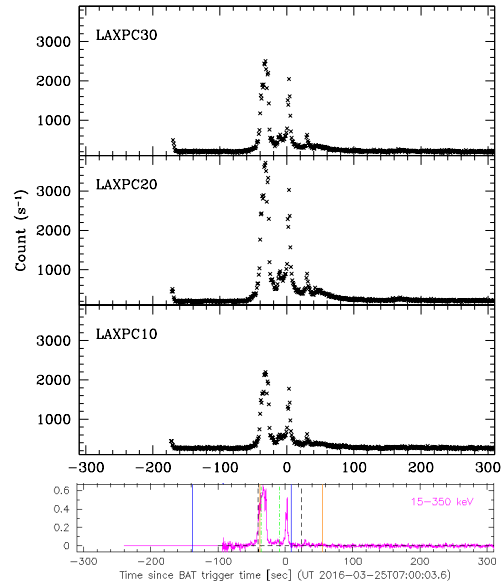}
\caption{The light curve for GRB160325A in the 3 LAXPC detectors.
For comparison the lowest panel shows the light
curve in Swift-BAT from
http://gcn.gsfc.nasa.gov/notices\_s/680436/BA/}
\label{fig:grb}
\end{figure}

AstroSat has completed more than 620 days of operation and more than
9200 orbits so far. More than 600 pointings for about 250 distinct sources have
been carried out. The response of the instrument is reasonably understood
and has been used to produce scientific results \citep{yad16b,mis17,jai17}.
The energy resolution of the detector at 30 and 60 keV is shown in
Figure~\ref{fig:a8} which shows that LAXPC10 energy resolution
is steady at $(15\pm2)$\%, while that in LAXPC20 is degrading slowly from
about 12\% just after launch to about 16\% currently at 30 keV. On the other
hand, the energy resolution of LAXPC30 has improved from about 11\% just after
launch to better than 10\% currently. The effective area of the detectors
is shown in Figure~\ref{fig:eff} and its normalization depends on the
pointing direction. The uncertainties in normalization
can be estimated by the results for
Crab shown in Figure~\ref{fig:crabres} which shows a variation
of $\pm5$\%. The LAXPC10 detector has a higher background counts as one of
the veto anodes is not functioning, while LAXPC30 has developed a leak and
hence its sensitivity is decreasing at high energies as shown in
Figure~\ref{fig:eff30}. The gain of all detectors is shifting with time and
is controlled in a reasonable range by adjusting the high voltage from time
to time. This shift
has to be accounted for in spectral analysis.

The background model
has uncertainties of up to 5\% because of variation between different
regions or satellite environment. The relatively large background in
LAXPC detectors and its variation with time limits the capability to observe faint sources.
The background uncertainty corresponds to about 10 counts per second, while
the Crab yields about 3000 counts per second in each detector,
which limits the study of faint sources.
It may not be possible to study sources fainter than a few mCrab. Similarly,
the variation in gain with time and understanding of detector response
limits the scope of spectral studies. At the same time, a large area and
consequently higher count rate and event analysis mode data allows
detailed timing studies to be carried out. While interpreting the power
spectrum possibility of instrumental features around 8--11 Hz in all
detectors and around 50 Hz in LAXPC20 should be considered.

\acknowledgments

We acknowledge the strong support from Indian Space Research Organization (ISRO) in various aspect of instrument building, testing, software development and mission operation during payload verification phase.
We  acknowledge support of TIFR central workshop during the design and testing of the payload.
We thank Dipankar Bhattacharya for arranging simultaneous observations with
NuSTAR to calibrate the LAXPC instrument. We thank Tomaso Belloni for pointing
out the feature around 50 Hz in the power spectrum of LAXPC20.



\clearpage

\begin{deluxetable}{lccc}
\tablecaption{Parameters, $e_0,e_1$ and $e_2$ (Eq.~\ref{eq:gain}) for the
three LAXPC detectors}
\tablewidth{0pt}
\tablehead{\colhead{Detector}&\colhead{$e_0$}&\colhead{$e_1$}&\colhead{$e_2$}}
\startdata
LAXPC10&$-3.46\pm0.05$&$8.936\pm0.010$&$-0.00198\pm0.00010$\\
LAXPC20&$+0.08\pm0.10$&$7.564\pm0.015$&$-0.00267\pm0.00010$\\
LAXPC30&$-7.67\pm0.05$&$9.022\pm0.015$&$-0.00197\pm0.00010$\\
\enddata
\label{tab:ei}
\end{deluxetable}

\begin{deluxetable}{lcccc}
\tablecaption{Pointing direction of each LAXPC detector as determined from the
Crab scan and the estimated loss of efficiency due to pointing offset
and collimator quality}
\tablewidth{0pt}
\tablehead{\colhead{Detector}&\colhead{RA}&\colhead{Dec}&\colhead{Offset}&
\colhead{loss of}\\
&\colhead{$(^\circ)$}&\colhead{$(^\circ)$}&\colhead{$(^\circ)$}&\colhead{efficiency}}
\startdata
Crab source&83.63&22.01\\
LAXPC10&$83.78\pm0.01$&$22.01\pm0.01$&0.15&21\%\\
LAXPC20&$83.63\pm0.01$&$22.08\pm0.01$&0.07&25\%\\
LAXPC30&$83.74\pm0.01$&$22.03\pm0.01$&0.11&23\%\\
Mean&83.72&22.04&0.09\\
\enddata
\label{tab:crab}
\end{deluxetable}

\begin{deluxetable}{lrrccccc}
\tablecaption{Blank sky coordinates observed by LAXPC}
\tablewidth{0pt}
\tablehead{\colhead{Target}&\colhead{RA}&\colhead{Dec}&\colhead{Date}
&\colhead{Exposure}&\multicolumn{3}{c}{{Count rate (s$^{-1}$)}}\\
&\colhead{$(^\circ)$}&\colhead{$(^\circ)$}&&(s)&\colhead{LAXPC10}&
\colhead{LAXPC20}&\colhead{LAXPC30}}
\startdata
Sky9&237.22&46.92&19 Oct 15&37465&195&205&195\\
Sky5&57.37&$-47.09$&21 Oct 15&18643&199&186&180\\ 
Sky2&136.35&26.46&23 Nov 15&43326&239&197&192\\ 
Sky8&236.80&70.35&05 Jan 16&12124&259&190&193\\
Sky9&237.22&46.92&14 Mar 16&48250&263&201&114\\ 
Sky9&237.22&46.92&23 Mar 16&48062&263&205&202\\ 
Sky9&237.45&47.26&16 Aug 16&39297&288&210&160\\ 
Sky5&57.37&$-47.09$&30 Aug 16&33809&290&212&161\\ 
Sky6&77.42&12.42&16 Sep 16&47162&296&212&143\\ 
Sky10&321.13&-48.53&16 Oct 16&42760&293&217&137\\ 
Sky6&76.13&12.71&03 Dec 16&45812&285&204&150\\
Sky3&129.32&-27.97&25 Dec 16&36986&280&205&117\\
Blank Sky&183.47&22.81&23 Jan 17&45312&293&206&130\\
Sky8&237.33&70.20&13 Feb 17&52701&286&209&129\\
Blank Sky&183.47&22.81&13 Apr 17&59094&290&191&111\\
Blank Sky2&180.01&35.19&29 May 17&60314&277&189&094\\
\enddata
\label{tab:back}
\end{deluxetable}

\begin{deluxetable}{lccccccc}
\tablecaption{Background count rates in LAXPC detectors in different
layers and energy bins during March 2016}
\tablewidth{0pt}
\tablehead{\colhead{Layer}&\colhead{Energy bin}&\colhead{LAXPC10}&\colhead{LAXPC20}&\colhead{LAXPC30}&\colhead{Total}\\
&\colhead{keV}&(s$^{-1}$)&(s$^{-1}$)&(s$^{-1}$)&(s$^{-1}$)}
\startdata
All&All&264&205&202&668\\
1&All&92.3&68.0&64.5&224.8\\
2&All&49.5&36.0&35.8&121.3\\
3&All&31.1&34.0&34.5&99.6\\
4&All&47.4&33.5&33.9&114.8\\
5&All&43.7&33.7&33.4&110.8\\
All&3--20&95.7&34.4&34.9&165.0\\
1&3--20&35.6&17.0&17.3&69.9\\
All&20--40&47.5&35.7&38.3&121.5\\
All&40--80&102&112&107&321\\
All&3--80&246&184&180&606\\
\enddata
\label{tab:backlay}
\end{deluxetable}

\begin{deluxetable}{lcccc}
\tablecaption{The background count rate in different ULD bins for each
detector on 14 March 2016}
\tablewidth{0pt}
\tablehead{\colhead{Detector}&\colhead{ULD-1}&\colhead{ULD-2}&\colhead{ULD-3}
&\colhead{ULD-4}\\
&\colhead{(s$^{-1}$)}&\colhead{(s$^{-1}$)}&\colhead{(s$^{-1}$)}&\colhead{(s$^{-1}$)}}
\startdata
LAXPC10&248&262&275&299\\
LAXPC20&201&203&214&230\\
LAXPC30&190&204&217&230\\
\enddata
\label{tab:uld}
\end{deluxetable}

\begin{deluxetable}{lcccc}
\tablecaption{High voltage adjustments carried out
on LAXPC detectors}
\tablewidth{0pt}
\tablehead{\colhead{Detector}&\colhead{operation}&\colhead{Date}&\colhead{Voltage}&\colhead{remarks}\\
&&&\colhead{(Volt)}}
\startdata
LAXPC10&HV&19 Oct 15&$2369\pm15$&First ON\\
LAXPC10&HV&23 Oct 15&$2334\pm15$&After purification\\
LAXPC10&HV&24 Oct 15&$2345\pm15$&After purification\\
LAXPC10&HV&28 Nov 15&$2331\pm15$&After purification\\
LAXPC10&HV&02 Jan 16&$2341\pm15$\\
LAXPC10&HV&21 Apr 16&$2328\pm15$\\
LAXPC10&HV&27 Jul 16&$2310\pm15$\\
LAXPC10&HV&06 Jan 17&$2300\pm15$\\
LAXPC10&HV&15 Mar 17&$2290\pm15$\\
LAXPC20&HV&23 Oct 15&$2608\pm15$&First ON\\
LAXPC20&HV&17 Oct 16&$2618\pm15$\\
LAXPC20&HV&15 Mar 17&$2628\pm15$\\
LAXPC30&HV&19 Oct 15&$2331\pm15$&First ON\\
LAXPC30&HV&23 Oct 15&$2317\pm15$&After purification\\
LAXPC30&HV&24 Oct 15&$2321\pm15$&After purification\\
LAXPC30&HV&28 Nov 15&$2314\pm15$&After purification\\
LAXPC30&HV&02 Jan 16&$2320\pm15$\\
LAXPC30&HV&17 Mar 16&$2260\pm15$\\
\enddata
\tablecomments{ After March 2016 LAXPC30 gain is adjusted regularly when required}
\label{tab:hv}
\end{deluxetable}


\clearpage

\clearpage

\clearpage

\clearpage

\clearpage

\begin{thebibliography}{}
\bibitem[Agostinelli et al.(2003)]{geant03} Agostinelli, S. et al.~2003, Nucl.\ Instrum.\ Meth.\ Phys.\ Res., A506, 250.
\bibitem[Agrawal(2006)]{agr06} Agrawal, P. C. 2006, Advances in Space Research, 38, 2989
\bibitem[Agrawal et al.(2017)]{agr17} Agrawal, P. C., Yadav, J. S., Antia, H. M. et al.~2017, J. Astrophy.\ Astron.\ (in press) arXiv:170506446
\bibitem[Antia(2012)]{ant12} Antia, H. M. 2012, ``Numerical Methods for Scientists and Engineers'',3rd edition, Hindustan book agency, New Delhi
\bibitem[Dean et al.(1991)]{dean91} Dean, A. J., Lei, F., Knight, P. J. 1991,
\ssr, 57, 109
\bibitem[Dias et al.(1991)]{dias91} Dias, T. H. V. T., Santos, F. P., Stauffer, A. D., Conde, C. A. N. 1991, Nucl.\ Instrum.\ Meth.\ Phys.\ Res., A307, 341
\bibitem[Dias et al.(1993)]{dias93} Dias, T. H. V. T., Santos, F. P., Stauffer, A. D., Conde, C. A. N. 1993, Phys.\ Rev.\ A, 48, 2887
\bibitem[Dias et al.(1997)]{dias97} Dias, T. H. V. T., dos Santos, J. M. F., Rachinhas, P. J. B. M., Santos, F. P., Conde, C. A. N., Stauffer, A. D., 1997, J. Appl.\ Phys., 82, 2742
\bibitem[Harrison et al.(2013)]{har13} Harrison, F. A., Craig, W. W., Christensen, F. E. et al.~2013, \apj, 770, 103
\bibitem[Jahoda et al.(2006)]{jah06} Jahoda, K., Markwardt, C. B., Radeva, Y., et al.~2006, \apj, 163, 401
\bibitem[Knoll(2000)]{kno00} Knoll, G. F. 2000, ``Radiation detection and measurement'', 3rd edition, Wiley, New York
\bibitem[Mandrou et al.(1979)]{man79} Mandrou, P., Vedrenne, G., Niel, M. 1979,
\apj, 230, 97
\bibitem[Misra et al.(2017)]{mis17} Misra, R, Yadav, J. S., Chauhan, J. V., et al. 2017, \apj, 835, 195
\bibitem[Rao et al.(1987)]{rao87} Rao, A. R., Agrawal, P. C., Manchanda, R. K., Shah, M. R. 1987, Adv.\ Space Res., 7, 129
\bibitem[Revnivtsev et al.(2003)]{rev03} Revnivtsev, M., Gilfanov, M., Sunyaev, R., Jahoda, K., Markwardt, C.~2003, \aa, 411, 329
\bibitem[Schonfelder et al.(1980)]{sch80} Schonfelder, V., Graml, F., Penningsfeld, F. P. 1980, \apj, 240, 350
\bibitem[Singh et al.(2014)]{sin14} Singh, K. P., Tandon, S. N., Agrawal, P. C. et al.~2014, Proc.\ SPIE, 9144, 91441S 
\bibitem[Verdhan Chauhan et al.(2017)]{jai17} Verdhan Chauhan, J., Yadav, J. S., Misra, R. et al.~2017, \apj, 841, 41
\bibitem[Yadav et al.(2016a)]{yad16a} Yadav, J. S., Agrawal, P. C., Antia, H. M., et al.~2016, Proc.\ SPIE, 9905, 99051D
\bibitem[Yadav et al.(2016b)]{yad16b} Yadav, J. S., Mishra, R., Chauhan, J. V. , et al.~2016, \apj, 833, 27
\bibitem[Yamaguchi et al.(2014)]{yam14} Yamaguchi, H., Badenes, C., Petre, R. et al.~2014, ApJ, 785, L27
\bibitem[Zhang et al.(1995)]{zh95} Zhang, W., Jahoda, K., Swank, J. H., Morgan, E. H., Giles, A. B. 1995, ApJ, 449, 930
\end{thebibliography}
\end{document}